﻿
 
\documentclass{aa} 

\usepackage{graphicx}
\usepackage{natbib} 
\usepackage{txfonts} 
\begin{document}

   \title{Data on 824 fireballs observed by the digital cameras \\ of the European Fireball Network in 2017--2018} 
   \subtitle{II. Analysis of orbital and physical properties of centimeter-sized meteoroids}

   \author{J.~Borovi\v{c}ka \inst{1} \and P.~Spurn\'y\inst{1} \and L.~Shrben\'y\inst{1}
          }

   \institute{Astronomical Institute of the Czech Academy of Sciences, Fri\v{c}ova 298, 25165 Ond\v{r}ejov, Czech Republic \\
              \email{jiri.borovicka@asu.cas.cz}      }

   \date{Received 7 June 2022, Accepted 6 September 2022}
   \titlerunning{Data on 824 fireballs observed by the European Fireball Network II}
   \authorrunning{Borovi\v{c}ka et al.}

 
  \abstract{Meteoroids impacting the Earth on a daily basis are fragments of asteroids and comets.
  By studying fireballs produced during their disintegration in the atmosphere, we can
  gain information about their source regions and the properties of their parent bodies. In this work, data on 824
  fireballs presented in an accompanying paper and catalog are used. We propose a new empirical  
  parameter for the classification of the physical properties of meteoroids, 
  based on the maximum dynamic pressure suffered by the meteoroid in the atmosphere. 
  We then compare the physical and orbital properties of meteoroids. We find that aphelion distance
  is a better indicator of asteroidal origin than the Tisserand parameter.  Meteoroids with aphelia
  lower than 4.9 AU are mostly asteroidal, with the exception of the Taurids and $\alpha$ Capricornids
  associated with the comets 2P/Encke and 169P/NEAT, respectively. 
  We found another population of strong meteoroids of probably asteroidal origin on orbits with either high eccentricities
  or high inclinations, and aphelia up to $\approx$\,7 AU. Among the meteoroid streams, the Geminids
  and $\eta$ Virginids are the strongest, and Leonids and $\alpha$ Capricornids the weakest.
  We found fine orbital structures within the Geminid and Perseid streams. Four minor meteoroid
  streams from the working list of the International Astronomical Union were confirmed. No meteoroid
  with perihelion distance lower than 0.07 AU was detected. Spectra are available for some of
  the fireballs, and they enabled us to identify several iron meteoroids and meteoroids deficient in sodium.
  Recognition and frequency of fireballs leading to meteorite falls is also discussed.}

{}
{}
{}
{}
{}

   \keywords{Meteorites, meteors, meteoroids -- Catalogs}

   \maketitle
%

\section{Introduction}

This paper is the second of two papers devoted to the fireballs observed by the
European Fireball Network in 2017 -- 2018. The first paper \citep[][hereafter Paper I]{paper1}
presented the motivation for the work, a brief history and the current status of the
European Fireball Network, a description of the main instrument of the network, the Digital
Autonomous Fireball Observatory (DAFO), the methods of data analysis, an explanation
of all catalog entries, and a statistical evaluation of the whole sample of 824 fireballs.
The catalog itself is available at the
\textit{Centre de Donn\'ees astronomiques de Strasbourg} (CDS).

As shown in Paper I, the network is capable of detecting meteors brighter
than absolute magnitude of about $-2$. Consequently, meteoroids of masses from about 5 grams
are detected for all entry velocities. In cases of high entry velocity ($\sim 70$ km s$^{-1}$),
meteoroids with masses as low as 0.05 g can be detected. On the other end of the mass range, 
the brightest observed fireball in
the two year period reached a magnitude of $-14$ and the largest meteoroid had 
a mass of slightly more than 100 kg. From the total sample of 824 fireballs, 222 belonged to one of 16
major meteor showers. Most represented were the Taurids, Perseids, and Geminids, which together
accounted for 164 fireballs.

The cataloged data for each fireball include date and time, atmospheric trajectory, 
initial and terminal velocity, radiant,
heliocentric orbit, suggested shower membership, possible related body (asteroid or comet),
maximum dynamic pressure along the trajectory, and terminal dynamic mass. 
Photometric data are given for all fireballs except for two that were observed during twilight. They include 
maximum brightness, total radiated energy, and derived quantities such as initial photometric mass 
and physical classification (see below). Information on the availability of spectral data is also provided
and unusual spectra (irons, and sodium-deficient and sodium-rich spectra) are marked.

The purpose of this paper is to analyze the obtained fireball data.
 The main goal is to compare the orbital and physical properties of meteoroids. First, we derive a new measure
of meteoroid physical properties based on the maximum dynamic pressure, which we compare
with the classical $P_E$ criterion of \citet{PE} based on fireball end height. We evaluate orbits
according to the Tisserand parameter and aphelion distance, but considering also other orbital elements.
We search the orbits occupied mostly by structurally strong (asteroidal) and structurally weak (cometary) meteoroids. 
We compare the physical properties of meteoroids from different showers, and we discuss the extreme orbits with low perihelion distances 
or eccentricities above unity.

To further exploit the data, we study the radiants and orbits of meteor showers. Previous
work on the Taurids \citep{Spurny_Taurids} have shown that the precision of data from DAFO is higher than 
that of other systems, and new structures within meteoroid streams can be revealed. We also examine
if some not yet established meteor showers can be confirmed. 
The last two discussed topics are the identification of probable meteorite falls  and relation of some meteoroids to specific asteroids.

   \begin{figure}
   \centering
   \includegraphics[width=1.0\columnwidth]{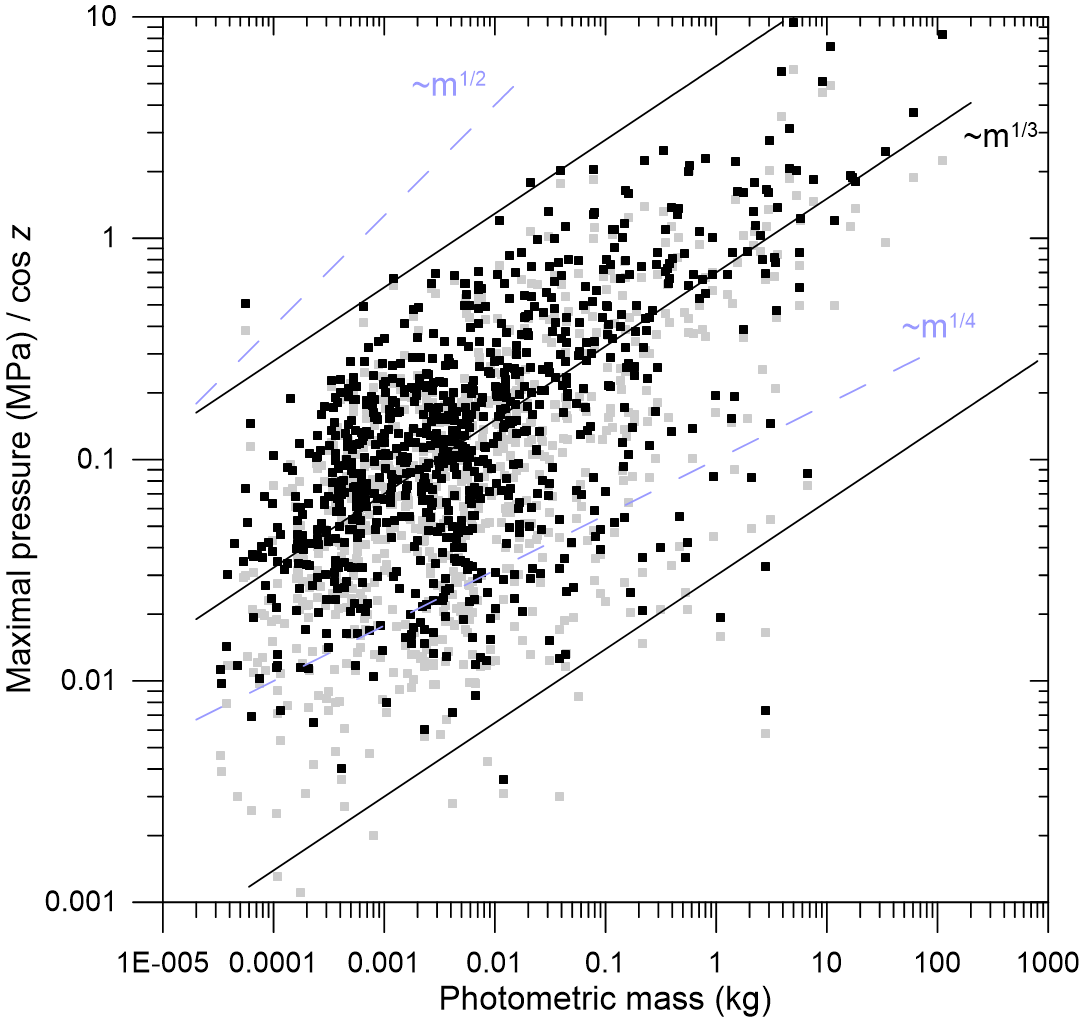}
   \caption{Maximum dynamic pressure divided by the cosine of radiant zenith distance as a function of photometric mass
   for 822 fireballs. 
   The light gray symbols represent the pressures without the $\cos z$ factor for the same fireballs. 
   Solid lines represent the dependency on the cube root of mass ($m^{1/3}$) for the bulk of the data, the
   upper envelope (the strongest meteoroids), and the lower envelope (the weakest meteoroids). Dashed lines
   show dependencies on $m^{1/2}$ and $m^{1/4}$ for comparison. These dependencies do not correspond to the data.}
   \label{mass-pres}
   \end{figure}

\section{Alternative physical classification of fireballs}
\label{Pfsection}

The $P_E$ criterion of \citet{PE}
has traditionally been used for fireball classification, that is for the evaluation of the physical properties of meteoroids.
As explained in Paper I, it is computed as
\begin{equation}
P_E = \log \rho_{e} -0.42 \log m_{\rm phot76} +1.49 \log v_\infty -1.29 \log \cos z_R,
\label{PE}
\end{equation}
where $\rho_{e}$ is the atmospheric density at the fireball end height in g cm$^{-3}$, 
$m_{\rm phot76}$ is the photometric mass in grams computed using
the original luminous efficiency, $\tau_{76}$ (see Paper I for details), 
$ v_\infty$ is the entry velocity in km s$^{-1}$, and 
$z_R$ is the zenith distance of the apparent radiant.
Four fireball types, I, II, IIIA, and IIIB (from the strongest to the weakest), were defined, the
boundary $P_E$ values being $-4.60$, $-5.25$, and $-5.70$. 

Other criteria were also proposed. \citet{Cep88} mentioned the $A_L$ criterion, which is equivalent to
$P_E$, but used the total radiated light instead of the photometric mass. He also listed the ablation coefficient as 
a possible classification criterion, since the ablation coefficient is smallest for type I fireballs and largest
for type IIIB fireballs. \citet{Cep93} proposed a two-dimensional classification with dynamic pressure at the
fragmentation point forming the second dimension. The fragmentation point was found from a careful
analysis of deceleration, requiring precise dynamic data. \citet{Trigo_strength}, on the other hand, 
used the dynamic pressure at the point of maximum brightness to estimate the strength of cometary
meteoroids. \citet{Taurid_phys} and \citet{2strengths} used deceleration and detailed radiometric light curves
to reveal multiple fragmentation points along fireball trajectories, and to study strength distribution inside 
meteoroids.

Fragmentation analysis is a time-consuming process and requires detailed data. The ablation coefficient was
computed for many fireballs during the physical fit of velocity 
 but it often has large 
uncertainty and for short fireballs it could not be computed at all. The maximum dynamic pressure, on the other hand,
was computed for all fireballs. We were therefore interested to know if a simple fireball classification can be build
on the basis of maximum dynamic pressure, $p_{\rm max}$, instead of end height. One reason is that
$p_{\rm max}$ is a more robust quantity, with end height being more dependent on observing circumstances.

   \begin{figure}
   \centering
   \includegraphics[width=0.95\columnwidth]{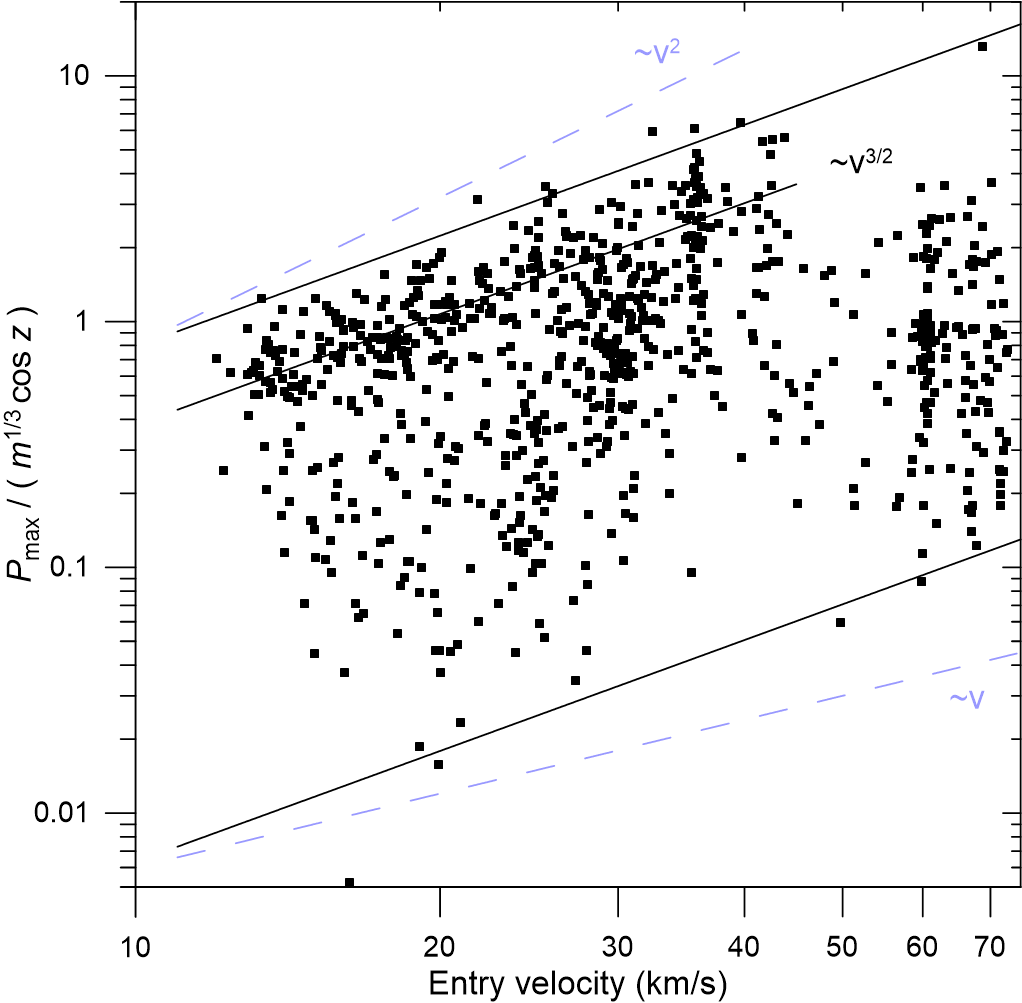}
   \caption{Maximum dynamic pressure divided by the cosine of radiant zenith distance and cubic root of mass as a function of 
   entry velocity for 822 fireballs. 
  Solid lines represent the dependency on the 1.5th power of velocity ($v^{3/2}$) for the bulk of slow fireballs, the
   upper envelope (the strongest meteoroids), and the lower envelope (the weakest meteoroids). Dashed lines
   show dependencies on $v^2$ and $v$ for comparison. These dependencies do not correspond to the data.}
   \label{vinf-pres}
   \end{figure}

It can be expected that $p_{\rm max}$, similarly to end height, will depend on the trajectory slope, since under
otherwise equal conditions, meteoroids on steep trajectories will penetrate deeper than meteoroids on shallow
trajectories. Therefore, $p_{\rm max}/\cos z$, where $z$ is the average zenith distance of the radiant,
is plotted in Fig.~\ref{mass-pres} as a function of photometric mass. It can be seen that the
maximum pressure reached, not surprisingly, increases with meteoroid mass, $m$. One can indeed expect that larger pieces of a given
material will penetrate deeper. The trend of the data, including the upper and the lower envelopes (not considering
a few outliers), follows a dependency  $p_{\rm max}/\cos z \sim m^{1/3}$. 
Figure \ref{mass-pres} also shows data without the $\cos z$ factor. It can be seen that the inclusion
of  $\cos z$ makes the range of pressures narrower. The lowest observed maximal pressures obviously belong
to fireballs on shallow trajectories.

To reveal material properties from dynamic pressure, meteoroid mass must, therefore, be taken into account.
The combined quantity  $p_{\rm max}/(m^{1/3}\cos z)$ is plotted in Fig.~\ref{vinf-pres} as a function
of entry velocity, $v$. One might expect no dependence on velocity because velocity is already contained
in the computation of dynamic pressure. The plot, nevertheless, shows a clear dependence on velocity,
well approximated by $\sim v^{3/2}$. 
It is not the purpose of this paper to explain the physics behind the dependency. 
It can be expected that the final fragmentation occurs when the dynamic pressure reaches
the material strength. It is possible that fireballs with higher velocities have a higher chance of reaching an even higher dynamic
pressure before the fragments are finally decelerated.

Nevertheless, by finding the dependence on mass and velocity, we can now define a new parameter
for the evaluation of material strength based on the maximal dynamic pressure. We will call it the ``pressure resistance factor''
or simply the ``pressure factor'', 
abbreviated as $P\!f$. It is defined as
\begin{equation}
P\!f = 100\ p_{\rm max} \,\cos^{-1}\! z\ \ m_{\rm phot}^{-1/3} \ \ v_\infty^{-3/2},
\label{Pf}
\end{equation}
where $p_{\rm max}$ is the maximal dynamic pressure in MPa, $z$ is the average zenith distance of the radiant
(i.e.,\ trajectory slope from the vertical), $m_{\rm phot}$ is the initial photometric mass in kg (see Paper I for the used
luminous efficiency), and $v_\infty$ is the entry velocity in km s$^{-1}$.
As an example, $P\!f$ is equal to one for a meteoroid with an initial mass of 1 kg, which entered the atmosphere on a vertical trajectory
with a speed of 21.5 km s$^{-1}$ and reached a maximal dynamic pressure of 1 MPa.

   \begin{figure}
   \centering
   \includegraphics[width=1.0\columnwidth]{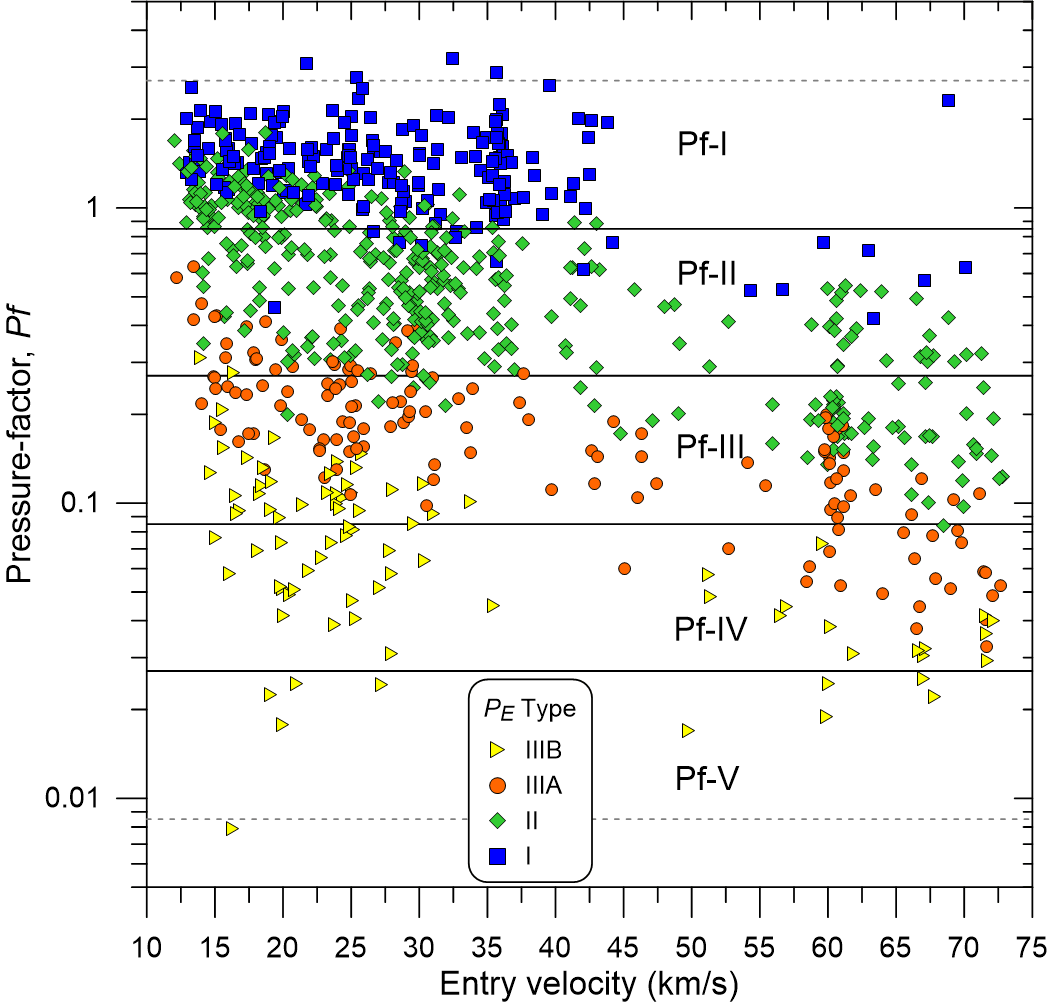}
   \caption{Pressure factor defined by Eq.~(\ref{Pf}) as a function of entry velocity. The division of fireballs into five strength
   categories Pf-I to Pf-V is shown. The color symbols indicate the $P_E$ type according to its definition.}
   \label{Pfgraf}
   \end{figure}
   
Figure~\ref{Pfgraf} shows the distribution of $P\!f$ values as a function of velocity. The $P\!f$ values are in the range from
about 0.008 to three. There is no obvious grouping of $P\!f$ values. Nevertheless, for easier referencing, we defined five
strength categories, designated Pf-I to Pf-V. The boundaries are marked in Fig.~\ref{Pfgraf} and their numerical values are as follows:
\begin{eqnarray}
\textrm{Pf-I} &:& 0.85 < P\!f,  \label{Pfvalues} \\
\textrm{Pf-II} &:& 0.27 < P\!f \leq 0.85, \nonumber \\
\textrm{Pf-III} &:& 0.085 < P\!f \leq 0.27, \nonumber \\
\textrm{Pf-IV} &:& 0.027 < P\!f \leq 0.085, \nonumber  \\
\textrm{Pf-V} &:& \hspace{1.2cm} P\!f \leq 0.027. \nonumber 
\end{eqnarray}
The sizes of categories Pf-II, Pf-III, and Pf-IV are equal in the logarithmic graph, and correspond to half of an order of magnitude. 
Most fireballs in categories Pf-I and Pf-V
also fall in the interval of the same size (marked by dashed lines in Fig.~\ref{Pfgraf}). There are only a few exceptions that lie 
outside. No special categories were defined for them. In fact, there are only a few fireballs in the weakest category, Pf-V.
Interestingly, there is a distinct upper limit of strength of fast fireballs ($>45$ km s$^{-1}$), which is exceeded in only one case. 
The boundaries of categories
were chosen so that this limit forms the upper boundary of category Pf-II. The width of the categories was set so that the
lower boundary of Pf-III corresponds to a $P\!f$ $10\times$ lower than the upper boundary of Pf-II. This way, almost the whole
group of fireballs with velocities about 60 km s$^{-1}$ (mostly Perseids) falls into Pf-III. By necessity, however, there are 
many fireballs near the boundaries. The $P_E$ classification has the same problem.

Figure~\ref{Pfgraf} can also be used to compare the $P\!f$ and $P_E$ classifications. Obviously, there is a different correction
for velocity. While there are eight fireballs with velocities above 50 km s$^{-1}$ classified as type I, there is only one Pf-I.
Similarly, many high-velocity fireballs classified as type II fall into Pf-III. On the contrary, there are a number of slow type II fireballs
that fall into Pf-I. We further investigate these differences in the next section, where physical and orbital
classifications are compared.

   \begin{figure*}
   \centering
   \includegraphics[width=0.992\columnwidth]{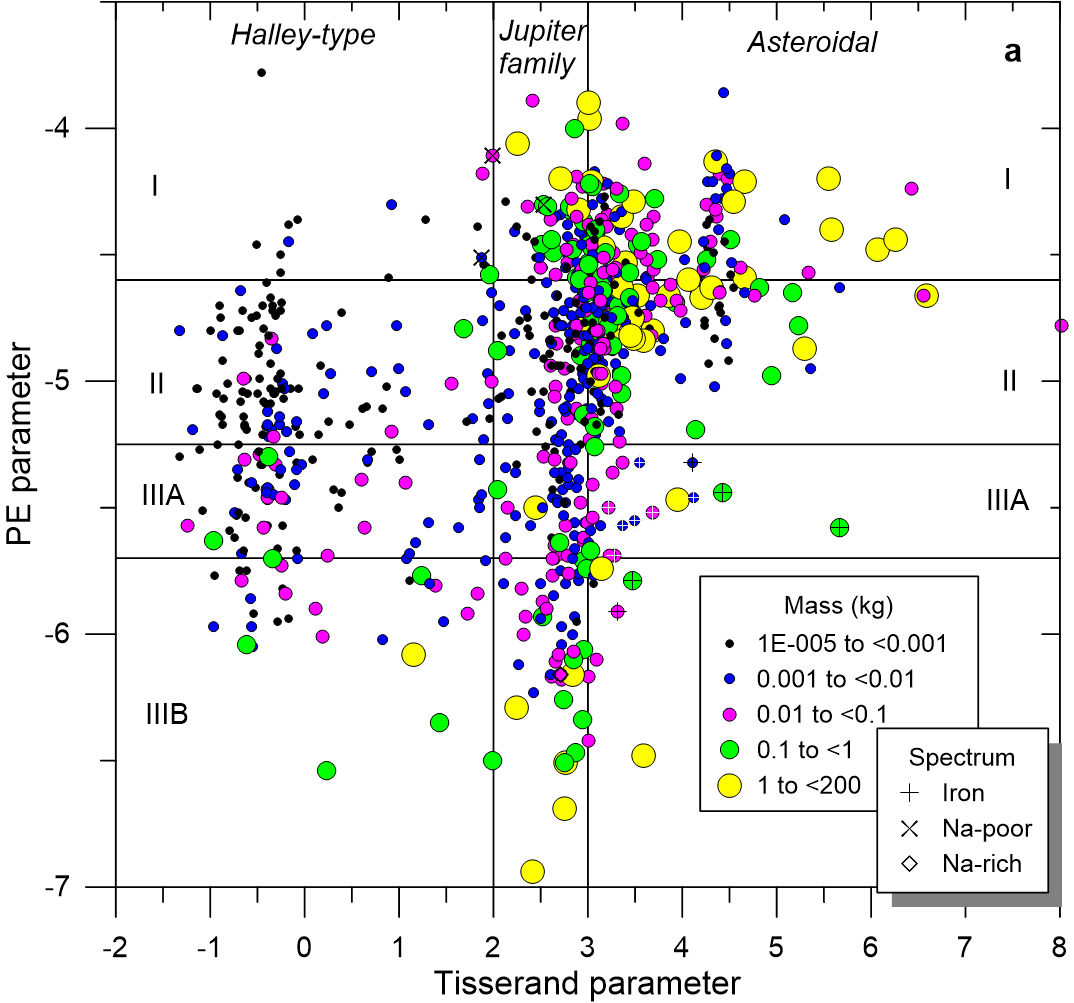}
     \includegraphics[width=1.0\columnwidth]{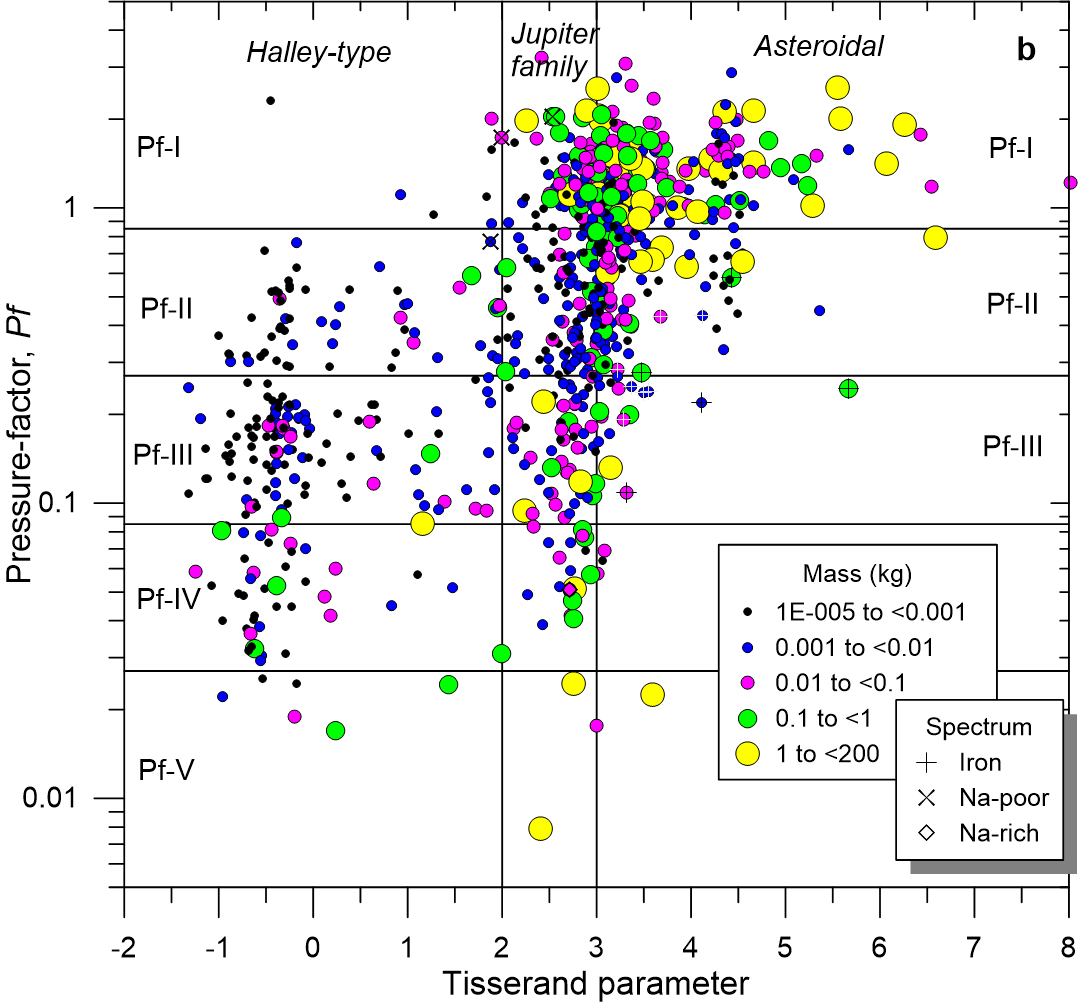}
   \caption{Comparison of orbital and physical classification of fireballs using two physical classification schemes. The Tisserand parameter
   with respect to Jupiter is compared with the $P_E$ parameter in panel {\textbf a} and with $P\!f$ in panel {\textbf b}. 
   Symbol sizes and colors distinguish five intervals of meteoroid masses. Fireballs with unusual spectra are also marked. For most
   fireballs, however, spectra are not available. Fireballs supposed to be irons on the basis of other criteria (see Sect.~\protect\ref{addirons})
   are marked by white crosses.}
   \label{Tisserand}
   \end{figure*}
   
      \begin{figure}
   \centering
   \includegraphics[width=1.0\columnwidth]{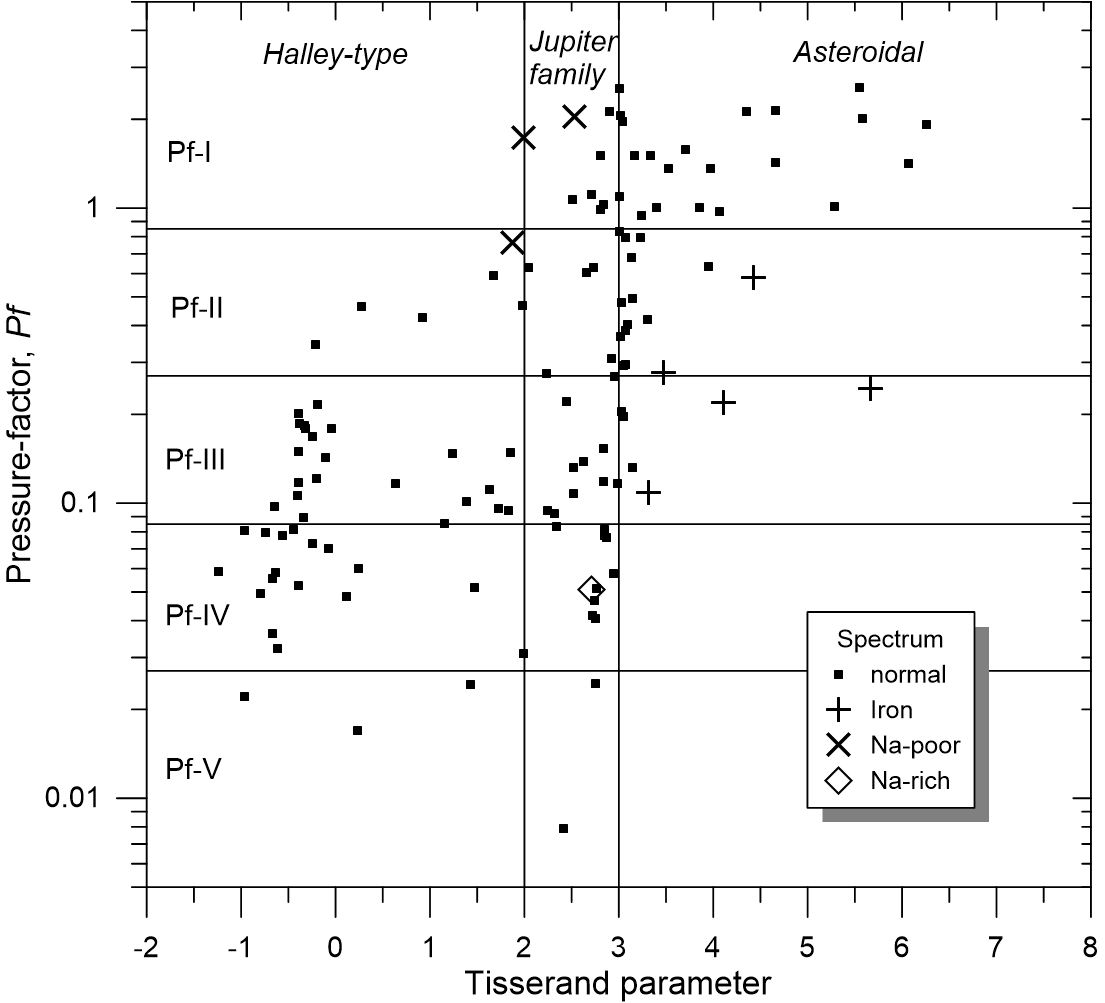}
   \caption{Pressure factor as a function of the Tisserand parameter for fireballs with good spectra. Unusual spectra are shown with different
   symbols.}
   \label{spectra}
   \end{figure}
   
    \begin{figure}
   \centering
   \includegraphics[width=1.0\columnwidth]{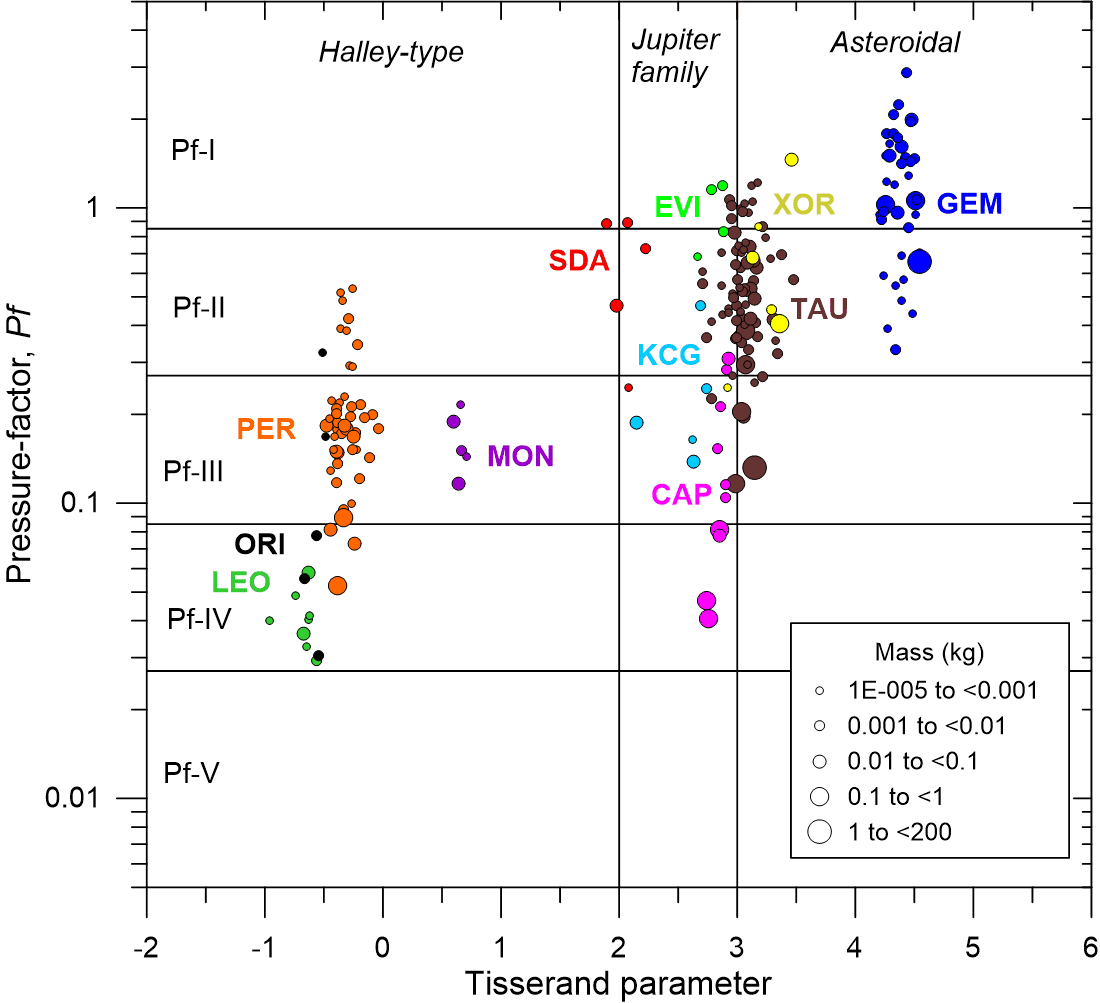}
   \caption{Pressure factor as a function of the Tisserand parameter for fireballs belonging to major meteor showers. 
   Only showers with at least four fireballs are shown. Colors are used to 
   discriminate between different showers. Symbol sizes mark five intervals of meteoroid masses.}
   \label{showers}
   \end{figure}
   
\section{Relation between orbital and physical properties}

\subsection{$P_E$ and $P\!f$ parameters as a function of the Tisserand parameter}

One of the most widespread methods for classifying orbits of Solar System bodies is that based on the
Tisserand parameter \citep{Tisserand}. Figure~\ref{Tisserand} shows both the $P_E$ parameter and $P\!f$ 
in relation to the Tisserand parameter, $T_J$. Both plots look similar but there are notable differences.
Using $P_E$, we would conclude the strongest observed body was a small meteoroid on a Halley-type
orbit ($T_J<2$). A number of Halley-type bodies have comparable  $P_E$ with meteoroids on evolved orbits
with $T_J>5$, which often fall into type II. Most of the weakest bodies with $P_E<-6$ are on Jupiter-family
orbits  ($2<T_J<3$). Although we do not know a priori the properties of meteoroids on different orbits,
the plot of $P\!f$ with $T_J$ better satisfies the expectation that bodies on cometary Halley-type orbits
are generally weaker than bodies on asteroidal obits with high $T_J$. The number of extremely weak 
Jupiter-family meteoroids is also lower in this plot. We therefore consider$P\!f$ to be a better proxy of
meteoroid physical properties 
(more specifically, resistance to ablation and fragmentation, which is probably connected 
to density, porosity, mechanical strength, or melting temperature),
and $P\!f$ will therefore be used from this point on, rather than $P_E$.

\subsection{$P\!f$ and spectra}

Figure~\ref{spectra} shows the same plot as in Fig.~\ref{Tisserand}b, but only for fireballs with good spectra. A remarkable
group is  five fireballs with iron spectra. All of them are on asteroidal orbits and have a significantly lower $P\!f$ than fireballs 
with normal (chondritic) spectra (and the same $T_J$). 
This fact was already stressed and explained by \citet{irons}. Metallic meteoroids with an iron-nickel composition
melt easily and efficiently lose mass in the form of liquid droplets. It is therefore somewhat inappropriate to call
meteoroids with a high $P\!f$ ``strong'' and those with a low $P\!f$ ``weak'', since irons are not weak. Better descriptions could  
perhaps be ``resistant'' and ``susceptible'' (to ablation).

In contrast to irons, there are three meteoroids deficient
in sodium (but containing magnesium, in contrast to irons) on cometary orbits, which have quite a high $P\!f$. Their nature is
currently unclear. Finally,
there is one fireball with an unusually bright sodium line on a Jupiter-family orbit. Its $P\!f$ is rather low but not unusual
for that type of orbit. A more detailed analysis of fireball spectra will be the subject of a future work.

\subsection{$P\!f$ and showers}
\label{shower_phys}
   
Figure~\ref{showers} shows the same plot as in Fig.~\ref{Tisserand}b, but for fireballs of major meteor showers.
An obvious fact is that all showers with a sufficient number of fireballs (Geminids, Taurids, $\alpha$ Capricornids,
Perseids) are inhomogeneous. The $P\!f$ values within a shower cover about an order of magnitude.
In Taurids, there is a mass-sorting effect, which was already noted by \citet{Spurny_Taurids} and studied in detail by
\citet{Taurid_phys}. Small meteoroids are more resistant than large ones. 
It was found that most of the Taurid material is porous and fragile
but it contains stronger inclusions, which can also exist separately as small meteoroids (while the porous material cannot).
It seems that the same effect is present in $\alpha$ Capricornids and possibly also in Perseids. But both
$\alpha$ Capricornids and Perseids are more susceptible to ablation than Taurids, and their $P\!f$ is shifted to lower values.
Geminids are the most resistant among shower meteoroids, a fact already known \citep[e.g.,][]{Babadzhanov02}.
Their $P\!f$ reaches roughly three in some cases, which is comparable to the highest values in our sample.
The mass sorting is not present, or is even opposite. As it can be seen in Fig.~\ref{showers}, the weakest
Geminids are the small ones. Detailed fragmentation modeling is need to investigate Geminid structure.

Other showers are represented by only a few fireballs in our sample. It is worth mentioning that the $\eta$ Virginids
are quite resistant, despite their cometary orbit. Using data from additional years, \citet{Brcek} have confirmed that
this is also the case for large meteoroids. The physical properties of $\eta$ Virginids are, therefore, more similar to Geminids
than Taurids. Southern $\delta$ Aquariids are also relatively resistant, which may be connected with their low perihelion
distance. While $\chi$ Orionids may be similar to Taurids, $\kappa$ Cygnids seem to be less resistant (but more than 
$\alpha$ Capricornids). December Monocerotids are similar to Perseids. Leonids are more susceptible 
than Perseids on average \citep[but resistant Leonids also exist, see][]{Kokhirova_Leo}. 

Although it is simplified and preliminary, average shower meteoroids can be assigned to the following Pf categories:
\begin{quote}
\begin{tabular}{l@{ :\ \ }l} 
Pf-I & GEM, EVI \\
Pf-II & TAU (except the large ones), XOR, SDA \\
Pf-III & PER, MON, KCG, TAU (large) \\
Pf-IV & ORI, LEO, CAP
\end{tabular}
\end{quote}

    \begin{figure}
   \centering
   \includegraphics[width=1.0\columnwidth]{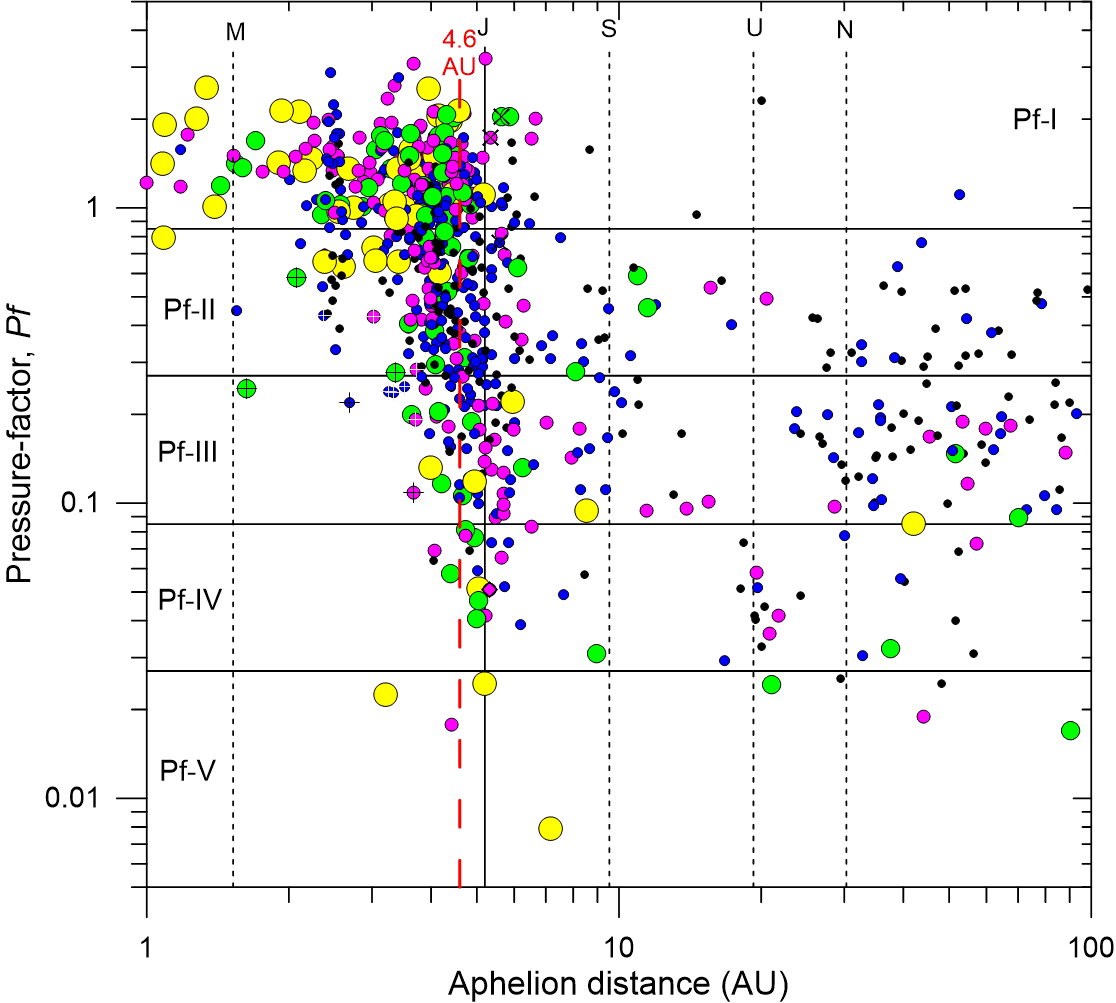}
   \caption{Pressure factor as a function of aphelion distance. Fireballs with an aphelion distance larger than 100 AU are not shown.
    Meteoroid masses and spectra are marked  as in Fig.~\protect\ref{Tisserand}. 
    Vertical black lines indicate the semimajor axes of planets from Mars to Neptune. The dashed red line is drawn at 4.6 AU.}
   \label{aphel}
   \end{figure}

\subsection{Orbital domains of asteroidal and cometary meteoroids}

Looking back at Fig.~\ref{Tisserand}b, we see that there are many fireballs concentrated along the
$T_J=3$ line, which is considered a boundary between asteroidal and cometary orbits. Fireballs with
a low $P\!f$ lie mostly on the cometary side of the boundary. Fireballs with
a high $P\!f$ are found on both sides of the boundary. It therefore seems that there are many
resistant meteoroids with orbits of Jupiter-family comet type. There are, however, other
criteria for distinguishing cometary and asteroidal orbits. One of the simplest was proposed by
\citet{Kresak}, which is based on the aphelion distance $Q$ and defines asteroidal orbits as those with $Q<4.6$ AU.
Figure~\ref{aphel} shows $P\!f$ as a function of $Q$. We can see here that the aphelion distance 
of the vast majority of  resistant meteoroids is lower than the semimajor axis of Jupiter (5.2 AU). The
most resistant ones indeed lie within 4.6 AU.

   \begin{figure*}
   \centering
   \includegraphics[width=1.3\columnwidth]{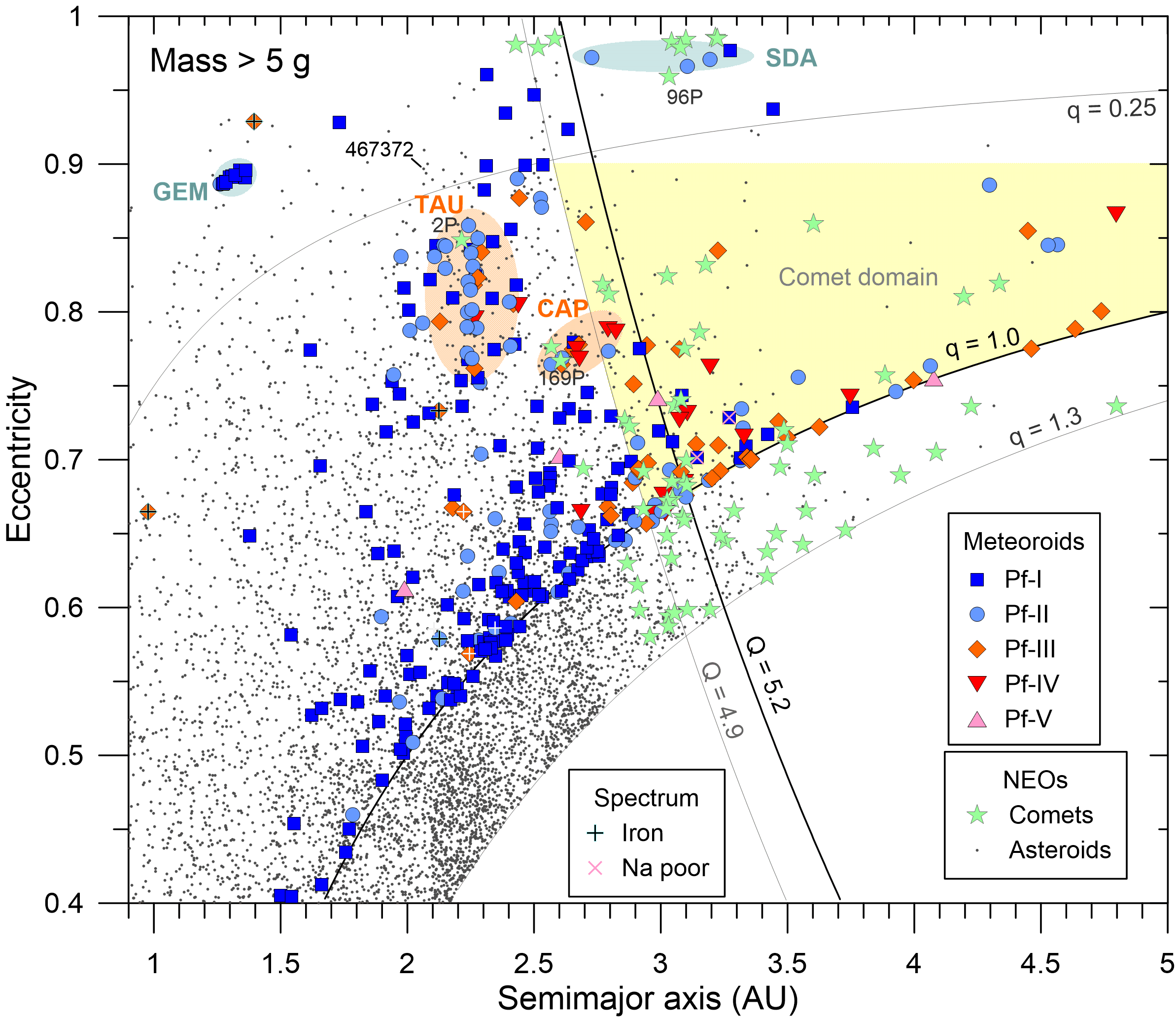}
   \caption{Eccentricities and semimajor axes of meteoroids of various Pf classes and near-Earth objects (asteroids and comets
   with perihelion distance $q<1.3$ AU). Semimajor axes ($a$) are restricted to 0.9--5 AU and eccentricities ($e$) to 0.4--1.
   All meteoroids with lower $a$ or $e$ are of class Pf-I, except two, which are Pf-II. Curves of constant perihelion ($q$) or
   aphelion ($Q$) distance are shown for selected values. Only meteoroids with masses larger than 5 g are shown. Regions 
   where members of meteor showers Taurids (TAU), $\alpha$ Capricornids (CAP), Geminids (GEM), and Southern $\delta$
   Aquariids (SDA) are found are highlighted (but not all meteoroids in that regions belong to the showers, and some shower members 
   can be found also  outside the regions). Meteoroids with iron or Na-poor spectra are marked. The region occupied primarily
   by comets and cometary meteoroids is highlighted in yellow.}
   \label{a-e}
   \end{figure*}
   
   \begin{figure}
   \centering
   \includegraphics[width=1.0\columnwidth]{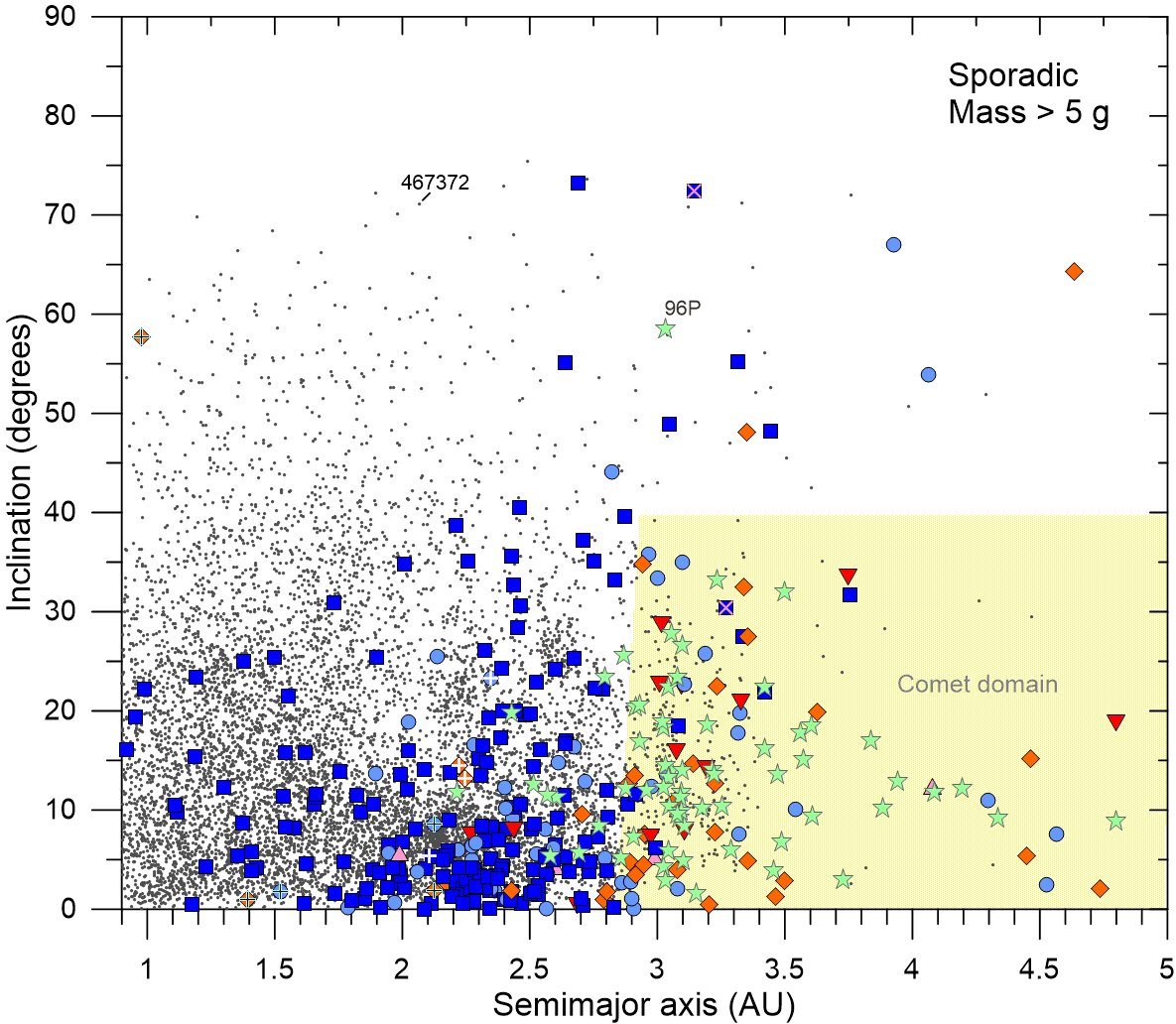}
   \caption{Inclinations and semimajor axes of meteoroids of various Pf classes and near-Earth objects (asteroids and comets
   with perihelion distance $q<1.3$ AU). Semimajor axes ($a$) are restricted to 0.9--5 AU. 
   Only sporadic meteoroids (i.e.,\ not belonging to any of the 16 major meteor showers)  with masses larger than 5 g are shown. 
   No meteoroid with $a<5$ AU has inclination lager than 90\degr.
   Meteoroids with iron or Na-poor spectra are marked. See Fig.~\protect\ref{a-e} for legends. The region occupied primarily
   by comets and cometary meteoroids is highlighted in yellow.}
   \label{a-i}
   \end{figure}

Figures~\ref{a-e} and \ref{a-i} further explore the orbital domains of resistant and susceptible meteoroids.
Figure~\ref{a-e} shows the eccentricities, $e$, and semimajor axes, $a$, around the boundary between asteroidal and 
Jupiter-family cometary orbits for meteoroids larger than 5 grams. Near-Earth asteroids and comets taken from
the Jet Propulsion Laboratory database\footnote{https://ssd.jpl.nasa.gov/tools/sbdb\_query.html, accessed October 26, 2021}
are shown for comparison. Only objects with a data-arc span larger than 90 days have been included.

We can see that meteoroids of various Pf classes are mixed, as there are asteroids and comets. In the whole sample,
meteoroids of class Pf-I prevail (54\%). Meteoroids of susceptible classes Pf-III to Pf-V form only 21\% of all
meteoroids (the rest is the intermediate type Pf-II). 
There is, however, an orbital region, highlighted in yellow, where susceptible meteoroids form a majority (56\%)
and the Pf-I class is represented by only 16\% of meteoroids. Comets are also more numerous than asteroids in this region. We therefore call it the comet domain.

The comet domain is  defined by $Q>4.9$ AU and $e<0.9$ (for $a<5$ AU). The other boundary for
meteoroids, $q<1$ AU, is due to the necessity of crossing the Earth's orbit to be observable. The boundaries are only approximate.
In particular, the $e$-limit is poorly defined because there is little data near the boundary. In any case, high-eccentricity
meteoroids with semimajor axes up to 3.5 AU are resistant, despite the fact that comets prevail over asteroids in this region. 
Comet 96/P Machholz~1 and several SOHO comets are here. The meteoroid resistance may be connected with their low perihelion
distances of $q<0.25$ AU. Some of these near-Sun meteoroids belong to the Southern and Northern $\delta$ Aquariid meteor
showers. 

The $Q$-limit nearly corresponds to the perihelion distance of Jupiter (4.95 AU). We note that there are susceptible cometary meteoroids
below this limit also. The $\alpha$ Capricornids lie on both sides of the boundary but the majority are on the asteroidal side, similar
to the comets 169P/NEAT and P/2003 T12 (SOHO). Taurids have even lower aphelia and also contain susceptible meteoroids,
especially the large ones. They are related to comet 2P/Encke but occupy a large orbital space. The resonant branch contains 
a wide range of eccentricities at $a\sim 2.25$~AU while other Taurids have a variety of semimajor axes \citep{Spurny_Taurids}.

Some of the susceptible sporadic meteoroids in the asteroidal orbital region are irons, but there are also truly weak,
low-density meteoroids in asteroidal orbits. On the other hand, there seem also to be quite resistant meteoroids in the comet
domain. Nevertheless, we also have to look at the orbital inclination, $i$, in Fig.~\ref{a-i}. There are several resistant meteoroids (Pf-I)
with semimajor axes 2.5 -- 3.5 AU and large inclinations 40\degr\ -- 75\degr. Susceptible cometary meteoroids are rare in this region
(only one with mass $>5$ g, two in the total sample). We therefore consider the comet domain to be restricted to $i<40\degr$ at
semimajor axes around 3.5~AU. Then, only a few Pf-I meteoroids remain in the comet domain. Remarkably, most of them still 
have relatively high inclinations $\geq 20\degr$. Moreover, if spectra are known, they are deficient in sodium in these cases.
Figure~\ref{a-i} shows only sporadic meteoroids. 

   \begin{figure}
   \centering
   \includegraphics[width=1.0\columnwidth]{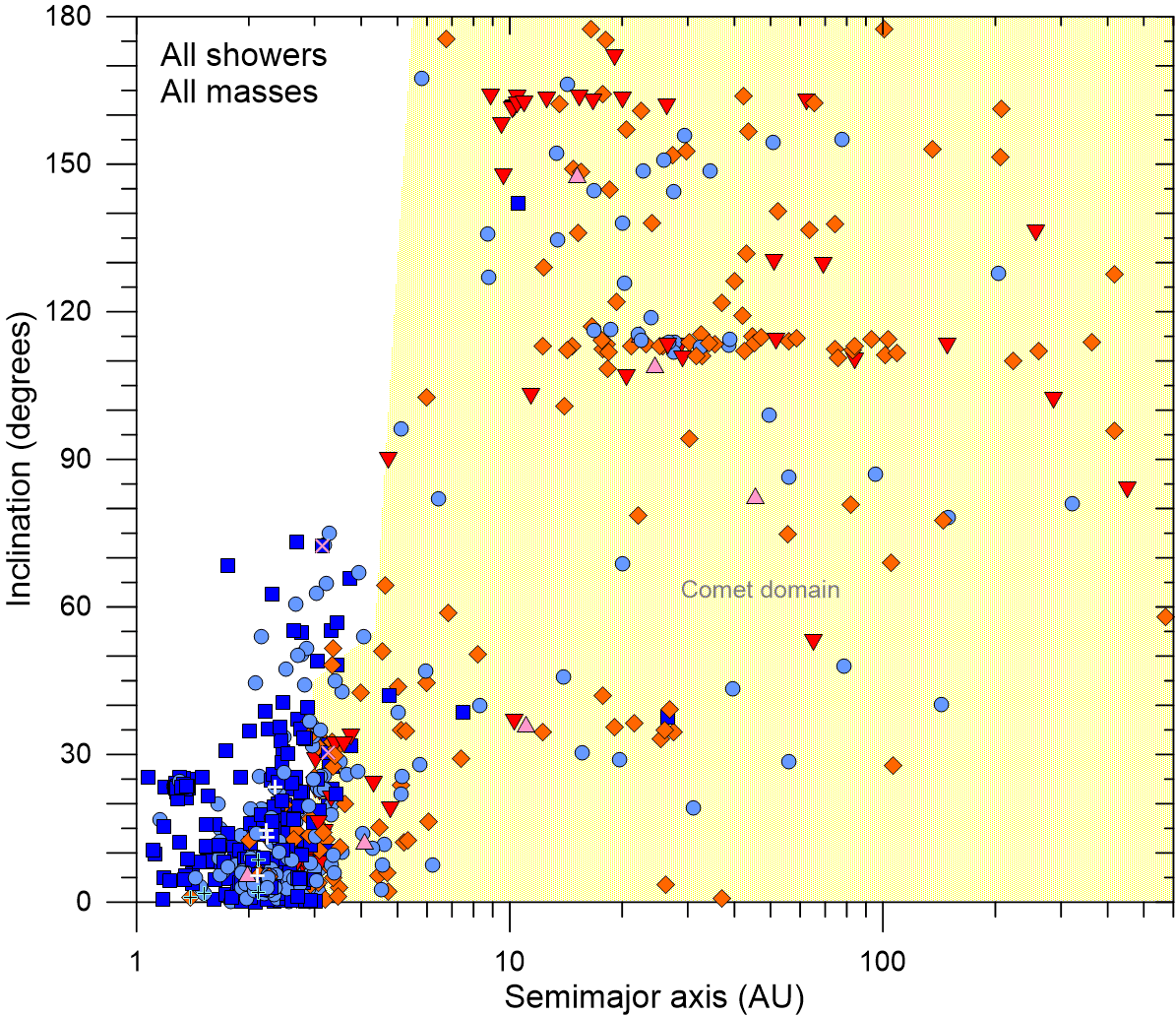}
   \caption{Inclinations and semimajor axes of meteoroids of various Pf classes. 
   Semimajor axes are shown up to 600 AU. 
   Meteoroids with iron or Na-poor spectra are marked. See Fig.~\protect\ref{a-e} for legends. The region occupied primarily
   by comets and cometary meteoroids is highlighted in yellow.}
   \label{a-i_Halley}
   \end{figure}

Figure~\ref{a-i_Halley} extends Fig.~\ref{a-i} up to 600 AU in $a$ and 180\degr\ in $i$. All observed meteoroids are shown.
It is important to note that large semimajor axes have high uncertainties. Most meteoroids with $i \sim 113\degr$ are Perseids. The comet domain
extends to all inclinations for $a\geq5$~AU. There are only a few meteoroids classified as Pf-I with large semimajor axes ($a>4$~ AU).

We can conclude that asteroidal and cometary meteoroids are partly mixed in orbital space, as are asteroids and comets. Nevertheless,
there are regions in the orbital space where mostly meteoroids susceptible to ablation and fragmentation are encountered.
Since these regions overlap with those occupied primarily by comets, we can be confident that these meteoroids originate from comets.
The domain of cometary meteoroids can be divided into a short-period one with:
\begin{quote}
$Q>4.9$ AU, \\ $e<0.9$ (or, possibly, $q>0.25$ AU), \\ $i<40\degr$, \\ $a<5$ AU,
\end{quote}
and a long-period one with:
\begin{quote}
$a>5$ AU. 
\end{quote}
Asteroidal meteoroids are primarily encountered in orbits with lower aphelia ($Q<4.9$ AU), or in orbits with somewhat larger aphelia
(up to about 7~AU) but with either low perihelia ($q<0.25$ AU) or high inclinations ($i>40\degr$). A special sort
of asteroidal meteoroids are irons, which are susceptible to ablation because they are readily melted. 

Meteoroids considered as asteroidal are primarily those of class Pf-I  with $P\!f>0.85$. Meteoroids of classes Pf-III to Pf-V
with $P\!f<0.27$ are considered as cometary (except irons). Meteoroids of class Pf-II can be of both asteroidal and cometary origin. 
In comets, they are mostly, but not exclusively, encountered as small bodies with masses below 10 grams.
In fact, the distinction between cometary and asteroidal bodies according to $P\!f$ parameter is more evident for larger bodies.
But regardless of mass, looking at Fig.~\ref{aphel}, we can see that meteoroids with $P\!f>0.6$ are still mostly asteroidal. 
Nevertheless, as noted above, 
asteroidal meteoroids in cometary orbits and cometary meteoroids in asteroidal orbits can be encountered.

\subsubsection{Additional irons}
\label{addirons}

\begin{figure}
   \centering
   \includegraphics[width=1.0\columnwidth]{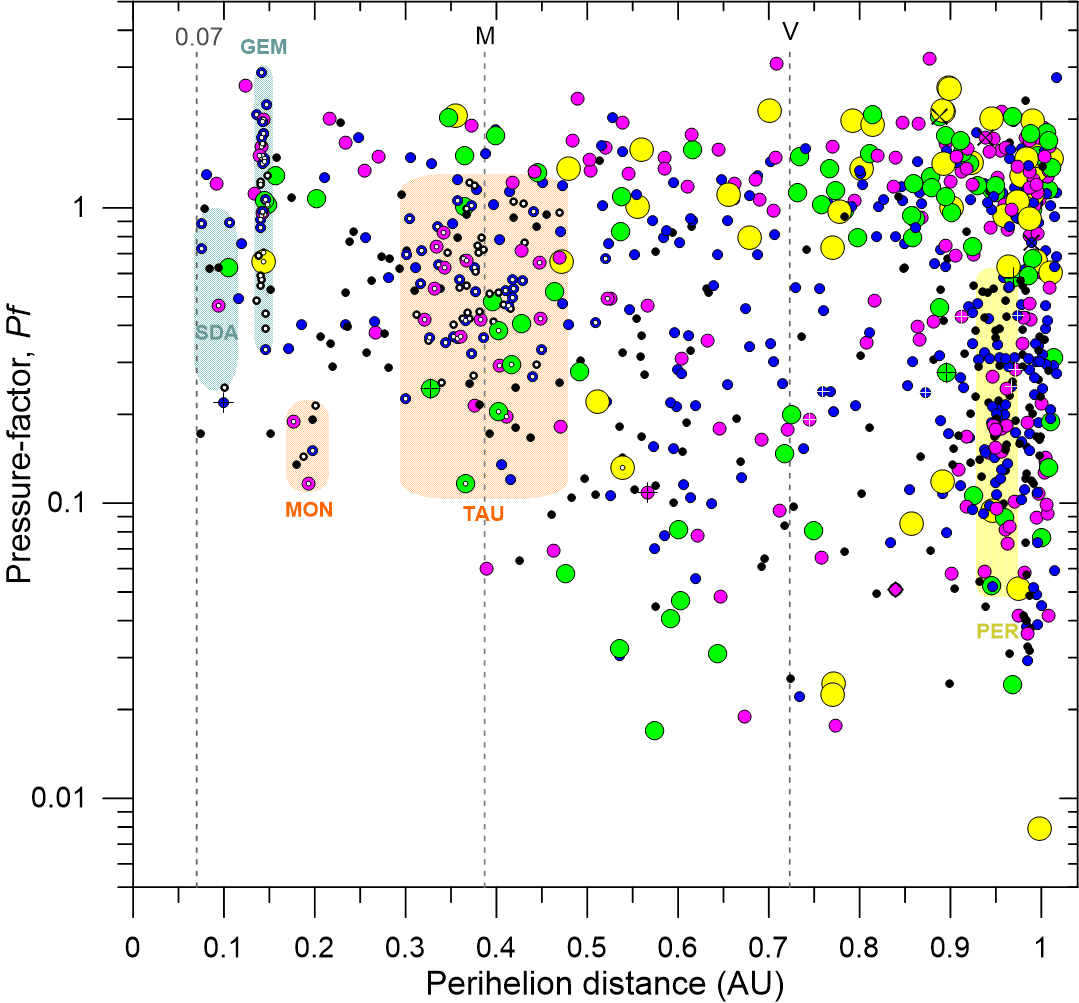}
   \caption{Pressure factor as a function of perihelion distance. 
    Meteoroid masses and spectra are marked as in Fig.~\protect\ref{Tisserand}. 
    Vertical dashed lines indicate the semimajor axes of Venus, Mercury, and the distance of 0.07 AU.
    Regions where members of the Southern $\delta$ Aquariids, 
    Geminids, December Monocerotids, Taurids and Perseids meteor showers are found are highlighted.
    Actual meteoroids belonging to the first four showers are marked by white dots.}
   \label{perihel}
   \end{figure}

The five identified irons have a $P\!f$ in a relatively wide range, 0.1 -- 0.6. Irons can be unambiguously distinguished according to their
spectra. They are also characterized by smooth radiometric light curves with sudden ends \citep{irons}. Because of their high meteoroid 
density, deceleration is low. Because of the lack of Na and Mg lines, fireballs appear rather bluish in color photographs. 
Based on these criteria, we identified seven fireballs with unavailable spectra that were likely caused by iron meteoroids.
They are listed in Table~\ref{ironcandidates} and marked by white crosses in Figs.~\ref{Tisserand}, \ref{aphel}, \ref{a-e}, 
\ref{a-i}, \ref{a-i_Halley},
\ref{perihel}, and \ref{ecc}. These suspected irons occupy the same orbital region and $P\!f$ range as the confirmed irons.
Because of their sudden ends, irons are better separated from other asteroidal meteoroids using the $P_E$ criterion
than the $P\!f$ value. 

There are 473 fireballs in our sample with aphelia less than 4.9~AU; 336 of them are sporadic (or
belonging to minor showers). Probable irons represent 2.5\% of all of them and nearly 3.6\% of sporadic ones.
The largest iron has a photometric mass of 0.25 kg, corresponding to a diameter of 4~cm.
The iron fraction is probably higher among small asteroidal meteoroids \citep{Mills}.

\begin{table}

\caption{Fireballs without spectra suspected to be irons}
\label{ironcandidates}
\vspace{-1ex}
\begin{tabular}{ll} \hline  \noalign{\smallskip}
EN280817\_233341& EN170418\_194938 \\
EN161017\_185044& EN121118\_185325 \\
EN171017\_205152& EN141118\_195214 \\
EN180318\_001818 \\
\hline
\end{tabular}
\end{table}

\subsection{Extreme orbits}

In this section we check how perihelion distance and orbital eccentricity relates to the physical classification
of meteoroids. We also check the resonances with Jupiter.

\subsubsection{Perihelion distances}

Figure~\ref{perihel} shows $P\!f$ plotted against perihelion distance, $q$. All very susceptible
meteoroids with $P\!f<0.1$ have perihelion distances higher than 0.35~AU. There are, nevertheless, some susceptible 
Pf-III meteoroids with $0.1<P\!f<0.27$ even at $q<0.2$~AU. 
Some of them belong to the December Monocerotid shower and are supposedly relatively young.
Another one is an iron, and thus, in fact, a quite compact body. The others are small meteoroids. 
But most meteoroids with low perihelia are resistant. Those with masses larger than 100 grams all have a $P\!f>0.6$.
Many observed resistant low-perihelion bodies are Geminids, and some are Southern $\delta$ Aquariids, which have even
lower $q$. However, no meteoroid was observed at $q<0.07$~AU. This seems to be significant (see also the
histogram of $q$ in the Appendix of Paper I)
and corresponds with the lack of asteroids with $q<0.076$~AU \citep{Granvik,Wiegert}. 

\begin{figure}
   \centering
   \includegraphics[width=1.0\columnwidth]{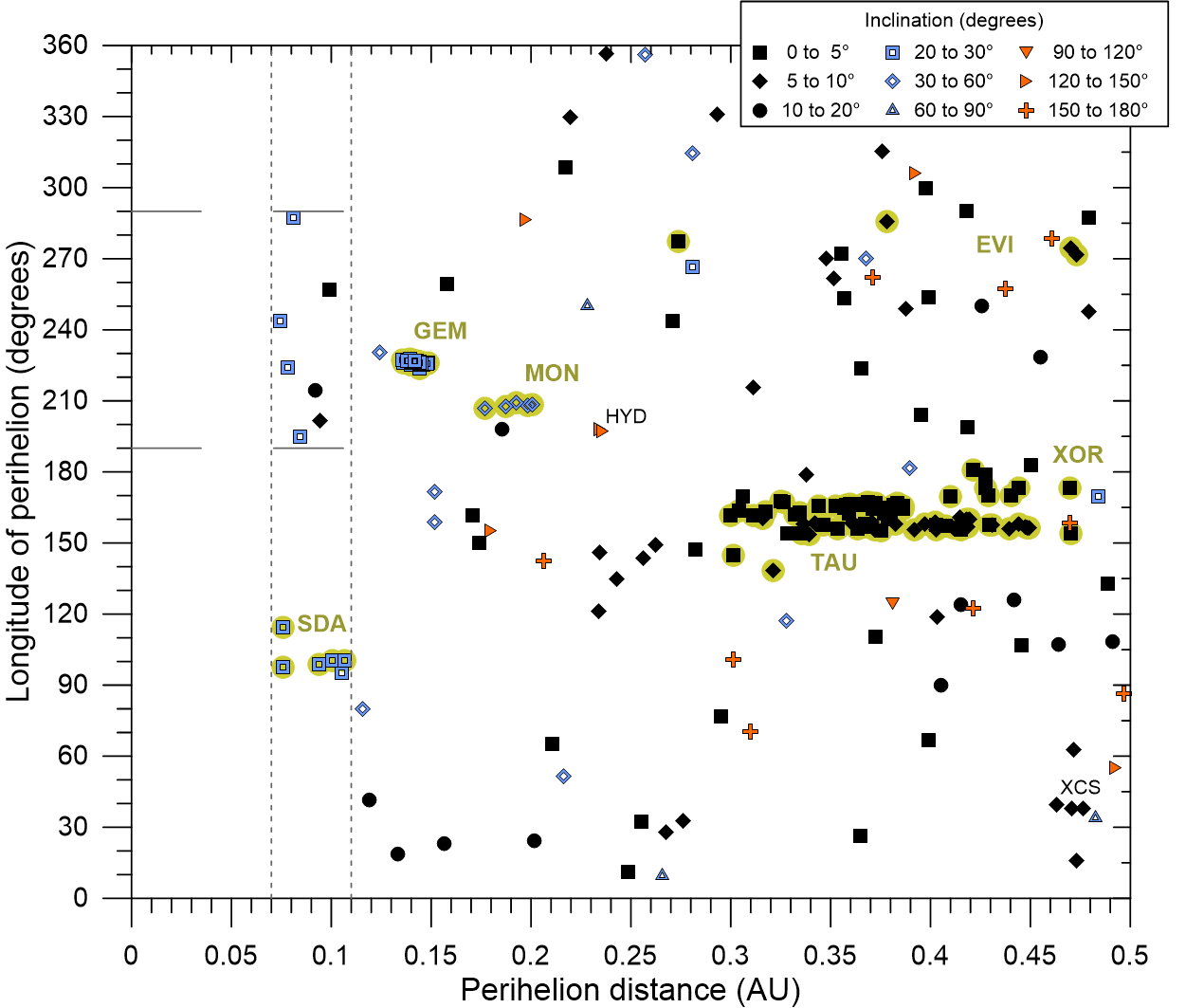}
   \caption{Longitude of perihelion as a function of perihelion distance. Different symbols
    correspond to different inclination intervals (see the legend). Meteoroids belonging to major
    meteor showers (Southern $\delta$ Aquariids, 
    Geminids, December Monocerotids, Taurids, $\eta$ Virginids, and $\chi$ Orionids) are highlighted. Minor showers
    $\sigma$ Hydrids (HYD) and $\xi^2$ Capricornids (XCS) are also marked. The excess of sporadic
    meteoroids with $q<0.11$ AU at $190\degr < \tilde{\omega} < 290\degr$ is notable.}
   \label{lperi}
   \end{figure}

It is worth mentioning that the longitudes of perihelia, $\tilde{\omega}$, 
of all seven sporadic meteoroids with $q<0.11$ AU, lie between 190\degr\ and 290\degr\ (Fig.~\ref{lperi}). 
\citet{Wiegert} found the region $180\degr < \tilde{\omega} < 270\degr$ and $i<12\degr$ to be 
least contaminated by small cometary meteoroids, which may serve as impactors destroying larger bodies. 
At higher inclinations, this region may be linked with Geminids.
The only other meteoroids with very low perihelia are five Southern $\delta$ Aquariids
and one Northern $\delta$ Aquariid, all concentrated at $95\degr < \tilde{\omega} < 115\degr$. 
At larger perihelion distances, especially $q >0.2$ AU, longitudes of perihelia of sporadic meteoroids are more
randomly distributed. 
We note that only five retrograde meteoroids have $q<0.3$ AU, two of them
belonging to the $\sigma$ Hydrid shower.

In general, the largest concentration of perihelion distances is near 1 AU. Meteoroids in such orbits have the highest
probability of collision with the Earth.

   \subsubsection{Eccentricities}
   
\begin{figure}
   \centering
   \includegraphics[width=1.0\columnwidth]{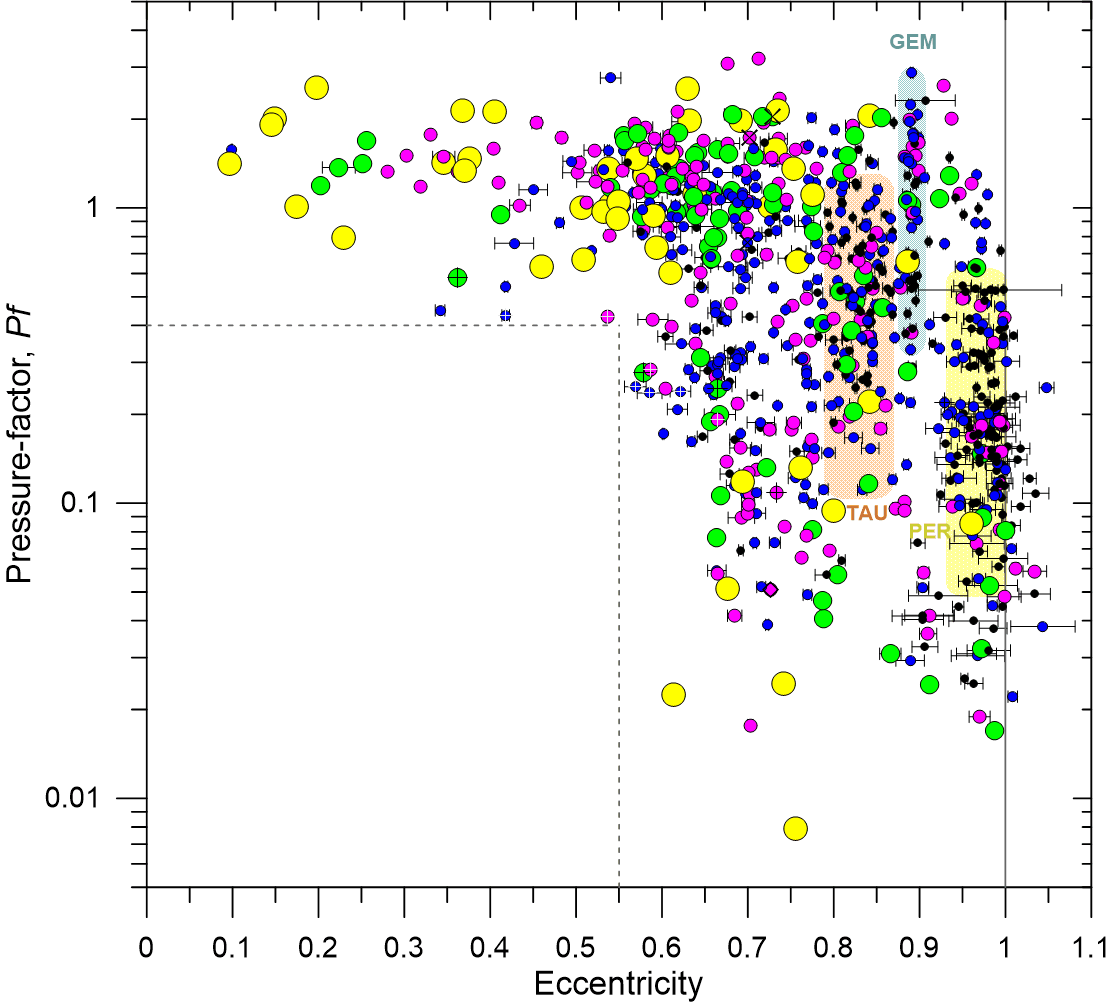}
   \caption{Pressure factor as a function of eccentricity. 
    Meteoroid masses and spectra are marked as in Fig.~\protect\ref{Tisserand}. 
    Regions where members of the Taurids, Geminids, and Perseids meteor showers are found are highlighted.  
    Formal error bars of eccentricities are included. The vertical solid line marks the parabolic limit.
    The dashed lines border the region with no meteoroids.}
   \label{ecc}
   \end{figure}

  Figure~\ref{ecc} shows $P\!f$ plotted against eccentricity, $e$.  Only resistant meteoroids with 
   $P\!f>0.4$ were observed at $e<0.55$. These are typically asteroidal orbits (all of them are prograde). Semimajor axes
   are lower than 2.2 AU; otherwise Earth encounter would not occur.
   Mostly large meteoroids are encountered here. Because of low entry velocities, small meteoroids
   remain under the detection limit.
   
   There are 15 fireballs with hyperbolic orbits within one sigma of the formal error. 
   Only two remain hyperbolic within three sigma limits. The largest eccentricity, $e=1.048 \pm 0.009$, was measured 
   for EN271117\_012837. This fireball was nicely observed at one station but the other two records are from a large distance. 
   The  error may therefore be underestimated. It is possible that the fireball was a member of the minor
   (and unconfirmed) December $\epsilon$ Craterids meteor shower. With $P\!f=0.25$, it had a medium resistance.
   The second good candidate is EN190918\_213159 with $e=1.028 \pm 0.008$. It was observed very well
   and had a cometary $P\!f=0.11$.
   
   Although most of the hyperbolic orbits are probably the result of observational errors, it is possible that some of them were
 truly hyperbolic. Nevertheless, the eccentricity in all cases is only slightly above unity and there is no reason to consider
   the meteoroids to be of interstellar origin.
   Orbits that were originally highly eccentric elliptic orbits could be transformed to hyperbolic orbits by gravitational disturbances
   of planets, nongravitational forces, or during the ejection process from the parent body. All hyperbolic orbits, except one 
   with $i=59\degr$, are retrograde. Meteoroid masses range from subgram to about 20 grams. All $P\!f$ values
   are below 0.4, so as expected, the bodies are of cometary nature, similar to or slightly weaker than those encountered 
   among the Perseids.
   
   \begin{figure}
   \centering
   \includegraphics[width=1.0\columnwidth]{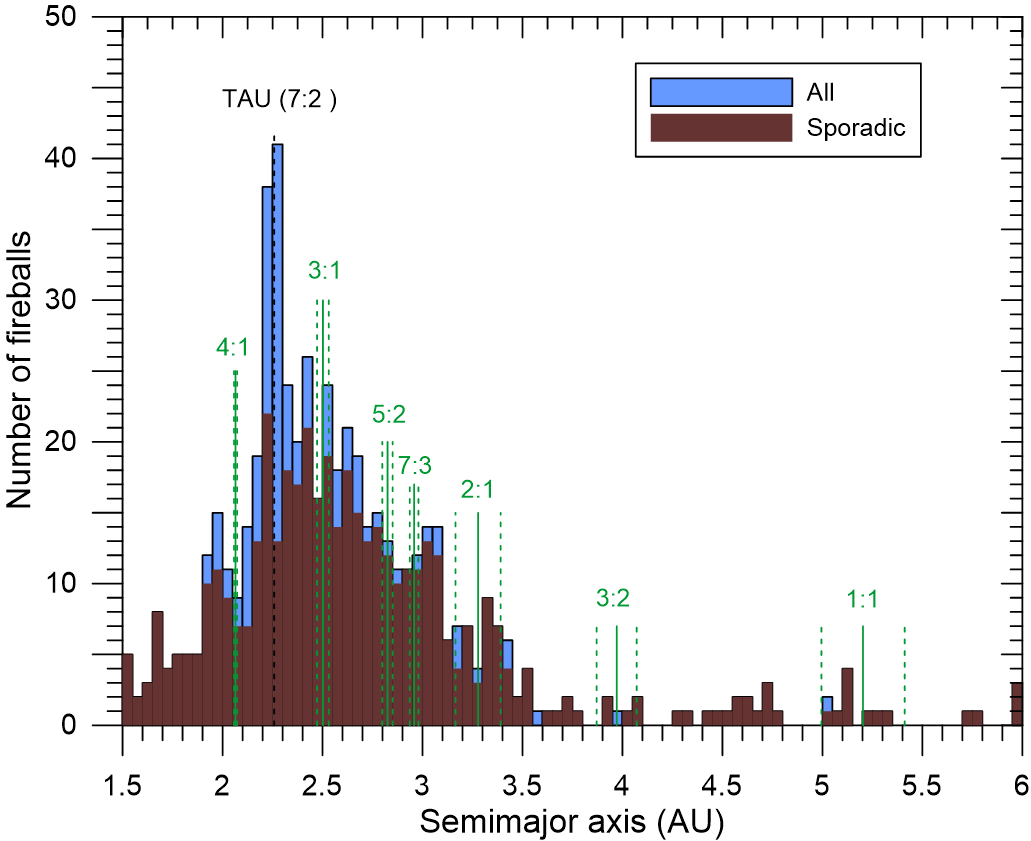}
   \caption{Histograms of semimajor axes in the range 1.5 -- 6~AU for all meteoroids and sporadic ones. 
   The positions and widths of mean motion resonances with Jupiter  \citep{Tisserand} are shown. 
   The location of Taurids is indicated. The sporadic sample was created by removing members of the 16 major meteor showers.}
   \label{hist-reso}
   \end{figure}

      \begin{figure}
   \centering
   \includegraphics[width=1.0\columnwidth]{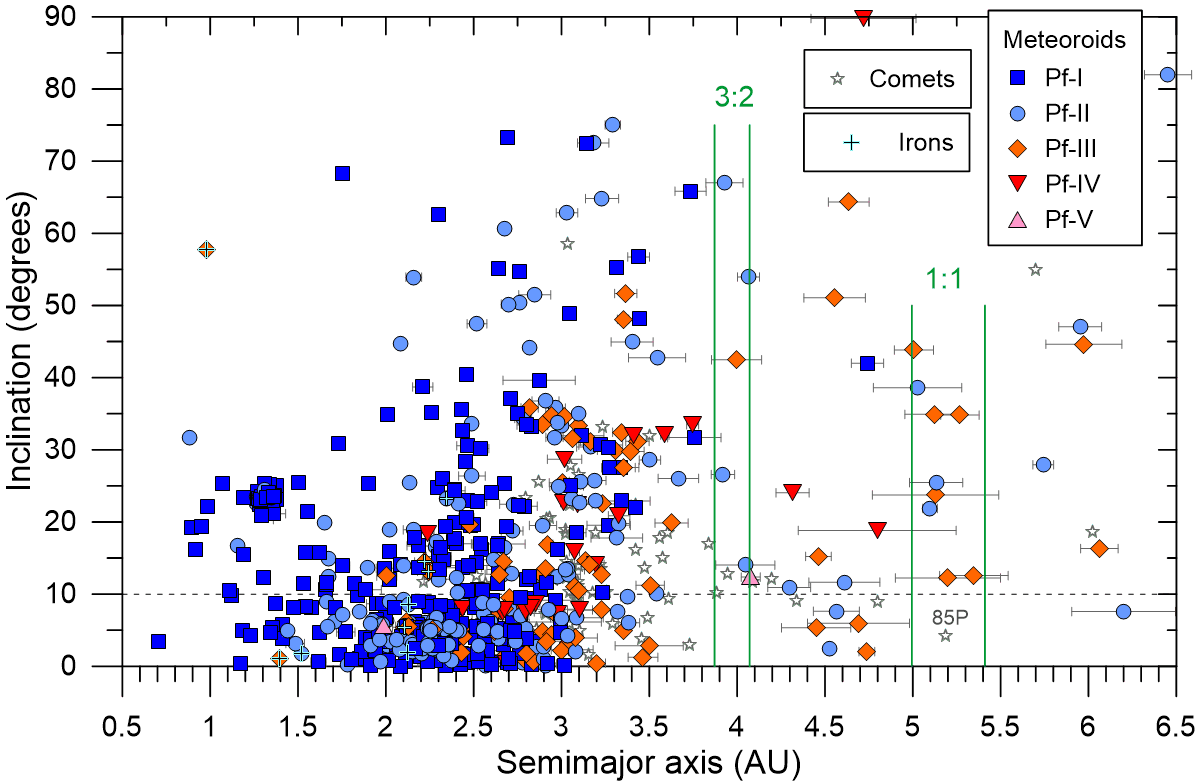}
   \caption{Inclination as a function of  semimajor axis in the range 0.5 -- 6.5~AU. Meteoroids of various Pf classes and 
   comets with perihelion distances $q<1.3$ AU are shown. Probable iron meteoroids are marked. Error bars of semimajor axes
   are shown (errors in inclination are negligible at this scale).
   The intervals of semimajor axes corresponding to 1:1 and 3:2
   resonances with Jupiter \citep{Tisserand} are indicated. No meteoroids with inclinations below 10\degr\ were detected
   inside these resonances. The same is valid for the 1:1 resonance with the Earth near 1~AU.}
   \label{resonant}
   \end{figure}
   
\subsubsection{Resonances}

Figure~\ref{hist-reso} suggests that there may be a small excess of meteoroids in the 1:1 resonance with Jupiter
near 5.2~AU.  Figure~\ref{resonant} shows that these meteoroids have
moderate inclinations between 10\degr\ -- 45\degr. Physical classification is consistent with a cometary origin.
No meteoroids were detected here with $i<10\degr$, despite
the presence of comet 85P/Boethin here. 
A similar situation may be present in the 3:2 resonance near 4~AU. In contrast to the resonant regions, there are low-inclination 
meteoroids around 4.6~AU. Comet 85P probably broke apart \citep{Meech_85P}.

At lower semimajor axes, there is no clear excess (or lack) of meteoroids in resonance with Jupiter (see Fig.~\ref{hist-reso}), 
except Taurids trapped in the 7:2 resonance \citep{Spurny_Taurids}. However, it can be noted that
no meteoroids with low inclination
were detected near the 1:1 resonance with the Earth (Fig.~\ref{resonant}).

\section{Meteor showers}
\label{showersection}

The major established meteor showers detected in our sample have been listed and 
the position of their radiants have been plotted in Paper~I. The physical properties of the meteoroids of these showers 
have been discussed in Sect.~\ref{shower_phys}. In this section, we first provide more details about the radiants and orbits
of the most frequently detected showers, the Geminids, Perseids, and $\alpha$ Capricornids. 
Taurids have been already discussed by \citet{Spurny_Bratislava} and will be presented in more detail elsewhere. 
Then, we discuss the detection of minor showers, whether established or not.

   \begin{figure}
   \centering
   \includegraphics[width=\columnwidth]{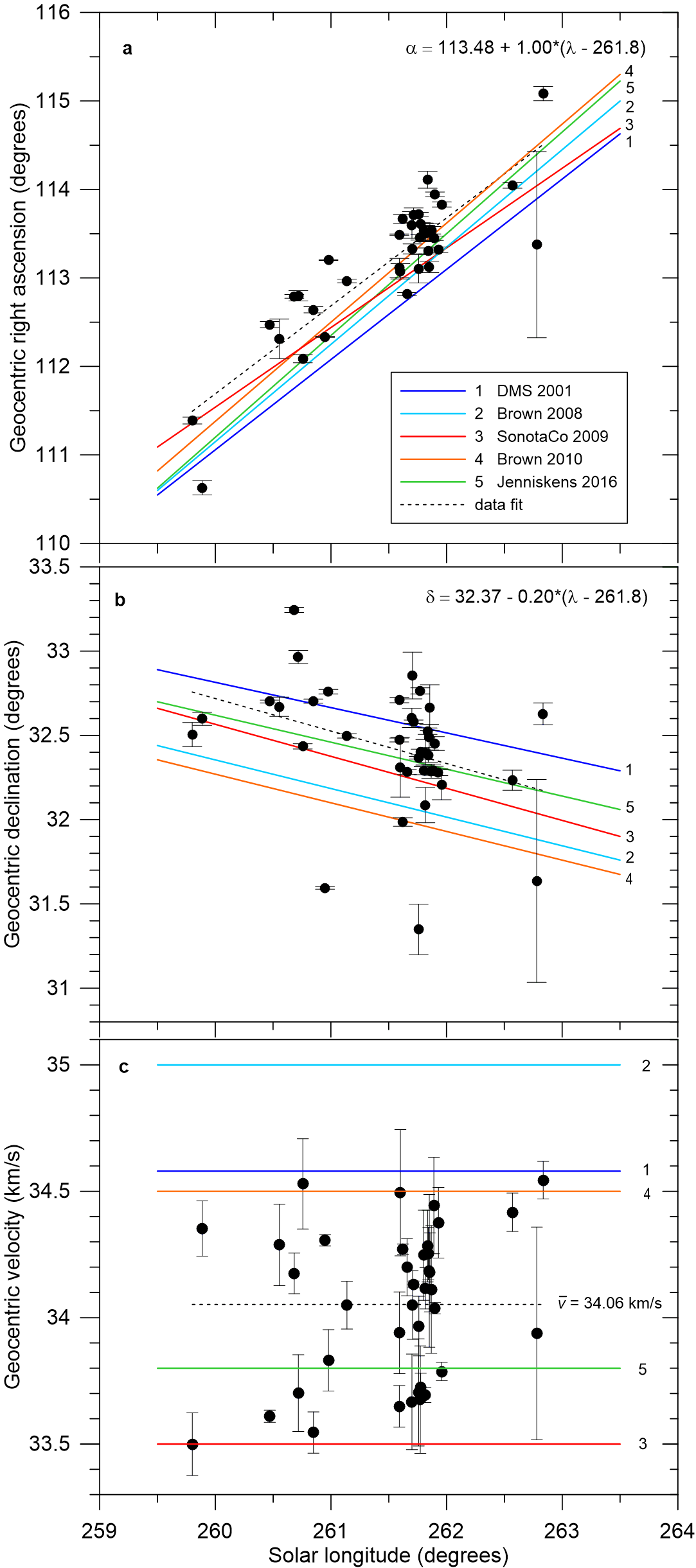}
   \caption{Motion of the Geminid geocentric radiant. Right ascension (panel \textbf{a}), declination (\textbf{b}), 
   and velocity (\textbf{c}) are plotted as a function of solar longitude (all in equinox J2000.0) for individual meteoroids
   with formal error bars.  The data fit is plotted as a dotted line and the corresponding equation is inserted
   (no change of velocity with solar longitude was assumed).
   The motion of the mean radiant as reported by authors cited in the IAU Meteor Data Center is plotted by solid lines
   as follows: 1 -- Dutch Meteor Society data \citep[see][]{Jopek2003}; 2 -- \citet{Brown2008}; 
   3 -- \citet{SonotaCo}; 4 -- \citet{Brown2010}; 5 -- \citet{CAMS}.}
   \label{GEMmotion}
   \end{figure}
   
\subsection{Major showers}

\subsubsection{Geminids}

Figure~\ref{GEMmotion} shows the radiant coordinates and geocentric velocities of 
Geminids as a function of solar longitude. The data are compared with mean radiant  positions
and motions reported by other authors as cited in the 
International Astronomical Union Meteor Data Center 
(IAU MDC)\footnote{https://www.ta3.sk/IAUC22DB/MDC2007/, accessed November 22, 2021}
\citep{MDC}.
We observed slightly larger right ascensions, although the reason for this is not clear. Other authors used various techniques
(photographic, video, radar) sensitive to meteoroids with various masses and having various precisions.
The declinations are in good agreement. As for velocities, the scatter of individual values in our data is
lower than the range of mean values from other sources. 

   \begin{figure}
   \centering
   \includegraphics[width=0.85\columnwidth]{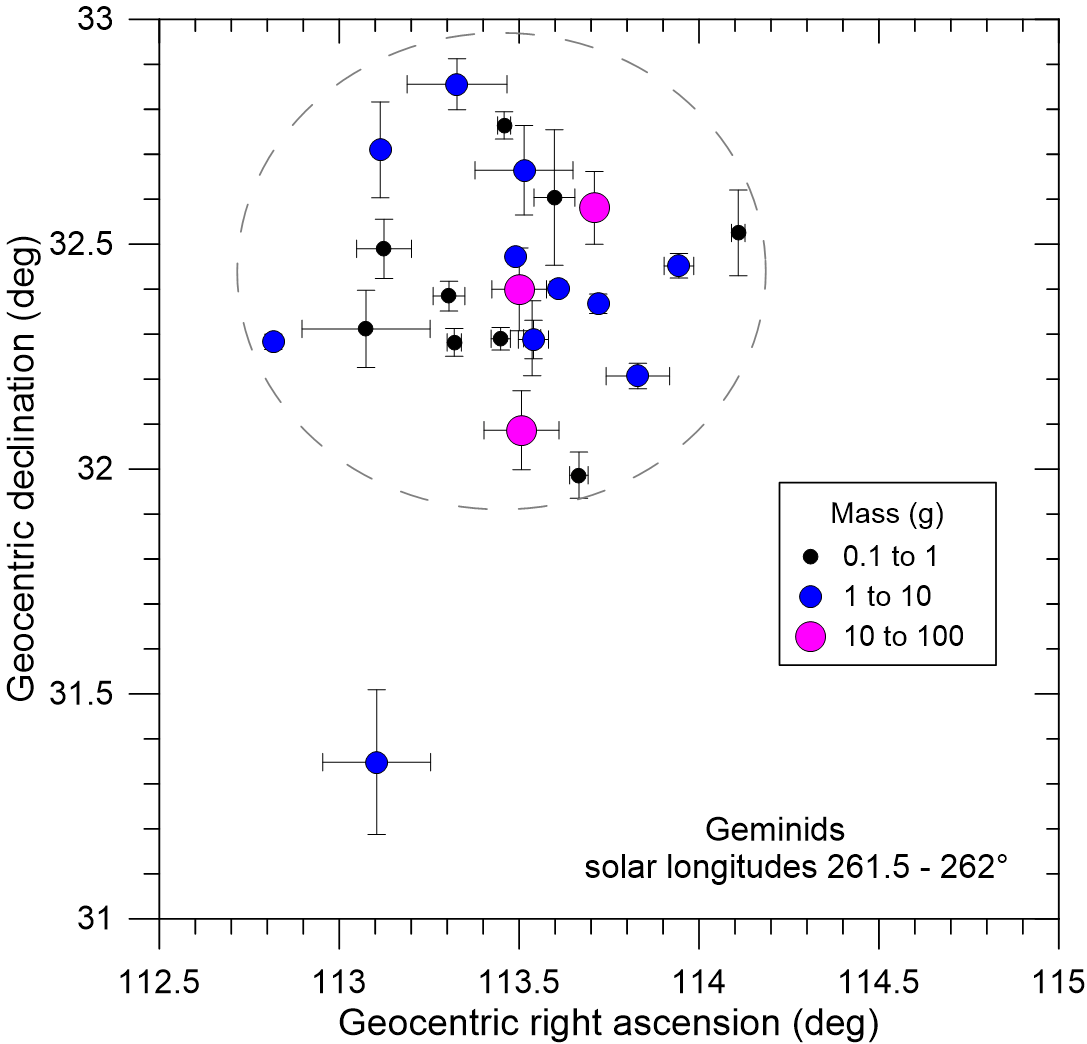}
   \caption{Geminid radiants during the shower maximum (solar longitudes 261.5\degr\ -- 262.0\degr).
   Actual radiants, not corrected for radiant motion, are shown. Symbol sizes and colors distinguish three intervals of meteoroid masses.}
   \label{GEMradiant}
   \end{figure}
   
Figure~\ref{GEMradiant} shows the radiant positions in a half-day interval during the shower maximum. 
Except for one outlier, the radiants are confined in an area of one degree in declination and one and a half degrees in right
ascension. \citet{Ryabova} computed the theoretical size and shape of radiant area under the assumption that 2000 years ago, meteoroids
were ejected from the parent body 3200 Phaethon by a cometary outgassing process. The computed
area is somewhat smaller than one degree for milligram meteoroids and is expected to be smaller for larger bodies.
Our radiants occupy a larger area, which may suggest that the ejection speeds were larger, or that the shower is older or 
was dispersed by some additional effects (e.g.,\ nongravitational forces). 

Our radiant is more compact than that reported by \citet{GMN_radiants} from video observations. 
Applying the methodology of subtracting radiant motion to all Geminds, we obtained 
a  median deviation from the mean radiant of 0.29\degr, while \citet{GMN_radiants} found 0.38\degr. 
This either suggests that larger meteoroids are indeed less dispersed, or that our data are more precise. \citet{KresakPorubcan}
obtained 0.49\degr\ from old photographic data.

   \begin{figure}
   \centering
   \includegraphics[width=\columnwidth]{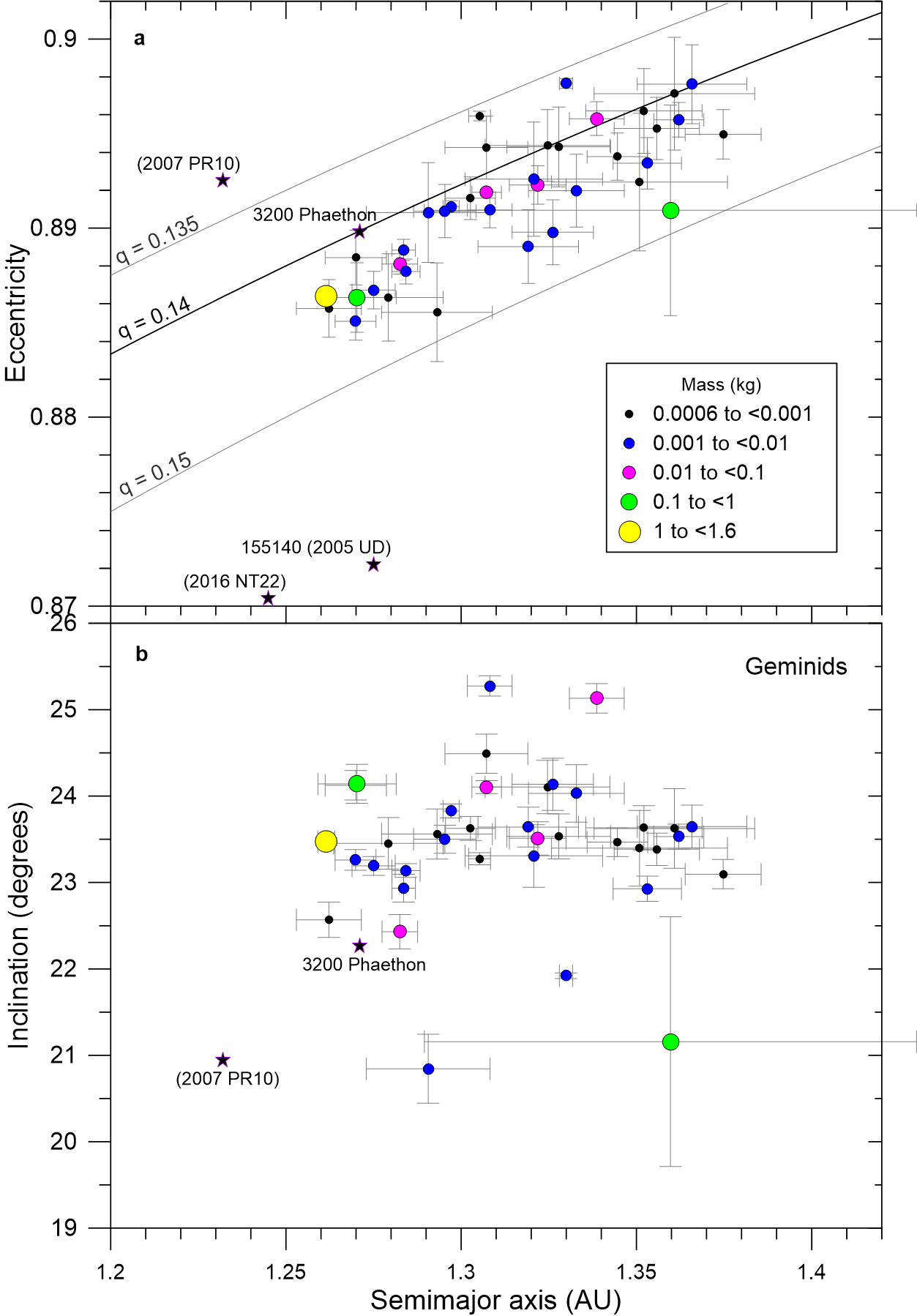}
   \caption{Eccentricity (\textbf{a}) and inclination (\textbf{b}) as a function of semimajor axis for Geminid
   meteoroids. Lines of constant perihelion distance (in AU) are shown in panel \textbf{a}.
    Symbol sizes and colors distinguish five intervals of meteoroid masses. Four NEO asteroids fall in the $a$-$e$ range of panel
\textbf{a} and two of them have inclinations in the range of panel \textbf{b}.}
   \label{GEMorbit}
   \end{figure}

   \begin{figure}
   \centering
   \includegraphics[width=\columnwidth]{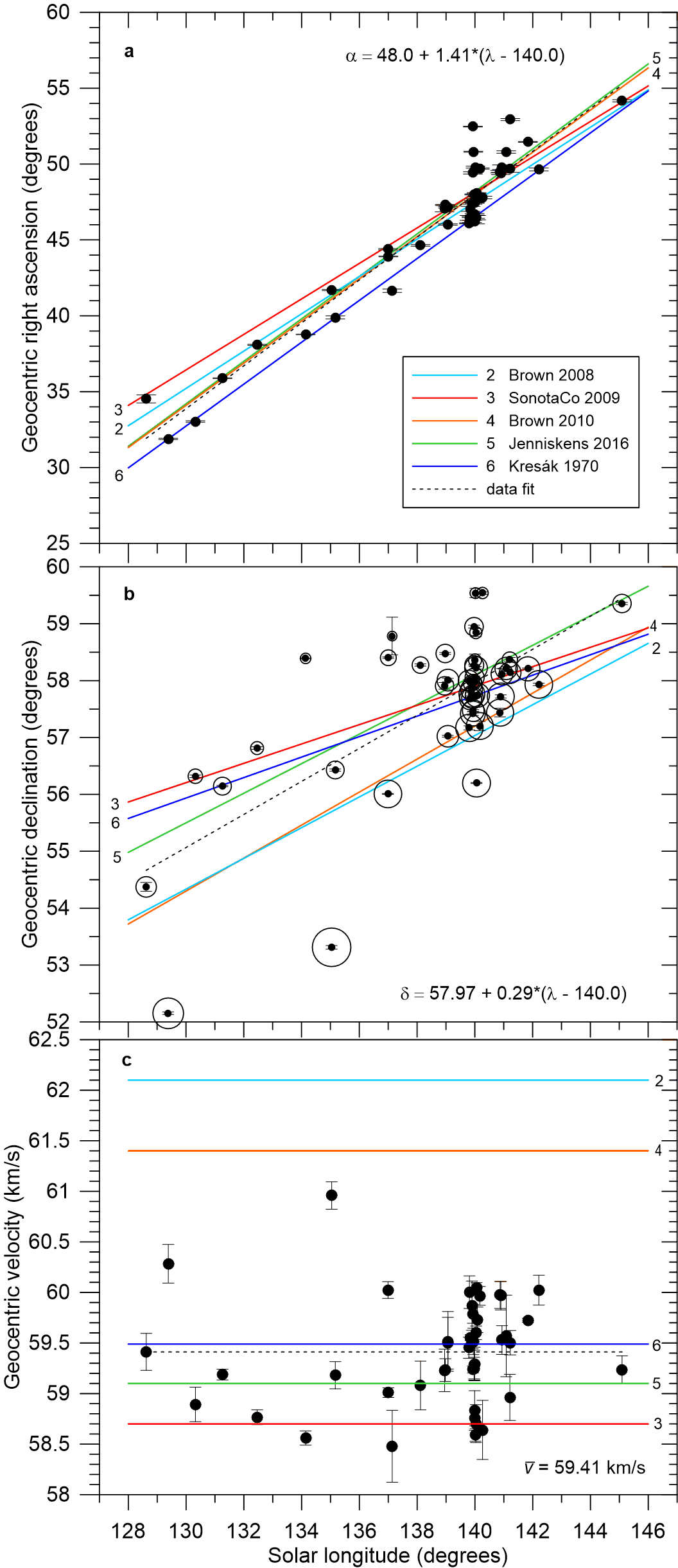}
   \caption{Motion of the Perseid geocentric radiant. Right ascension (panel \textbf{a}), declination (\textbf{b}),
   and velocity (\textbf{c}) are plotted as a function of solar longitude (all in equinox J2000.0) for individual meteoroids
   with formal error bars.  The sizes of the additional circles in plot \textbf{b} are proportional to velocity to show the correlation
   between declination and velocity.
   The data fit is plotted as a dotted line in all plots and the corresponding equation is inserted
   (no change of velocity with solar longitude was assumed).
   The motion of the mean radiant as reported by authors cited in the IAU Meteor Data Center is plotted by solid lines
   as follows: 2--5 see caption of Fig.~\protect\ref{GEMmotion}; 6 -- \citet{KresakPorubcan}.}
   \label{PERmotion}
   \end{figure}
   
An interesting pattern is seen in Fig.~\ref{GEMorbit}, where eccentricity and inclination are plotted against
the semimajor axis. There seems to be a core and a wing of the stream. The core has a semimajor axis similar to that of Phaethon
and a somewhat lower eccentricity, and thus a larger perihelion distance than Phaethon ($\sim$ 0.145 AU vs. 0.140~AU). 
The wing extends from the core to larger semimajor axes and somewhat lower perihelion distances (but with a larger scatter).
The core is rather extended in inclination, from that of Phaethon to about two degrees larger.
The wing is more concentrated in inclination but with some outliers. Core and wing meteoroids are mixed
in the stream, and meteoroids of both components are encountered throughout the duration of the shower.
The core meteoroids are those with lower geocentric velocities (cf. Fig.~\ref{GEMmotion}c). 
Comparing the data with the computations of \citet{Ryabova2}, we can see that 
the core consists of meteoroids of smaller masses in her model than the wing meteoroids. Her model, however, does not produce any Geminids 
with semimajor axes larger than 1.32~AU, while in reality 
the wing extends up to 1.375 AU.

\subsubsection{Perseids}

The radiant motion of Perseids (Fig.~\ref{PERmotion}) in our data  agrees with those of other authors, and most closely with that of \citet{CAMS}. 
The average geocentric velocity is closest to that of \citet{KresakPorubcan}. The velocities reported from radar \citep{Brown2008, Brown2010}
are significantly higher. There is a clear correlation between velocity and declination of the radiant in our data (Fig.~\ref{PERmotion}b): 
the lower the declination (in comparison with the value expected for the given solar longitude), the higher the velocity.

   \begin{figure}
   \centering
   \includegraphics[width=0.85\columnwidth]{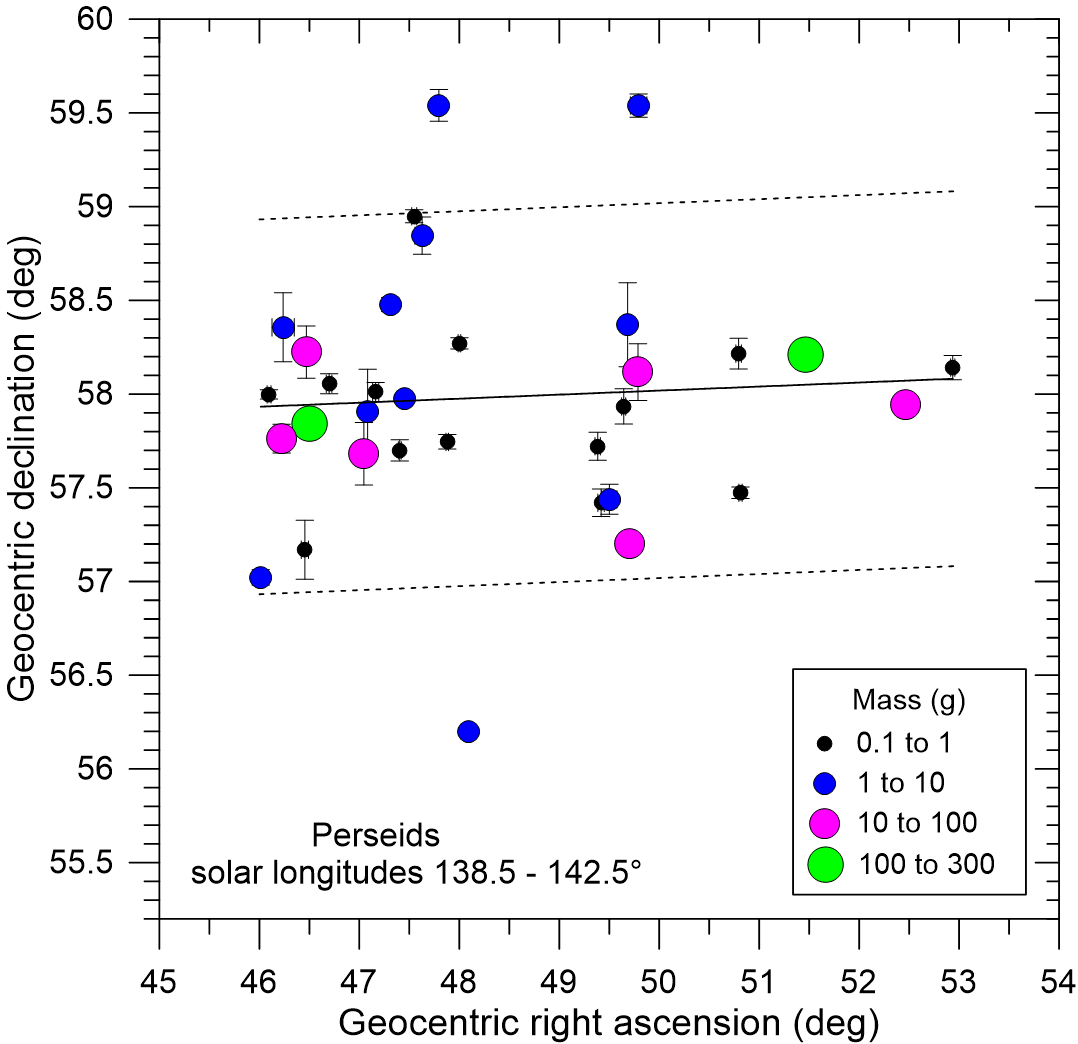}
   \caption{Perseid radiants during four days of high activity of the shower (solar longitudes 138.5\degr\ -- 142.5\degr).
   Actual radiants, not corrected for radiant motion, are shown. Symbol sizes and colors distinguish four intervals of meteoroid masses.
   The solid line is a linear fit to the data. Dashed lines indicate the interval $\pm$1\degr\ in declination around the fit.}
   \label{PERradiant}
   \end{figure}
   
      \begin{figure}
   \centering
   \includegraphics[width=\columnwidth]{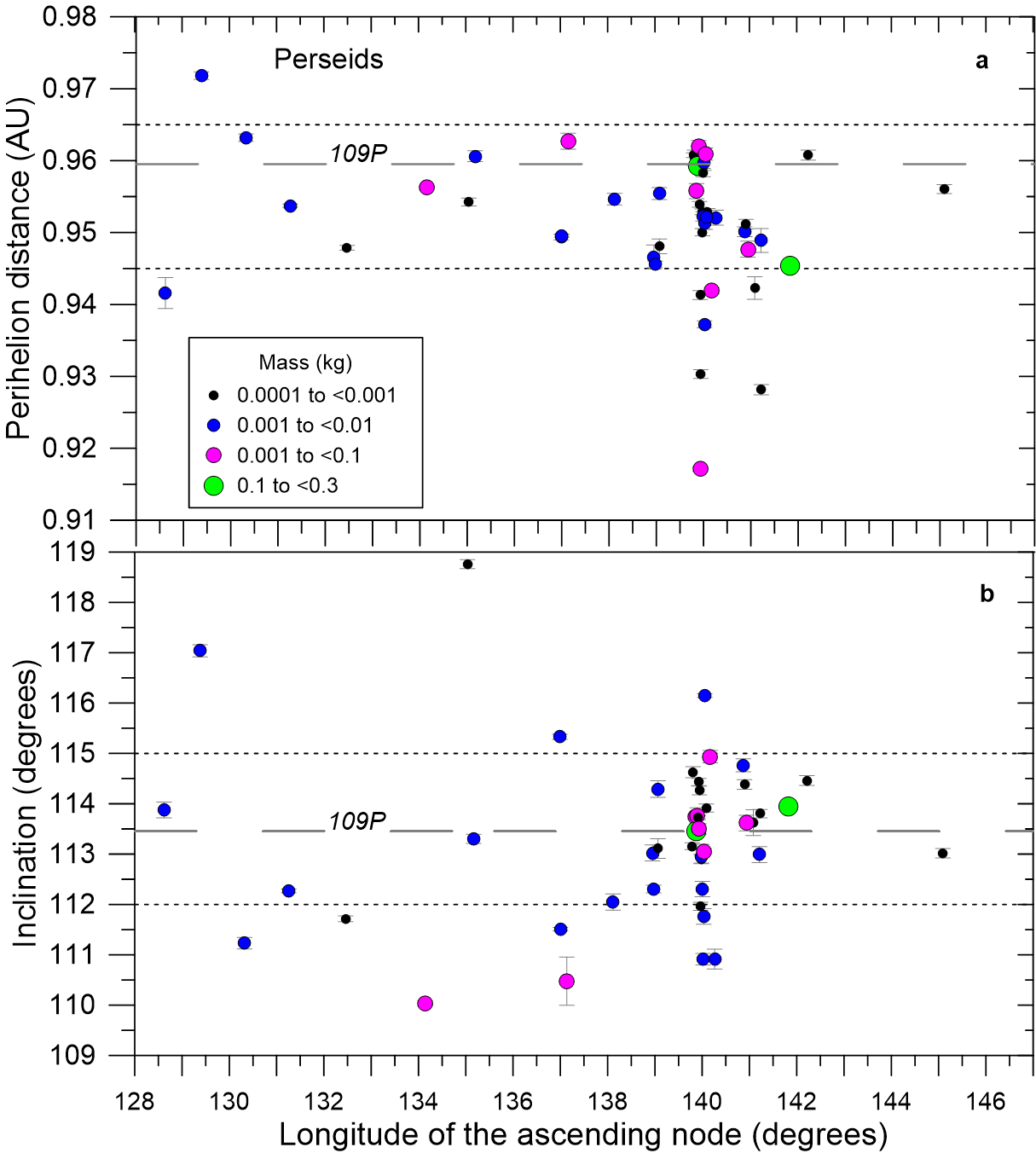}
   \caption{Perihelion distance (\textbf{a}) and inclination (\textbf{b}) as a function of the longitude of ascending node for Perseid
   meteoroids. Symbol sizes and colors distinguish four intervals of meteoroid masses. The dashed gray lines show the values for the parent comet
   109P/Swift-Tuttle. The dotted lines mark the intervals where most Perseids are confined.}
   \label{PERorbit}
   \end{figure}

Figure~\ref{PERradiant} shows the radiant positions in a four-day interval during the shower maximum. 
Nearly 90\% of radiants are confined to a band two degrees wide in declination. The extent in right ascension is
partly due to radiant motion and is partly real (cf. Fig.~\ref{PERmotion}a). If all Perseids are used and radiant motion is subtracted,
the median deviation from the mean radiant is 0.88\degr. \citet{GMN_radiants} obtained 1.14\degr\ and \citet{KresakPorubcan}
reported 1.26\degr.

Individual eccentricities and semimajor axes are difficult to study for high-velocity meteors such as Perseids, because they are very sensitive
to velocity determination. The average eccentricity of our 46 Perseid fireballs is $0.970 \pm 0.003$. The eccentricity of the parent comet 
109P/Swift-Tuttle is 0.963. Therefore, the eccentricities and semimajor axes of meteoroids in our size range probably do not differ much from
the parent comet. A similar result ($e=0.96$) was obtained from 254 photographic Perseids by \citet{KresakPorubcan} 
and from 4367 video Perseids by \citet{CAMS}, who listed $e=0.95$. 
The value $e=0.896$ reported recently from video data (10424 meteors) by \citet{Vida_GMN} is probably too small.

Perihelion distances and inclinations can be reliably computed for individual meteoroids and are plotted in Fig.~\ref{PERorbit} as a function
of the longitude of ascending node (virtually equal to the solar longitude). Perihelion distances of most Perseids across the whole stream
are between 0.945 -- 0.965 AU, and are therefore similar to, or slightly lower than that of the parent comet ($q=0.9595$ AU). 
Nevertheless, in the dense part of the stream around solar longitudes of 140\degr, about 20\% of meteoroids have lower
perihelion distances, down to 0.915 AU. They are those with the right ascension of the radiant larger than expected from 
the general trend (cf. Fig.~\ref{PERmotion}a). Low-perihelion Perseids seem to be absent, or at least less numerous 
in the outer parts of the stream, although the statistics are rather poor there. 

In contrast, orbital inclinations are concentrated
around the inclination of the parent comet ($i=113.45\degr$) in the dense part of the stream and are much more scattered
at lower solar longitudes. Outlying inclinations correspond to outlying declinations of radiants (cf. Fig.~\ref{PERmotion}b).
The lowest inclination, 110.0\degr, had fireball EN060818\_221424, which was observed very well and also had a high eccentricity, $0.995 \pm 0.006$. 
Our data suggest that meteoroids with lower inclinations have high eccentricities and 
that meteoroids with higher inclinations have lower eccentricities, but this cannot be proven because the uncertainties of
eccentricities are too high in most cases.

     \begin{figure}
   \centering
   \includegraphics[width=0.87\columnwidth]{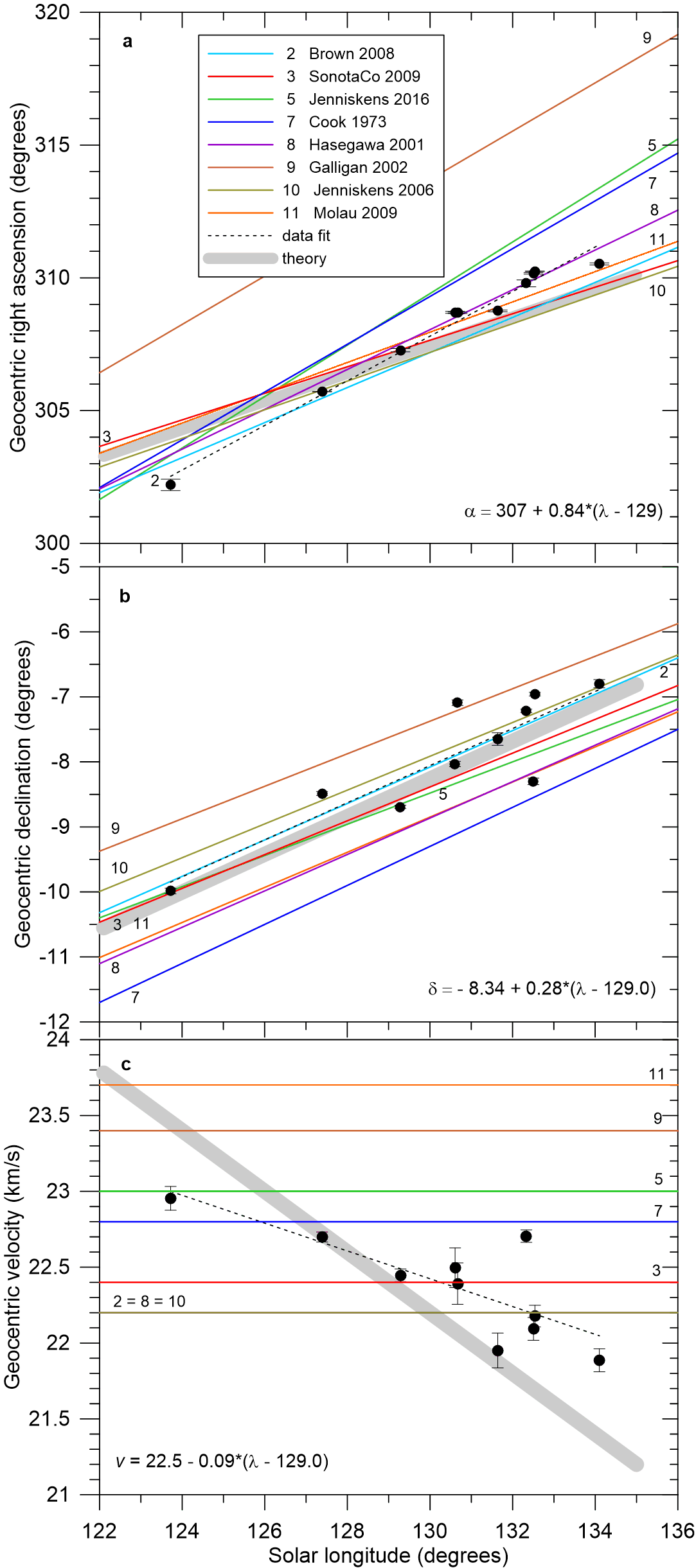}
   \caption{Motion of the $\alpha$ Capricornid geocentric radiant. Right ascension (panel \textbf{a}), declination (\textbf{b}), 
   and velocity (\textbf{c}) are plotted as a function of solar longitude (all in equinox J2000.0) for individual meteoroids
   with formal error bars.  
   The data fit is plotted as a dotted line in all plots and the corresponding equation is inserted.
   The motion of the mean radiant as reported by authors cited in the IAU Meteor Data Center is plotted by solid lines
   as follows: 2, 3, 5 see caption of Fig.~\protect\ref{GEMmotion}; 7 -- \citet{Cook};  8 -- \citet{Hasegawa}; 
   9 -- \citet{Galligan}; 10 -- \citet{Jennbook}; 11 -- \citet{MolauRendtel}. The thick gray lines show the motion of theoretical
   radiants for particles ejected from comet 169P/NEAT between 2000--5000 years ago \citep{Jenn_CAP}.}
   \label{CAPmotion}
   \end{figure} 

\subsubsection{$\alpha$ Capricornids}
\label{aCAP}

There are only ten $\alpha$ Capricornids in our sample but the radiant motion is well defined (Fig.~\ref{CAPmotion}).
Our data for right ascension are closest to the drift proposed by \citet{Hasegawa}. The data for declination
are in agreement with most authors cited in the IAU MDC. The velocities from single station video data \citep{MolauRendtel}
and from the AMOR radar \citep{Galligan} seem to be too high. Our data suggest a slight decrease in geocentric velocity 
during the shower activity. A similar result was obtained by \citet{GMN_radiants}.

Eight of the ten $\alpha$ Capricornids have semimajor axes between 2.55\,--\,2.7 AU, and eccentricities between 0.76\,--\,0.78, 
corresponding to perihelion distances 0.59\,--\,0.62 AU (the other two have a similar $q$ but $a$\,$\sim$\,2.8 AU). 
These parameters perfectly match comet 169P/NEAT (see Fig.~\ref{a-e}), which is likely the parent body of the stream \citep{Jenn_CAP}. 
Comet P/2003 T12 (SOHO), also mentioned in connection with $\alpha$ Capricornids \citep{CAMS}, is not far in this respect. 
But the inclinations and other angular elements of both comets are different. Nevertheless, as shown by \citet{Jenn_CAP},
meteoroids released from 169P/NEAT between 2000--5000 years ago could evolve into the current $\alpha$ Capricornid orbit.
The modeled theoretical radiants match the observations; the velocity match is somewhat worse (Fig.~\ref{CAPmotion}). 

More recently, it has been proposed that asteroid 2017 MB1 is associated with the stream \citep{Wiegert_CBET, Ye}. The matches of the angular elements and the perihelion distance are good,
but the semimajor axis of 2017 MB1 is only 2.374 AU. There are other, though smaller, asteroids with similarly good orbital matches:
2015 DA54, 2016 BN14, or 2019 CZ1. The latter two have short data-arc spans and rather uncertain orbits.

\subsection{Minor showers}

In addition to the 16 major and well-known meteor showers discussed so far in this paper, the associations of some
fireballs with minor showers are provided in the catalog. These associations should be considered as a suggestion. 
In this section we discuss some of these associations,
where at least three or a pair of closely related fireballs were detected.
Established showers are discussed first, followed by showers on the working list 
(as of November 2021). The tables for this section are given in Appendix~\ref{atables}.

\subsection{Established minor showers}

Five fireballs may be associated with either the Northern or Southern $\delta$ Cancrids (NCC and SCC, respectively).
They are listed in Table~\ref{NCC}. Only the selected orbital elements without errors and the $P\!f$ value characterizing
the physical properties of the meteoroids are given. Complete data can be found in the catalog. The $\delta$ Cancrids
are a diffuse ecliptical shower overlapping with the antihelion source, and are therefore subject to sporadic contamination.
None of our five fireballs can be firmly classified as a shower member. Three of them were observed within six hours on
January 20, 2017, and have mutually similar orbits, but these are somewhat different from the nominal orbit of $\delta$ Cancrids.
The perihelion distance was larger. All of them had $P\!f>1$, and were therefore resistant bodies likely of asteroidal
origin. Also, their orbits were asteroidal. It is possible that it was a random association.

Two fireballs were detected from the $\sigma$ Hydrids (HYD), July $\gamma$ Draconids  (184 GDR),
and $\kappa$ Ursae Majorids (KUM).
All these showers have long-period orbits, are well defined, and there is no doubt 
about the fireball shower membership (Table~\ref{HYD}). 
Our data suggest a larger eccentricity, and thus a longer orbital period of July $\gamma$ Draconids than previously thought.
Both $\sigma$ Hydrids and July $\gamma$ Draconids  
had  $P\!f\sim0.4$, which ranks them among the most resistant meteoroids on Halley-type orbits.
    The $P\!f$ of $\kappa$ Ursae Majorids was one order of magnitude smaller and they belong, on the contrary,
to the weakest meteoroids.

Two fireballs were possibly detected also from the x Herculids (XHE) and one from each of 13 other established
minor showers (AVB, NDA, TAH, COR, SSG, SZC, AUD, OCC, OCT, XUM, PPS, LUM, and SLD). 
The association with the shower is, however, uncertain in some cases.

\subsection{Showers on the working list}

Among the showers on the working list of the IAU MDC, the most associations were obtained
for the $\nu$ Draconids (NDR). This toroidal shower has a dispersed radiant \citep{CAMS2}
and random associations are, therefore, possible. Five possible $\nu$ Draconids  were
identified (Table~\ref{NDR}). Meteoroid EN180918\_030212 had high ablation resistance ($P\!f=1.4$) but also
rather low eccentricity. Its aphelion was at 4.1~AU, while the aphelia of the other four meteoroids 
were at 4.9\,--\,5.7 AU. It is quite possible that EN180918\_030212 did not belong to the $\nu$ Draconids. 
If so, the range of $P\!f$ of the $\nu$ Draconids was similar to the $\kappa$ Cygnids.

Table~\ref{SPI} provides a list of six fireballs detected over a period of one month
but with radiants close together in the ecliptic coordinates related to the Sun 
(near $\lambda-\lambda_\sun=198\degr$, $\beta=-5\degr$, see the plot in Paper I).
Their perihelion distances are also similar, so the fireballs may be related.
The associated showers may be Southern $\delta$ Piscids (SPI),
Southern October $\delta$ Arietids (SOA), or $\xi$ Arietids (XAR).
These showers are probably also related and differ mainly by the period of activity.
According to \citet{CAMS}, they may belong to the Encke Complex.
The fireball orbits are distinct from those of the Southern Taurids by smaller
perihelion distances and, except one, also by smaller semimajor axes.
Inclinations are also slightly higher in most cases. Fireballs EN150918\_231506 
and EN131018\_020534 may be random
interlopers (their semimajor axes and inclinations differ from the other four fireballs)
but the other four may belong to a single shower active at 
the end of September and in the first half of October. All four were 
small bodies with masses up to one gram and their $P\!f$ was similar to
those of Taurids of the same sizes, in other words relatively high.

Four fireballs could be associated with each of these showers on the working list:
$\lambda$ Ophiuchids (460 LOP), December $\zeta$ Taurids (638 DZT), and
$\lambda$ Leonids (733 LAL). All these proposed showers have indistinct orbits
with low inclinations and a Tisserand parameters near three. 
Since there are many sporadic meteors with this type of orbit and the possibly associated
fireballs are not clustered, we cannot confirm the detection of any of these showers. 
The situation is the same for two showers with three fireballs: 
June $\epsilon$ Ophiuchids (459 JEO) and February $\pi$ Leonids (501 FPL).
In fact, four associations were also detected for the April $\beta$ Sextantids (449 ABS)
and three for May $\alpha$ Comae Berenicids (455 MAC),
which have both been removed from the working list and are now considered nonexistent.
This demonstrates that most associations with these proposed ecliptical showers are probably random.

There is a somewhat different situation with the April $\psi$ Ursae Majorids (133 PUM), with three possible
fireballs. The orbits of the fireballs are quite different but there are also several different orbits of this shower
listed in the IAU MDC. Different fireballs were therefore paired with orbits of different authors. 
Thus, the detection of this shower cannot be confirmed.

The detection of the toroidal shower August $\mu$ Draconids (AMD) is more probable. 
Three candidate fireballs are listed in Table~\ref{AMD}. The shower radiant
is close to that of established showers $\kappa$ Cygnids (KCG) and August
Draconids (AUD), and all three showers may be related \citep{CAMS}. Two
fireballs had $P\!f$ values similar to $\kappa$ Cygnids but
EN260817\_185317 was much more resistant. Since it also had a larger eccentricity,
it may in fact not be a member of the shower.

The detection of $\xi^2$ Capricornids (XCS) is even better. Three candidate fireballs with quite
similar radiants and orbits were detected within one degree of solar longitude (Table~\ref{AMD}).
The shower may be related to $\alpha$ Capricornids and comet P/2003 T12 (SOHO) \citep{CAMS}.
$\xi^2$ Capricornids have smaller perihelion distances and semimajor axes than
 $\alpha$ Capricornids. All three fireballs
exhibited bright flares and the $P\!f$ values were similarly low as for $\alpha$ Capricornids.
The photometric masses ranged from 30 grams to almost a kilogram.
Other showers from the working list have, at most, two possible detections.

    \begin{figure}
   \centering
   \includegraphics[width=0.95\columnwidth]{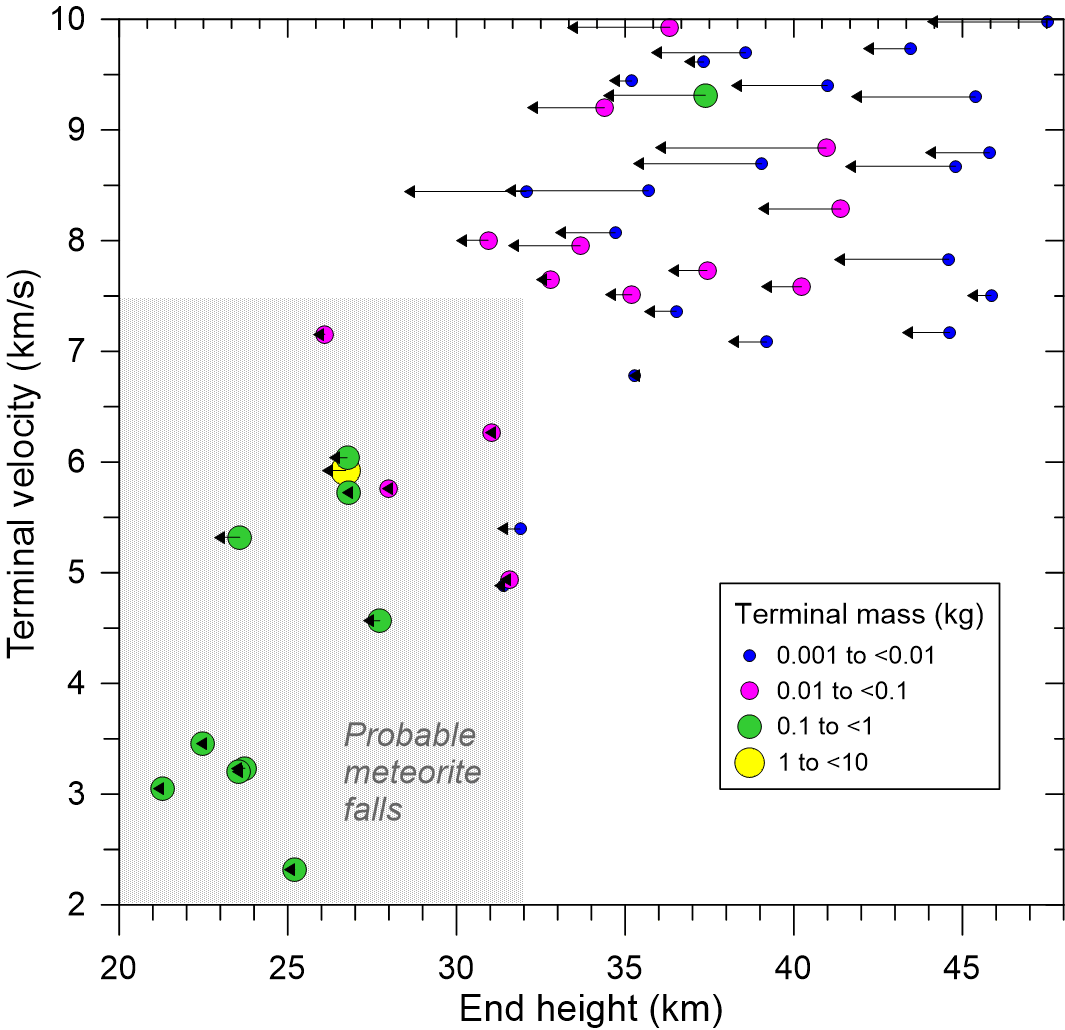}
   \caption{Terminal velocities and end heights for fireballs with computed terminal masses larger than one gram. 
   The terminal mass is coded by the symbol size and color. The symbols are placed at the heights where terminal
   velocities were measured. The arrows point to the actual end heights, which are lower in cases
    where velocity was difficult to measure toward the fireball end. The velocity at the end is expected to be lower 
   than shown in these cases. The shaded area encompass fireballs with a high probability of dropping a meteorite
   larger than one gram.}
   \label{termass}
   \end{figure} 

\section{Terminal masses and meteorite falls}
\label{meteoritesection}

One of the purposes of fireball networks is the recovery of meteorites. In 2017--2018, the European
Fireball Network recovered one meteorite -- Renchen \citep{Spurny_Bratislava_talk,Bischoff_Renchen}. That fireball
and meteorite fall occurred in western Germany near the French border. It was captured by
two digital cameras close to the horizon but the velocity could be measured only on the
cameras of the German part of the network. The fireball is therefore not included in this catalog. 
A paper devoted to the Renchen fall will be published elsewhere.

The catalog contains terminal masses indicating a possibility of meteorite fall. However, the masses are 
only approximate and result from the four-parameter velocity fitting of the whole fireball, ignoring 
any fragmentation. Terminal masses are listed only if they are larger than one gram and if the
terminal velocity is lower than 10 km s$^{-1}$.

Figure~\ref{termass} shows the end heights and terminal velocities for fireballs with nonzero terminal masses.  In some cases, the terminal velocity could not be measured at the end height because
the shutter breaks were either faint and noisy, or not well separated. Therefore, both the height of the last velocity
measurement and the end height are plotted. The measured terminal velocity is given, while 
the terminal mass is computed for the end height as a result of extrapolation of the four-parameter fit. 

We can see that there are many fireballs with terminal velocities above 7 km s$^{-1}$ measured
at heights above 30 km. Terminal masses are below 0.1 kg, except for one case where the terminal mass is slightly above
this value (0.13 kg for EN071118\_010142). Since ablation continues down to $\sim3$  km s$^{-1}$, it can be
expected that most of these fireballs did not drop any significant meteorite. In an exceptional case, that of
EN071118\_010142 
over northern Poland, the relatively high end height and terminal mass may have been caused by the large distance to the fireball, 
and a non-negligible meteorite fall cannot be excluded.

    \begin{figure}
   \centering
   \includegraphics[width=0.75\columnwidth]{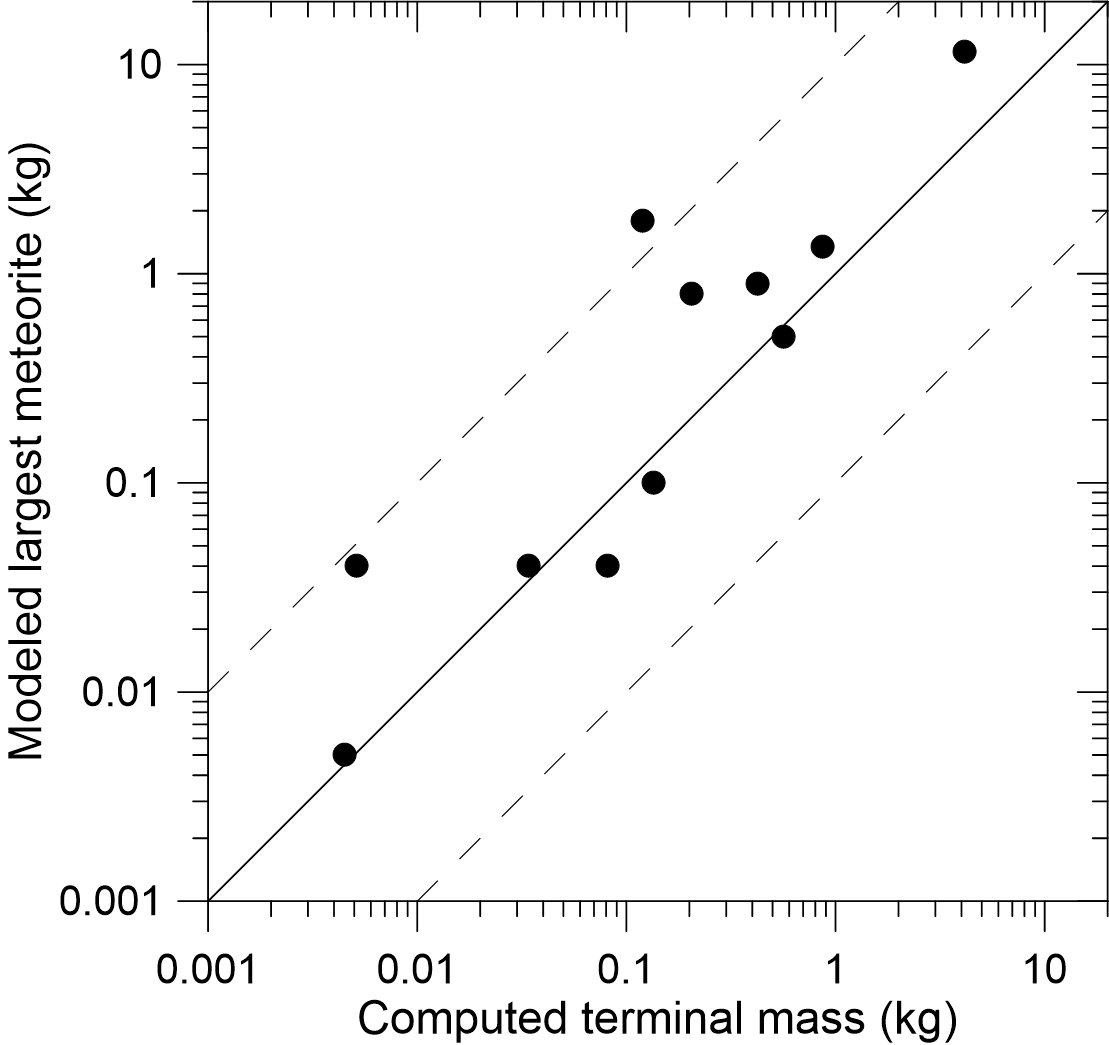}
   \caption{Comparison of terminal masses from the present catalog with the mass of the largest expected meteorite
   resulting from fragmentation modeling by \citet{2strengths} for 11 fireballs. The solid line marks
   equality of both approaches and the dashed lines mark an order of magnitude difference.}
   \label{termass-model}
   \end{figure} 

The much more likely meteorite falls are those with end heights below 32 km and terminal velocities
below 7.5 km s$^{-1}$. This area is shaded in Fig.~\ref{termass} and contains 16 fireballs. Eleven
of them have been included in the fragmentation modeling of \citet{2strengths}. Fragmentation
modeling provides more reliable estimates of meteorite masses by considering the fragmentation points
and fitting the dynamics from the last fragmentation toward the end point. Figure~\ref{termass-model}
shows that the fragmentation model expects a larger meteorite than the routine procedure in most cases,
and even by an order of magnitude in two cases. Nevertheless, even the result of fragmentation modeling is
just an estimate. Not only does the mass depend on the assumed meteorite density and shape, but the meteorites
sometimes break-up further during the dark flight, as evidenced by incomplete coverage by the fusion crust
\citep[see e.g.,][]{Bischoff_Renchen}. We note that the mass of the largest meteorite is discussed here
but in many cases the fall also includes numerous small meteorites, and it may be more likely to find some of them
\citep[see e.g.,][]{Zdar}. Searches of various intensities were performed for ten meteorite falls discussed here,
without success.
   
     \begin{figure}
   \centering
   \includegraphics[width=\columnwidth]{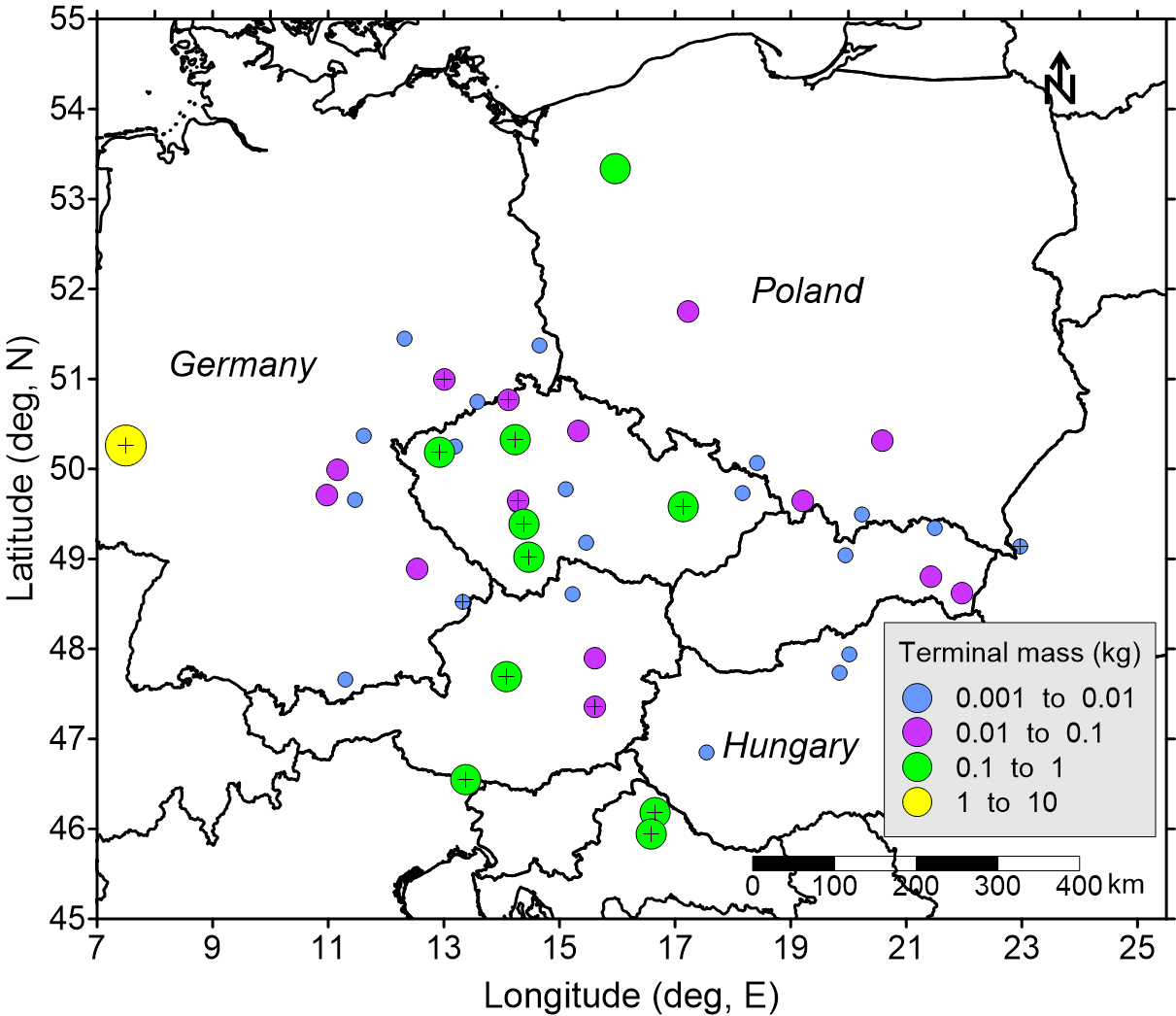}
   \caption{Geographic distribution of fireballs with computed nonzero terminal masses. Those fireballs that probably ended with
   meteorite falls are marked by a cross. Most of the others probably ablated out after the last velocity measurement.
   The plotted positions are for the fireball end.}
   \label{meteoritemap}
   \end{figure}

    \begin{figure}
   \centering
   \includegraphics[width=\columnwidth]{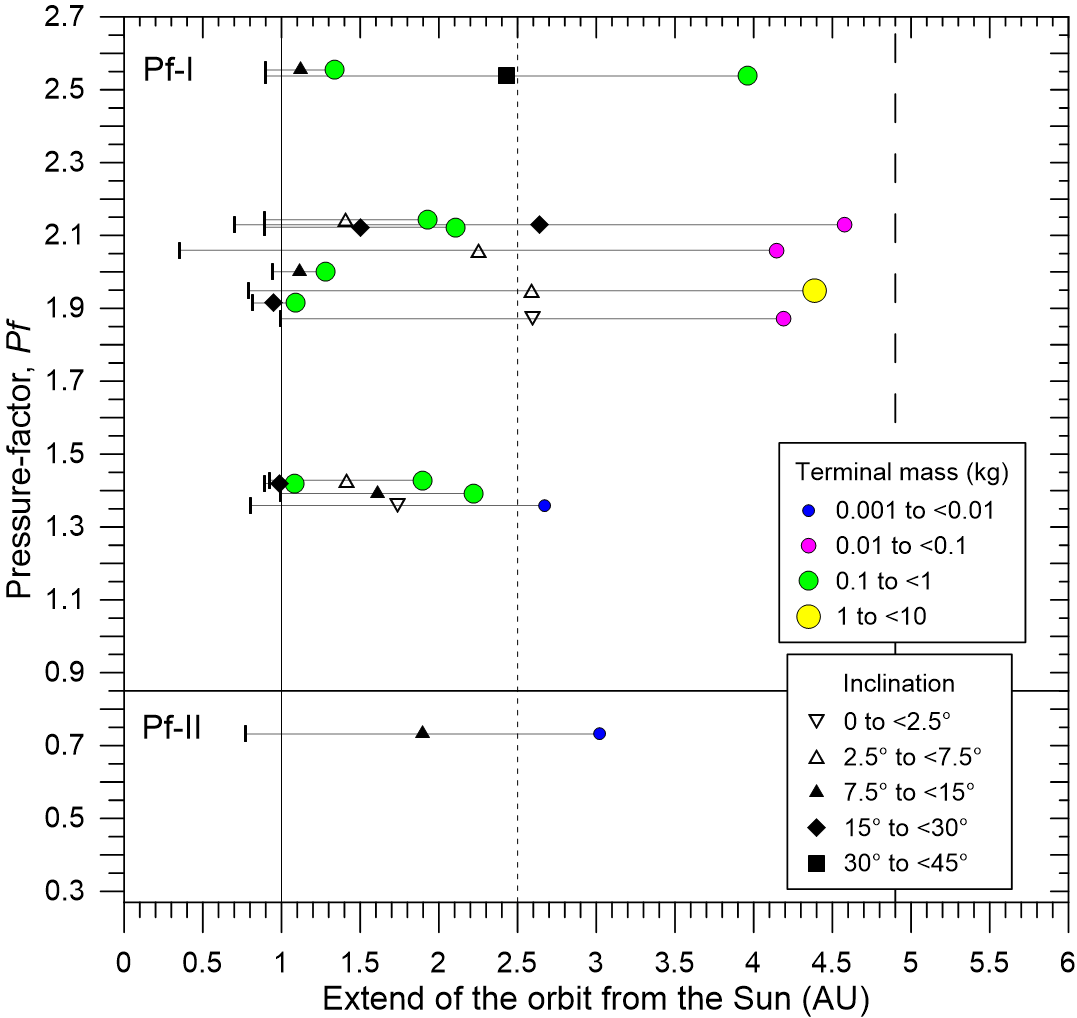}
   \caption{Pressure factors (vertical axis) and sizes of the orbits (horizontal axis) for 15 probable
   meteorite falls. Each fireball is characterized by three symbols connected with a horizontal line.
   The vertical bar is plotted at the perihelion distance ($q$), a symbol encoding the orbital inclination ($i$) is plotted
   at the semimajor axis ($a$), and a colored circle encoding the terminal mass computed from the whole-trajectory fit
   is plotted at the aphelion distance ($Q$). Vertical lines indicate the Earth orbit (1 AU), 3:1 resonance with Jupiter (2.5 AU),
   and the limit for asteroidal orbits according to this paper (4.9 AU). The horizontal line indicates the boundary between
   the Pf-I and Pf-II classes (0.85). Fireball EN311018\_161746 could not be included since it was observed during twilight
   and has no photometry or $P\!f$. The orbital parameters are $q=0.925$ AU, $a= 2.45$ AU, $Q= 3.98$ AU, and $i= 8.6\degr$.}
   \label{meteoriteorbit}
   \end{figure} 

Satisfying the conditions $h_{\rm e}<32$ km and $v_{\rm e}<7.5$ km s$^{-1}$, though stricter than that of
\citet{Halliday_fall} who gave rough limits 35 km and 10 km s$^{-1}$, cannot be considered as
sufficient for a meteorite fall. It is necessary to evaluate also the deceleration and light curve.
Nevertheless, all 16 fireballs from the present sample seem to be good candidates for dropping
meteorites (although quite small meteorites in some cases). 
It is therefore worthwhile to take them as a special class of fireballs -- likely meteorite droppers -- and evaluate
their orbits and pressure factors.
Their geographic distribution is shown in Fig.~\ref{meteoritemap}, together with the less promising candidates.

The most important orbital parameters and $P\!f$ values of meteorite droppers are shown
in Fig.~\ref{meteoriteorbit}\footnote{The orbits of EN141117\_164658, EN080418\_184736, and EN230518\_194647 have been
slightly changed in the catalog in comparison with those published in \citet{2strengths} since the method of
computing velocities from supplementary video cameras has been improved.}.
Not surprisingly, all orbits are asteroidal with aphelia $Q<4.9$ AU and low to moderate inclinations (up to 35\degr).
Two were on Aten-type orbits with low eccentricities and inclinations about 20\degr. As expected, the pressure factors
were mostly high, between 1.3 and 2.6. Only fireball EN160517\_205435 had $P\!f$=0.7. It was probably a carbonaceous body
according to \citet{2strengths} and the modeling showed that it dropped only a few gram-sized meteorites.

We can also roughly evaluate meteorite fall statistics.
\citet{Halliday_flux} concluded, on the basis of the observation of the Canadian Meteorite Observation and Recovery Project (MORP),
that there are about nine meteorite falls per $10^6$ km$^2$ per year with a total fallen mass larger than 1 kg, and 58
falls dropping more than 0.1 kg. We detected 12 fireballs in two years
with the computed mass of the main meteorite larger than 0.05 kg. The total fallen mass can be reasonably expected to
exceed 0.1 kg in these cases. 
Our covered area is $7\times10^5$ km$^2$. Due to weather limitations, cameras exposed 
40 -- 60\% of the prescribed dark time at different stations (52\% on average). The covered fraction of the
total time (day+night) was 18 -- 26\%. Since this time also includes the times when the skies were mostly
cloudy and cameras were running, we use the lower limit. It gives us 48 meteorite falls per $10^6$ km$^2$ per year, 
in reasonable agreement with \citet{Halliday_flux}. 

There were six fireballs with main mass exceeding 0.5~kg (see Fig.~\ref{termass-model}), 
with the total fallen mass thus probably exceeding 1~kg.
It gives a rate about two times that of \citet{Halliday_flux} even when we take into account the fact that these brighter events
can also be observed in poor conditions. 
But here we enter low number statistics and it seems that the covered years were somewhat richer
in larger meteorite falls than average years.

\section{Possible related asteroids}

Some of the major meteor showers, such as the Perseids, Geminids, Leonids, or Lyrids, have well known parent bodies with orbits
similar to the orbits of the meteors. They do not need to be discussed here. The situation may be more complicated
for some other showers. The Taurids are a good example. Their parent body is most likely comet 2P/Encke.
\citet{Spurny_Taurids} have, nevertheless, identified several asteroids, most notably 2015 TX24, with orbits more
similar to the resonant branch of the Southern Taurids. These asteroids can be considered as large members of the
stream rather than parent bodies. \citet{Egal} confirm their orbital convergence with 2P/Encke and also with 2004 TG10, which is
now related to the Northern Taurids, about 5000 years ago. Here we note the close orbital similarity of another asteroid, 2014 NK52,
with the Northern Taurids, namely with those observed around mid-November (while the orbit of 2004 TG10 is more similar to NTA meteors
observed earlier in November). 2014 NK52 was not included in the work of \citet{Egal}. That work did include 2003 WP21,
which shows orbital similarity with the Southern Taurids observed by us toward the end of November, but orbital convergence 
with 2P/Encke was not found in this case.

    \begin{figure*}
   \centering
   \includegraphics[width=1.4\columnwidth]{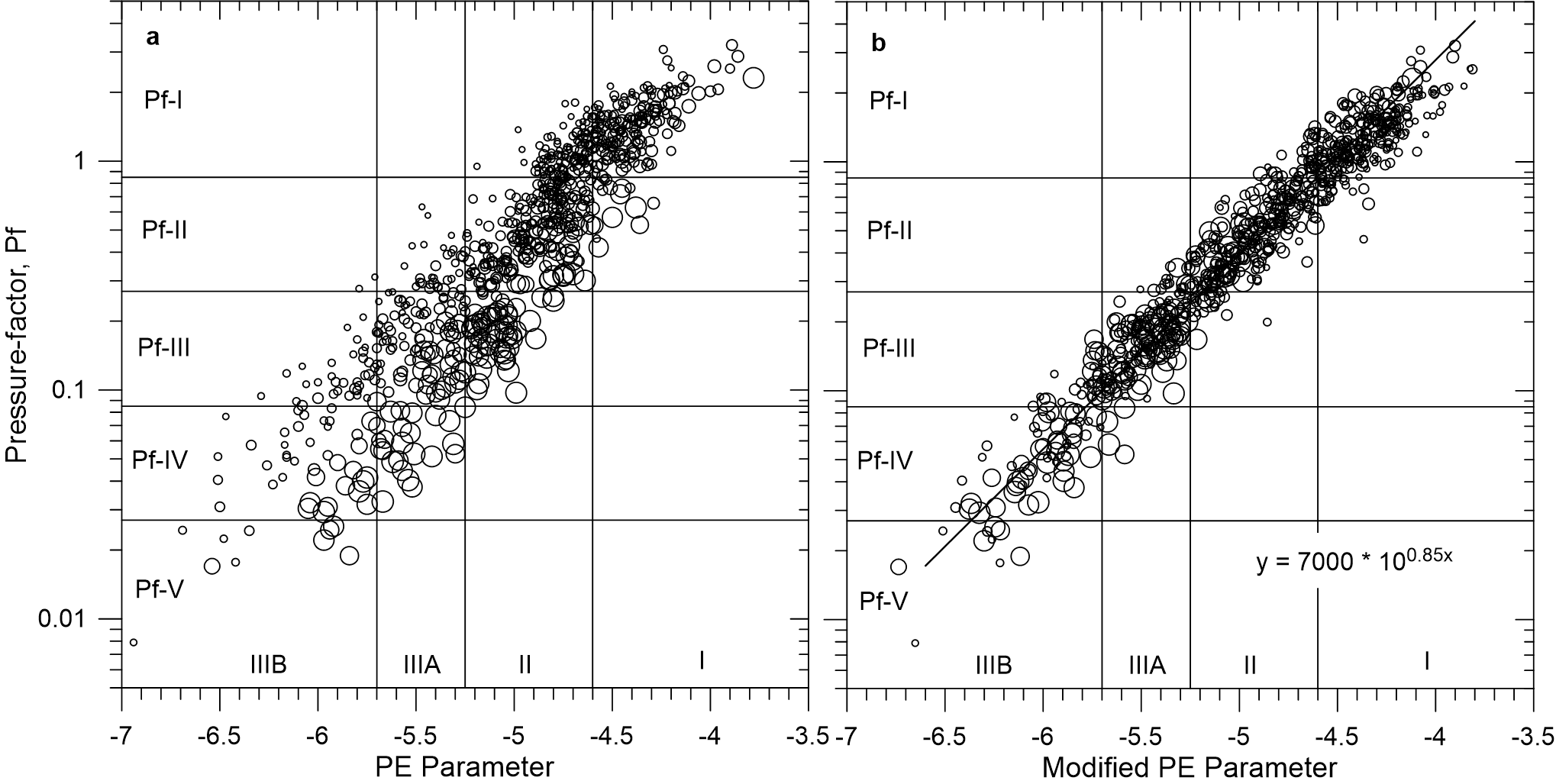}
   \caption{Relation between the $P_E$ and $P\!f$ parameters for all fireballs (panel \textbf{a}). The symbol size is
   proportional to fireball initial velocity. Panel \textbf{b} shows the same relation for the $P_E$ parameter modified according to 
   Eq.~(\ref{PEmodification}). A simple function is drawn through the data.}
   \label{PE-Pf}
   \end{figure*}

A situation similar to the Taurids is encountered in $\alpha$ Capricornids. Several asteroids, discussed in Sect. \ref{aCAP}, have
more similar orbits to the fireballs than the likely parent comet 169P/NEAT. The parent comets of both showers,
2P/Encke and 169P/NEAT, have orbits
classified as asteroidal (see Fig.~\ref{a-e}). The physical properties of Taurid and $\alpha$ Capricornid meteoroids are, 
nevertheless, different (see Sect. \ref{shower_phys}).

The orbits of two of the three observed August $\mu$ Draconids (Table~\ref{AMD}) were found to be close to the orbit
of asteroid 2002 GJ8. That asteroid is on unstable cometary orbit \citep{Fernandez} and has a low albedo 
of 0.018 \citep{SpitzerNEO}\footnote{Data acquired from http://nearearthobjects.nau.edu/spitzerneos.html on January 6, 2022}. 
It was discussed in connection with $\kappa$ Cygnids and related showers by \citet{Shiba_KCG}.
As he noted, the current orbit was quite different before the encounter with Jupiter in 2015. Thus, the current
orbital similarity with August $\mu$ Draconids is probably due to chance.

Asteroid 2003 AA83 has been associated with three fireballs (EN200117\_181106, EN200117\_192130,
EN230117\_202629). Two of them are possible $\delta$ Cancrids, but the orbital similarity is not particularly high.
Moreover, asteroid 2003 AA83 was observed for only 6 days and its orbit is uncertain.

An additional run was done at the beginning of 2022, comparing the orbits of all known NEAs with sporadic
fireballs on asteroidal orbits. The only asteroid having $D_{\rm SH} < 0.05$ with three or more fireballs,
where $D_{\rm SH}$ is the dissimilarity criterion of \citet{SH}, was found to be 2019 DN. The only asteroid with two associated fireballs was 2020 GV1. 
The comparison of the orbits for 2019 DN with the three fireballs is provided in Table~\ref{2019DN}. 
Given that the fireball appearances span almost the whole month of March,
this also is probably a chance coincidence. 
For two of the fireballs, asteroids with orbits even more similar than the orbit of 2019 DN exist.

We can conclude that, except for the Taurids and $\alpha$ Capricornids (and the well-known cases of Geminids), 
we have not found any convincing orbital similarity between multiple fireballs observed in 2017--2018
and asteroids. It does not mean that some individual links cannot exist.

\section{Discussion}
\label{discussion}

\subsection{The pressure factor, $P\!f$}

We have proposed a new one-dimensional metric to characterize meteoroid physical properties, which seems to be 
better than
the $P_E$ criterion of \citet{PE}. It is based on the maximal dynamic pressure encountered along the meteoroid 
trajectory, and we therefore called it the pressure resistance factor, or pressure factor for short, and abbreviated it as $P\!f$.
Provided that the fireball velocity could be measured along most of the trajectory, dynamic pressure is less dependent 
on observing conditions than the end height, which is the basis of the $P_E$ criterion. Both metrics take into account
meteoroid initial mass, velocity, and trajectory slope (in the case of $P_E$, unrealistically high initial mass resulting from the
use of historical values of luminous efficiency must be entered). Both also need the knowledge of atmospheric density
as a function of height.

Figure~\ref{PE-Pf}a compares the values of $P_E$ and $P\!f$ for our sample of fireballs. The main difference
turns out to be the velocity dependence. If the $P_E$ is modified as follows:
\begin{equation}
P_E \textrm{(modified)} = P_E - \log v_\infty + 1.5,
\label{PEmodification}
\end{equation}
($v_\infty$ is in km s$^{-1}$), much better correspondence between the two metrics is achieved (Fig.~\ref{PE-Pf}b).
The remaining differences can be ascribed to the difference between considering end height and maximum dynamic
pressure. For example, the modified  $P_E$ ascribes very high strength ($P_E \ga -4$) to some very slow
fireballs while $P\!f$ does not reach any extreme values ($P\!f \la 2$).

\begin{table*}
\caption{Pressure factors, $P\!f$, for bright meteorite-dropping fireballs and superbolides without meteorites.}
\label{Pf-large}
\begin{tabular}{l l l l l l l l}
\hline \hline  \noalign{\smallskip}
Fireball & Meteorite type  & $m_\infty$ & $v_\infty$ & $p_{\rm max}$ & $z$ & $P\!f$ & References   \\
 & or Bolide date& kg &  km/s & MPa & \degr  \\
\hline \hline  \noalign{\smallskip}
&\multicolumn{5}{l}{\textit{Ordinary and enstatite chondrites}} \\[0.25ex]
Chelyabinsk & LL5 & 1.2$\times10^7$   & 19  & 18   &  72   & 0.31  & [1] \\
Ko\v{s}ice & H5 & 3500   & 15  & 6   &  30   & 0.78  & [2] \\
Mor\'avka & H5 & 1500 & 22.5 & 5 & 70 & 1.2 & [3,4] \\
Neuschwanstein & EL6 & 300 & 21.0 & 9.6 & 40 & 2.0 & [5,6] \\
Hamburg & H4 & 200 & 15.8 & 7 & 66 & 4.7 & [7] \\
\v{Z}\v{d}\!\'ar nad S\'azavou & L3.9 & 150 & 21.9 & 2.7 & 65 & 1.2 & [8] \\[0.5ex] 
&\multicolumn{5}{l}{\textit{Mixed breccias}} \\[0.25ex]
Almahata Sitta & Ureilite+E+H+..  & 50,000 & 12.8 & 1.3 & 70 & 0.23 & [9,10,6] \\
Bene\v{s}ov & LL3.5+H5+.. & 3500 & 21.3 & 9 & 9 & 0.61 & [11,12] \\[0.5ex]
&\multicolumn{5}{l}{\textit{Carbonaceous chondrites}} \\[0.25ex]
Tagish Lake &  C2 & 90,000 & 15.8 & 2.2 & 72 & 0.25 & [13,6] \\
Flensburg & C1 & 20,000 & 19.4 & 2& 65  & 0.20 & [14] \\
Maribo & CM2 & 2000 & 28.3 & 5 & 59 & 0.5 & [15] \\[0.5ex]
&\multicolumn{5}{l}{\textit{Large fragile bodies without meteorites}} \\[0.25ex]
\v{S}umava & 1974-12-04 & 5000 & 26.9 & 0.14 & 63 & 0.013 & [16] \\
Romanian & 2015-01-07 & 4500 & 27.8 & 3 & 47 & 0.18 & [17] \\
Taurid & 2015-10-31 & 650 & 33.1 & 0.19 & 78 & 0.055 & [18,19] \\
\hline \hline \noalign{\smallskip}
\end{tabular}
\tablefoot{[1] \citet{Chelyabinsk}, [2] \citet{Kosice}, [3] \citet{Moravka1}, [4] \citet{Moravka4},
[5] \citet{Neuschw}, [6] \citet{Popova2011}, [7] \citet{Hamburg}, [8] \citet{Zdar}, 
[9] \citet{Shaddad}, [10] \citet{Meteosat}, [11] \citet{Benesov}, [12] \citet{Benesov98},
[13] \citet{Hildebrand}, [14] \citet{Maribo}, [15] \citet{Flensburg},  [16] \citet{icarus96}, 
[17] \citet{Romanian}, [18] \citet{Spurny_Taurids}, [19] \citet{Taurid_phys}.}
\end{table*}

The pressure factor definition (Eq.~\ref{Pf}) was found empirically using the present sample of fireballs, which covers
meteoroid masses from about $10^{-4}$  to $10^2$ kg. In order to see if $P\!f$ values are reasonable for
much larger meteoroids, we computed them in Table~\ref{Pf-large} for some more massive meteorite falls 
and very bright fireballs (superbolides) that disintegrated in the atmosphere completely. The input data
including dynamic pressures were compiled from the literature. Velocities and trajectory slopes are 
well known in all cases, but dynamic pressures and especially meteoroid masses may be less certain.

The fireballs are divided into four groups according to meteorite type. Within the group, they are sorted 
according to initial mass. For ordinary and enstatite chondrites, we expect $P\!f>0.85$ corresponding to class
Pf-I. It is so for most of them but not for Chelyabinsk, by far the largest body of all. It is reasonable 
to expect that above a certain mass limit, the maximum dynamic pressure does not depend, or depends 
only weakly, on mass. Thus, the pressure factor cannot be used for the classification of Chelyabinsk class bodies 
(diameter $>10$~m). The $P\!f$ is also somewhat lower than expected for Ko\v{s}ice. This meteoroid 
larger than 1 meter fragmented heavily in the atmosphere \citep{Kosice}, so the lower $P\!f$ can be ascribed to
many internal cracks. On the contrary, an unusually large $P\!f=4.7$ was obtained for Hamburg. In this case,
flares indicating fragmentation were observed only deep in the atmosphere 
\citep[below heights of 26.5 km, see][]{Hamburg}. It seems that Hamburg was indeed a quite resistant meteoroid. Still,
it is also possible that its initial mass was underestimated.
We estimated the $P\!f$ also for the Carancas crater forming event, considered to be caused by
an unusually strong meteoroid \citep[see e.g.,][]{Brown_Carancas}. Two sets of possible entry parameters 
from \citet{Carancas} were used: [1500 kg, 15 km s$^{-1}$, 16 MPa, 0\degr] and
[10,000 kg, 20 km s$^{-1}$, 40 MPa, 30\degr]. In both cases the result was $P\!f=2.4$, which is not an extreme value.
Nevertheless, the lack of observations makes the Carancas data uncertain.

Next there are two mixed breccias, namely meteorite falls that produced meteorites of various types. The corresponding
meteoroids were probably formed by the accumulation of fragments of several parent asteroids and can be expected
to be less coherent than single-type meteoroids. Their $P\!f$ was indeed lower than those of ordinary chondrites.
In the case of Almahata Sitta, it was near the boundary of the Pf-III and Pf-II classes, which may be connected with the fact that this
body was composed mainly from ureilites, which are achondrites with a lower density than ordinary chondrites. 
\citet{Shaddad} classified the body as ``cometary'' of type IIIA/B but the Pf-classification does not point 
to such a low resistance.

For carbonaceous chondrites, we expect a $P\!f$ between 0.27 and 0.85, corresponding to the class Pf-II. It was so for
Maribo. For Tagish Lake and Flensburg, both quite massive bodies, the $P\!f$ was somewhat lower. The reason for this may be that 
either the $P\!f$ is underestimated for large bodies or that large bodies are indeed composed of weaker
material than smaller bodies of the same origin, as has been observed for some meteor showers (see Sect.~\ref{shower_phys}).

Finally, there are three large meteoroid entries not accompanied with meteorite falls. The Romanian superbolide 
\citep{Romanian} was caused by a relatively homogeneous body that disintegrated completely at dynamic pressures 1--3 MPa. 
The corresponding $P\!f$ is somewhat lower than for carbonaceous chondrites, which is what may be expected.
The brightest Taurid EN311015\_180520 observed by \citet{Spurny_Taurids} in 2015 falls into the Pf-IV class,
which corresponds with the Taurid trend of decreasing strength with increasing size. The internal structure of this
meteoroid was investigated by \citet{Taurid_phys}. The 1974 \v{S}umava superbolide was caused by one of the
most fragile meteoroids ever observed and falls into the class Pf-V. It belonged to the Northern $\chi$ Orionids, part
of the Taurid complex. We note that no massive meteoroids ($>100$ kg) have been observed on Halley-type orbits so far.

In summary, we conclude that the pressure factor $P\!f$ provides a reasonable quick classification of meteoroid
physical properties for stony and cometary meteoroids of masses from 0.1 gram up to about $10^5$ kg. 
As is also the case  for the $P_E$ parameter,
iron meteoroids cannot be identified using $P\!f$ only. Additional information, ideally the spectrum, is needed.

\subsection{Physical properties of meteoroids from various sources}

It is not a new result, but the $P\!f$ classification confirms that physical properties vary among
meteoroids of common origin. The $P\!f$  values span an order of magnitude for all four of the
best represented meteor showers in our sample (Taurids, Geminids, Perseids, and $\alpha$ Capricornids).
This reflects the heterogeneity of their parent bodies.
Still, we can observe clear differences between the showers. In Taurids, there is a pronounced trend of 
decreasing compactness with increasing size of the meteoroid. If we accept that the \v{S}umava
meteoroid was Taurid-related, the whole range from meter-sized extremely fragile
bodies (probably porous dustballs) to centimeter-sized compact objects (stony projectiles) can be observed within that stream.
A similar mass trend may be present in the Perseids and $\alpha$ Capricornids, but centimeter-sized
meteoroids are less compact in these streams than in the Taurids and much larger meteoroids were not observed.
In the Geminids, the mass trend is not present and although there is some overlap with Taurids, they generally
contain more compact, probably denser, meteoroids.

Keeping in mind that neither asteroids nor comets are homogeneous bodies, we can still see
differences between the physical properties of sporadic meteoroids in asteroidal and cometary orbits.
In fact, the classification of physical properties allowed us to define approximate boundaries
between asteroidal and cometary orbital domains. The domains are discussed in more detail
in the following subsections.

\subsubsection{Classical asteroidal orbits, $Q< 4.9$ AU}

We have found that the aphelion distance, $Q$, is a better discriminator between asteroidal
and cometary material than the Tisserand parameter, $T$. Most meteoroids with $Q< 4.9$ AU
have asteroidal properties, although a significant fraction of them have $T<3$ because
they have either high eccentricity or high inclination. Thus, we defined classical asteroidal orbits
as those having  $Q< 4.9$ AU. Since their orbits must intersect the Earth's orbit, they have semimajor axes
$a<3$ AU. 
In our sample, all these orbits were prograde with inclination $i<75\degr$. Of course, retrograde
orbits could not be called classical asteroidal. 
\citet{Kresak} used a similar criterion: $Q< 4.6$ AU.

\begin{figure}
   \centering
   \includegraphics[width=0.875\columnwidth]{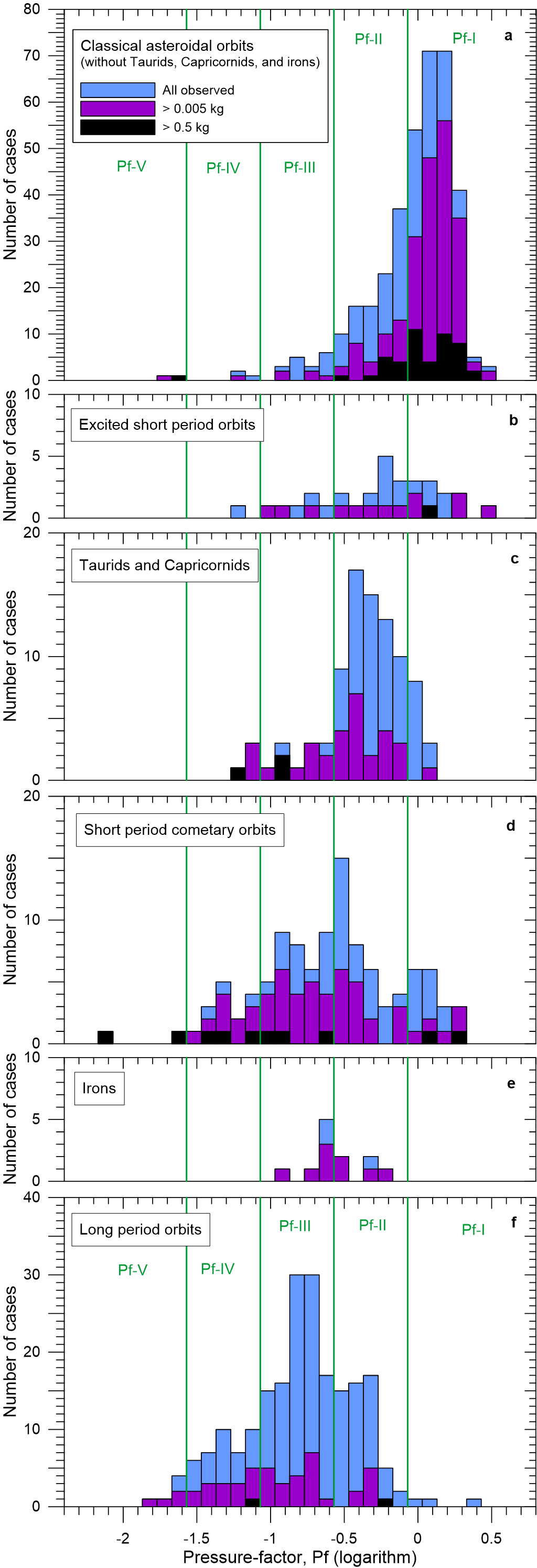}
   \caption{Histograms of pressure factors, $P\!f$, for six orbital classes. 
   Embedded are histograms for meteoroids above a certain mass.
   Division into five classes of ablation ability, Pf-I to Pf-IV, is marked.
   See the Appendix of Paper I for the combined histogram.}
   \label{Pfhists}
   \end{figure}

The histogram of $P\!f$ values for meteoroids on classical asteroidal orbits,
excluding  Taurids, Capricornids ($\alpha$ and $\xi^2$), and irons (known or suspected),
 is shown in Fig.~\ref{Pfhists}a.
Taurids and most Capricornids have classical asteroidal orbits, as have their parent comets.
This situation can be considered as the intrusion of cometary bodies into the asteroidal domain. 
They were therefore removed from the sample, together with irons, to see the properties of  other
meteoroids on classical asteroidal orbits. 
As is shown in Fig.~\ref{Pfhists}a, the majority of them indeed belong
to the most resistant class Pf-I. A significant fraction also fall into the class Pf-II, especially the part with higher $P\!f$ values. 
We suppose that most of them are less resistant asteroidal meteoroids such as carbonaceous chondrites, but
a contribution from comets is possible. Interestingly, there is a long tail of weaker bodies down
to the most susceptible class Pf-V. They are almost certainly of cometary origin.

The histogram for Taurids and Capricornids (dominated by Taurids) is in Fig.~\ref{Pfhists}c.
The majority is of class Pf-II. The larger bodies in particular tend to be weaker and belong
to Pf-III or Pf-IV. On the other hand, some small meteoroids enter the Pf-I region.

Irons, as already discussed, cannot be well classified using the $P\!f$ value. Formally, they
fall into the Pf-III or the Pf-II class (Fig.~\ref{Pfhists}e).

\subsubsection{Asteroidal material on excited orbits}

One of the most surprising findings of this work is the presence or even prevalence 
of resistant material on orbits that can be considered cometary. These are orbits
with aphelia $Q>4.9$~AU, semimajor axes $a<5$~AU, and either high
eccentricities (and thus low perihelia) or high inclinations. We have set the boundaries at 
$e>0.9$ or $i>40\degr$.  If both parameters are below these limits, we call
the orbit a short-period cometary orbit. If at least one of the parameters is above the limit,
we call the orbit an excited short-period orbit. The Tisserand parameters of the excited orbits
are $T<3$, and in many cases even $T<2$. 

In total, 31 meteoroids were observed on excited orbits.
The histogram of their $P\!f$ values is given in Fig.~\ref{Pfhists}b.
The distribution is flat but most meteoroids belong to classes Pf-I and Pf-II. 
From seven Pf-III or Pf-IV meteoroids, three have $a>4.5$~AU. 
All Pf-I meteoroids have $2.6<a<3.8$~AU. This is therefore the most interesting region.

It is remarkable that two of three meteoroids in the whole sample with sodium-deficient spectra
fall in this region. Both EN310718\_213217 and EN111118\_221402 had $Q>5$~AU, $a>3$~AU, and $i>70\degr$.
The third meteoroid deficient in sodium (EN140518\_005437) had a similar orbit but inclination was only 30\degr, 
so the orbit was classified as short-period cometary. Nevertheless, the meteoroid was very strong with $P\!f=2$ and 
can be considered as an intruder from the asteroidal population.
Its spectrum was dominated by iron, while the other two sodium-deficient spectra were dominated by magnesium.

   \begin{figure}
   \centering
   \includegraphics[width=1.0\columnwidth]{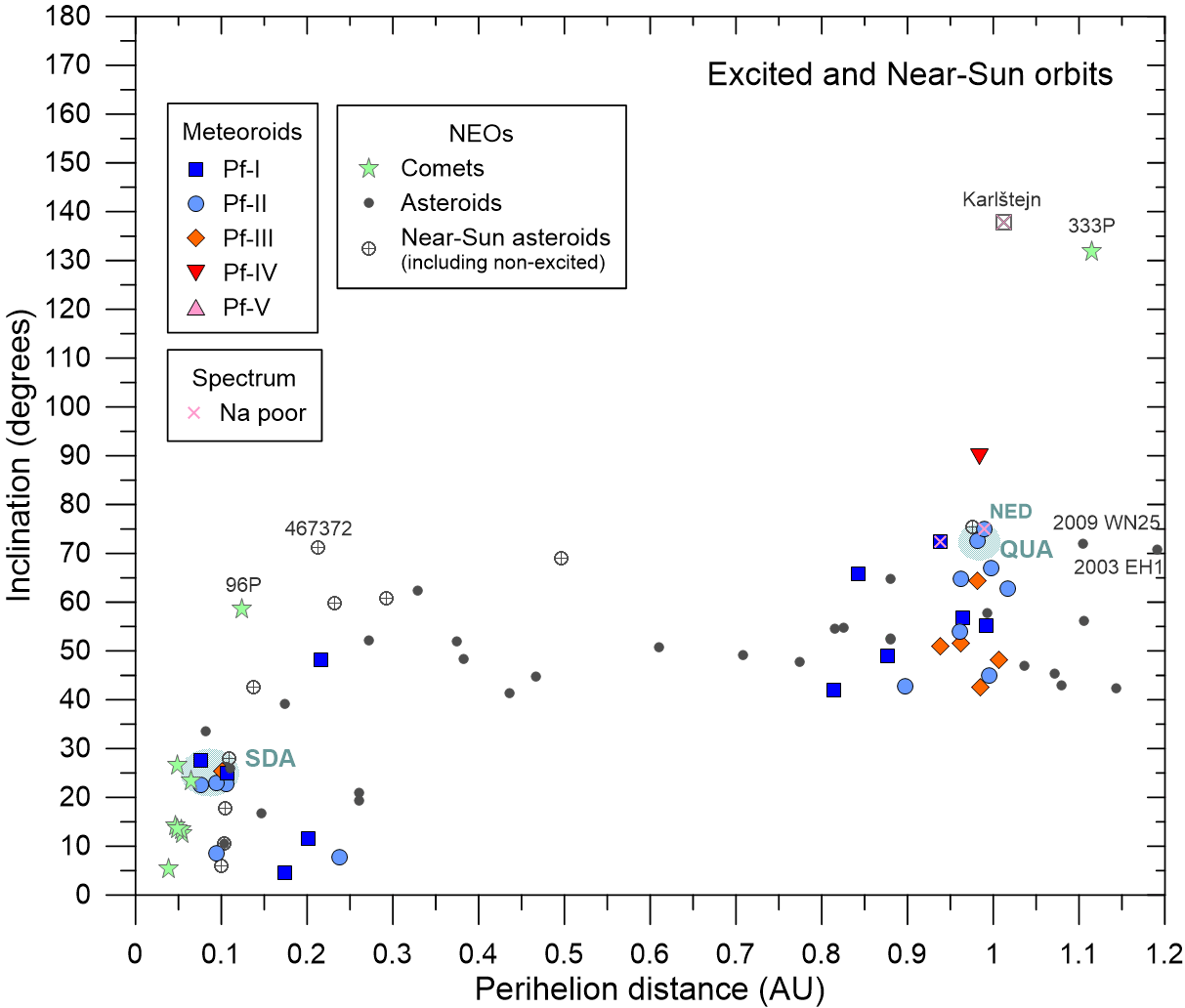}
   \caption{Inclinations and perihelion distances of meteoroids, comets, and asteroids on excited orbits 
   (i.e.,\ with $Q>4.9$~AU, $a<5$~AU, and either $e>0.9$ or $i>40\degr$). Meteoroid
   Pf classes are coded by different symbols, as shown in the legend. 
   Meteoroids with Na-poor spectra are marked. 
   In contrast to Fig.~\protect\ref{a-i}, meteoroids of all masses and showers are included. 
   The regions of Southern $\delta$ Aquariids (SDA) and Quadrantids (QUA), 
   which overlap with  November Draconids (NED), are highlighted. Near-Sun asteroids according to
   \citet{Emelya} are included and shown by different symbols, even if they have $Q<4.9$.
   The strong Karl\v{s}tejn meteoroid observed in 1997 \citep{Karlstejn2} is also included. For asteroid 2009 WN25,
   the orbit valid in 2018 was used.}
   \label{q-i}
   \end{figure}

The origin of the asteroidal material on excited orbits is not clear. 
Despite the wide range of eccentricities and inclinations, the material may form one population.
Figure~\ref{q-i} shows the inclinations plotted against perihelion distances. Comets and asteroids with
orbits satisfying our definition of excited orbits are plotted as well. Although asteroids in these orbits are not 
numerous, meteoroid orbits overlap better with asteroids than comets. There are only SOHO comets
with very low perihelion distances and comet 96P/Machholz 1, which has a higher inclination than meteoroids at
similar perihelion distances, in this region.

The meteoroids on excited orbits also include the Southern and Northern $\delta$ Aquariids and Quadrantids 
(there is only one Quadrantid in our sample).
With the exception of one small meteoroid, members of these showers belong to classes Pf-I and Pf-II. As shown by \citet{Sekanina},
the Southern $\delta$ Aquariids, Quadrantids, SOHO comets of Marsden and Kracht groups, comet 96P, and asteroid 2003
EH1 have a common origin and form the Machholz interplanetary complex. 

Our sample also contains one
possible November Draconid (753 NED), a shower possibly associated with asteroid 2009 WN25 \citep{Segon2015}.
The reflectance spectrum of 2009 WN25 suggests it has cometary properties \citep{Ieva2019} but no activity was observed.
\citet{Micheli2016} associated 2009 WN25 with another shower, the November i Draconids (392 NID), but the orbital correspondence
with NED is better. Interestingly, the meteoroid is one of those deficient in sodium (EN111118\_221402) and with
$P\!f=0.8,$ it was rather strong.

Since the Machholz interplanetary complex was probably formed by cometary fragmentation \citep{Sekanina}, and asteroids
such as  2003 EH1 and 2009 WN25 are supposed to be inactive cometary fragments, there is a conflict with the physical
properties of meteoroids.  Figure~\ref{Pfhists} shows that meteoroids in excited orbits (even with some cometary admixture)
are stronger than Taurids and meteoroids in cometary orbits. One possible explanation could be that 
close approaches to the Sun lead to the loss of volatile sodium and to compaction of the bodies 
(or destruction of those that are not compact). For comparison, Fig.~\ref{q-i} shows asteroids 
(including those with $Q<4.9$~AU) that
according to \citet{Emelya} have recently approached the Sun within 0.1~AU. Due to the Kozai–Lidov secular perturbations,
they can now have large perihelia but in that case they  have high inclinations. One of them, 467372 (2004 LG), survived a perihelion distance 
of only 5.6 solar radii (0.026 AU), the nearest distance of any known asteroid \citep{2004LG}, about 3000 years ago.
96P/Machholz was at $q\sim 0.07$ AU 4000 years ago \citep{Gonzi1992}.

The problem with solar heating is that many meteoroids on excited orbits now have large perihelia and only those with inclination
above 70\degr\ could have approached Sun within 0.1 AU in the past  \citep[using the formula for $q^\prime$ in][]{Emelya}.
Only two meteoroids with depleted sodium and one Quadrantid with an unknown spectrum have such large inclinations. 
Volatile sodium can be lost
near the Sun due to thermal desorption \citep{Capek}. But we also have spectra of one Southern and one Northern $\delta$ Aquariid 
with low perihelia, and they
do contain sodium. \citet{Rudawska_spectra} observed one moderately bright $\delta$ Aquariid meteor with a strong sodium line
and one with weak sodium. Fainter $\delta$ Aquariids were found to be completely Na-free by \citet{icarus05}, who observed
that many small meteoroids approaching the Sun within 0.2~AU have lost their sodium. The Geminids lose part of their sodium
and the depletion increases with decreasing meteoroid size \citep{iaus263,Abe_Gem}. Small Quadrantids have lost
sodium to lower extent than Geminids \citep{Koten_Qua, iaus263} while the larger ones have a normal sodium
content \citep{Madiedo_Qua}.

It is therefore common that small millimeter-sized shower meteoroids approaching the Sun are losing their sodium.
In larger centimeter-sized shower meteoroids, sodium loss was not detected. Sodium loss can be expected to depend on meteoroid size,
internal structure (grain size, porosity), distance to the Sun, and exposure time \citep{Capek}. Since the atmospheric behavior
of two large meteoroids depleted in sodium that we observed does not support their large porosity and the orbit does not suggest 
a particularly close approach to the Sun in the past (closer than for $\delta$ Aquariids), the only way to reconcile the difference seems
to be a long exposure time to solar heat. The shower meteoroids are probably younger (i.e.,\ were released from their parent bodies
more recently).

Still, it is not clear whether solar heating can explain asteroidal properties of many meteoroids on excited orbits. Some of them
probably never approached the Sun closer than 0.2 -- 0.3 AU. Since the semimajor axes of most meteoroids in question are within
2.5 -- 3.5 AU, their origin in the middle and outer asteroid belt seems possible. Their eccentricities could have been
pumped up by mean motion resonances with Jupiter and some of them could have then gained high inclination due to the Kozai-Lidov
perturbation. 

\citet{Wiegert} discussed the possibility that the Marsden and Kracht groups of SOHO 
comets are in fact asteroidal fragments subject to the process of ``supercatastrophic disruption'' proposed by \citet{Granvik}.
They finally preferred cometary origin because of the cometary nature of 96P/Machholz. Nevertheless, asteroids
in the outer belt also contain volatiles \citep{MBC} and can probably behave similar to comets when moved closer to the Sun.

In our view, the physical properties of meteoroids on excited orbits can be explained by the following hypotheses: 
(i) the parent bodies of a significant part or even the majority of the meteoroids on excited orbits 
came from the asteroid belt (middle or outer).  The rest originated in Jupiter-family comets; (ii) most of the meteoroids on excited orbits either currently approach
the Sun, or approached it in the past, within $\approx0.3$ AU. More fragile meteoroids had a higher chance of being destroyed in the
vicinity of the Sun \citep{Wiegert}, so meteoroids representing the stronger parts of the parent bodies prevailed. 
The existence of this process is supported by the general  lack of weak
centimeter-sized meteoroids with low perihelia (see Fig.~\ref{perihel}); and (iii) meteoroids exposed to strong solar heat for a long time 
can lose their sodium even if they are relatively large (centimeter-sized).

All orbits with $a<5$ AU in our sample are prograde. However, the European Fireball Network observed a retrograde
meteoroid, designated EN010697 Karl\v{s}tejn, with $a=3.49\pm0.09$~ AU in 1997 
\citep[][see also Fig.~\ref{q-i}]{Karlstejn1,Karlstejn2}. 
That unique fireball was classified as type~I, that is as a  strong body, and sodium was missing in its spectrum.
Its physical and chemical properties were, therefore, analogous to the extreme members of our population on excited
orbits. \citet{Greenstreet} found that a small fraction of near-Earth asteroids can evolve into retrograde orbits and the process
often involves periods of low perihelion distances and periods of high inclinations ($45\degr<i<80\degr$). The existence
of Karl\v{s}tejn is therefore in accordance with the hypothesis of asteroidal origin of meteoroids on excited orbits.

The question may arise why, between 3 and 5 AU, resistant meteoroids dominate at high inclinations. One reason is
that the low-inclination region contains not only asteroidal material but also young material from Jupiter-family comets. 
The high-inclination region contains older material that passed through the phase of low perihelion and then gained inclination 
through the Kozai-Lidov perturbation. Our data suggest that resistant asteroidal meteoroids have
a higher chance of enduring during this process.

\subsubsection{Jupiter-family orbits}

The domain of Jupiter-family or, equivalently, short-period cometary orbits is in our concept at $Q>4.9$ AU,
and $a<5$ AU, and $e<0.9$, and $i<40\degr$. The histogram of $P\!f$ values for meteoroids  on these
orbits is shown in Fig.~\ref{Pfhists}d. The distribution is wide, covering all classes from Pf-I to Pf-V.
For small meteoroids, the maximum of the distribution is at the boundary between Pf-II and Pf-III.
For meteoroids larger than 5~g, the maximum is shifted to Pf-III. The largest bodies ($> 0.5$ kg) are even 
more fragile. There are only two meteoroids in class Pf-V and both are large, but there are also meteoroids of various masses in the strongest class, Pf-I.

The $P\!f$ values demonstrate that most meteoroids on Jupiter-family orbits are of cometary origin. 
Therefore, the main meteoroid source is Jupiter-family comets. The trend of more massive
meteoroids being more fragile is the same as the trend observed for the Taurids. Both small and large meteoroids are, nevertheless,
shifted to lower $P\!f$ in comparison with Taurids. 
The material is therefore weaker than 2P/Encke but comparable to 169P/NEAT, the parent of $\alpha$
Capricornids. Some fraction of meteoroids on Jupiter-family orbits is strong and probably of mostly asteroidal origin. 
These orbits are intermediate between classical asteroidal and excited orbits. Asteroidal material
from both of them can enter the Jupiter-family domain. We note that strong meteoroid EN140518\_005437
with an unusual chemical composition, deficient in sodium and rich in iron, was on a Jupiter-family orbit but it
most likely originated in the asteroid belt.

\subsubsection{Long-period orbits}

We call long-period orbits those with semimajor axis $a>5$ AU. The histogram of $P\!f$ values
in Fig.~\ref{Pfhists}f shows that physical properties of meteoroids with these orbits are cometary. 
The maximum of the distribution is at class Pf-III and
the trend of large meteoroids being weaker is present as well. The material in the strongest class,
Pf-I, is rare and in this case may represent  small, strong inclusions in long-period comets.

\subsection{Comparison with other authors}

\citet{Kresak,Kresak2} looked for differences between orbital and physical characteristics of cometary
and asteroidal meteors. He considered the orbits with $i<30\degr$, $q>0.9$ AU, and $Q<4.6$ AU 
to contain the highest proportion of asteroidal meteors, and those with $i<30\degr$, $q>0.9$ AU 
and $4.6<Q<10$ AU 
to contain exclusively cometary meteors. He found statistically significant differences in deceleration
and beginning heights between these two samples. This corresponds with our finding that asteroidal
material prevails on orbits with lower aphelia, although we proposed a slightly different boundary, $Q=4.9$ AU,
and there is clearly overlap between the two populations.
The data used by \citet{Kresak2} suggested concentrations of orbits near mean motion resonances with Jupiter, 
especially 3:1 and 4:1. We detected possible concentrations near 1:1 and 3:2 resonances, but the other could not be confirmed.

\citet{Cep88} studied meteoroid populations defined, in the range of large meteoroids, by the PE criterion.
Only mean orbits were given for each  population. The largest difference was in the mean eccentricity,
which was found to be smallest ($\sim$\,0.6) for types I and II, intermediate ($\sim$\,0.7) for type IIIB, and largest for type IIIA.
This view is obviously too simplistic. The main purpose of the \citet{Cep88} paper was to provide absolute
fluxes for each population as a function of mass. He found that cometary material prevails at masses of $\sim$\,10
grams while carbonaceous material is more abundant for both smaller (down to about 0.1 gram) and larger
masses. The contribution of stony material was found to be negligible below 1 gram and most important between 1 -- 100 kg.
We did not compute absolute or even relative fluxes, but our data confirm that bodies $>0.5$~kg 
are mostly strong (see Fig.~\ref{Pfhists}).

\citet{Halliday} presented data for 259 fireballs observed by the Canadian camera network in 1971--1985.
In the data analysis, they concentrated mainly on the absolute flux estimates. Most meteoroids in the mass range 0.1 -- 10 kg
were classified as ``asteroidal''. In their definition, the ``asteroidal group'' included all fireballs with entry velocities lower than
25 km s$^{-1}$, with the exception of shower fireballs. For the part of the fireballs with sufficient deceleration and without
obvious fragmentation, they estimated meteoroid densities. From those estimates, they concluded that as many as a quarter of 
the members of their ``asteroidal group'' are probably of cometary origin. On the other hand, their ``cometary group''
with velocities above 25 km s$^{-1}$ contained some high-density objects with small perihelion distances. 
These observations do not contradict our results.

\citet{Matlovic} published an analysis of 202 bright meteors and fireballs observed by all-sky video cameras. They concentrated
mostly on spectra, but orbits and physical properties of 146 meteoroids were also determined. 
Their plot of the Tisserand parameter, $T$, against the $P_E$ parameter shows the same trends as our Fig.~\ref{Tisserand}a. 
Namely, strong type~I fireballs are encountered on orbits with $T>2$ and only rarely on Halley-type orbits with $T<2$. 
Jupiter-family orbits
with $2<T<3$ contain both strong meteoroids, some of which are deficient in sodium, and very weak meteoroids, which are
mostly large.  A more detailed orbital analysis was not done. The physical properties were also evaluated using the $K_B$ parameter,
but this parameter was originally developed for smaller meteoroids \citep[see][]{Cep88} and gave a more confusing picture.
\citet{Matlovic} also noted heterogeneity in the physical properties of Perseids and other meteoroid streams.
Generally, their data are in agreement with our results in the parts where both works overlap.

   \begin{figure}
   \centering
   \includegraphics[width=1.0\columnwidth]{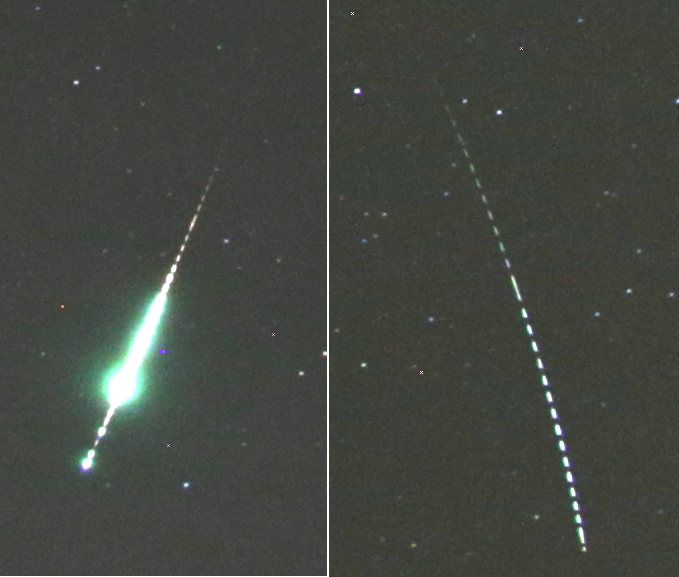}
   \caption{Comparison of the cometary fireball EN310317\_023002 (left) and the asteroidal fireball EN181017\_231532
   (right). Both meteoroids  were on cometary-type orbits with $(a,e,i) = (3.01, 0.68, 23\degr)$ and 
   (3.04, 0.71, 49\degr), respectively. The cometary fireball appeared at heights of 98 -- 61 km and reached
   a maximum absolute magnitude of $-13.2$. The meteoroid photometric mass was 6.7~kg. The asteroidal
   meteoroid was much smaller (0.04~kg) and the fireball was fainter ($-7.7$ mag) but longer (89 -- 37 km).
   The slopes of the trajectories were almost identical (28\degr\ and 29\degr, respectively, to the vertical).}
   \label{com-ast}
   \end{figure}

\citet{Wiegert} investigated meteoroids with low perihelia following the finding of \citet{Granvik} that there is a deficit of
 low-perihelion asteroids. \citet{Granvik} concluded that a super-catastrophic destruction of asteroids occurs 
 when they reach $q = 0.076$ AU. \citet{Wiegert} suggested an even stricter limit for smaller bodies, since in
the sample of  59 meter-sized meteoroids they used, 
none had perihelion lower than 0.35~AU. On the contrary,
an excess of low-perihelion millimeter-sized meteoroids (observed by meteor radar) was found.
They proposed that asteroids and large meteoroids disintegrate into millimeter-sized particles as a result of high-speed 
collisions with small meteoroids near the Sun. We observe nearly the same limit, 0.07 AU, for centimeter-sized meteoroids as the one
found by \citet{Granvik} for asteroids.
Meteoroids with perihelia below 0.35 AU are relatively strong and, 
interestingly, the sporadic ones with $q<0.11$ AU have perihelia
concentrated in a limited range of longitudes, corresponding to the region of lowest concentration of small
cometary meteoroids according to \citet{Wiegert}.
If meteoroid impacts are indeed eroding meter-sized objects below 0.35 AU, 
the medium-sized meteoroids we observe there may be intermediate products of this process. However, further work is needed
to address this question.

\citet{Shober} studied meteoroids on Jupiter-family orbits using data from the Desert Fireball Network.
In contrast with previous works and also our results, they conclude that less than 4\% of sporadic meteoroids larger
than 10 grams on these cometary-type orbits are of cometary origin. However, the methodology of \citet{Shober}
contains flaws. They used the $P_E$ criterion but substituted dynamic mass into the formula. More severely, the
necessity to compute dynamic mass necessarily prevented them from including fireballs with only mild deceleration 
in their sample. Their approach resulted in bias since it is the cometary meteoroids that exhibit little deceleration. 
Figure~\ref{com-ast} shows
two fireballs with similar orbits, one classified as cometary and one as asteroidal according to the pressure factor.
Despite having a larger initial mass, the cometary fireball was shorter, contained many flares, and the terminal velocity
was not very different from the entry velocity (18.9 vs. 20.0 km s$^{-1}$). 
The asteroidal fireball was longer, penetrated deeper into the atmosphere, and the
deceleration was significant (from 32.5 to 13.9 km s$^{-1}$). We conclude that the unexpected result of \citet{Shober} 
can be ascribed to the
bias in their data. In fact, most orbits they included would be classified as asteroidal in our view,
since despite having a Tisserand parameter $<3$, they had aphelia $<4.9$ AU and/or eccentricities $>0.9$.
It is not surprising that these meteoroids had asteroidal properties. However, the authors
almost completely missed the typical cometary population at $a\sim3 - 3.5$ AU and $e\sim 0.7$, which are clearly present in our 
data (see Fig.~\ref{a-e}).

In their review, \citet{Hajdukova} discussed  the challenge of identifying interstellar meteors, that is meteors
on hyperbolic orbits. They demonstrated that although meteor databases contain a significant fraction (up to 12\%) of
hyperbolic orbits, and many of them with large eccentricities, none can be confirmed to be truly interstellar 
because of observational errors. Our sample contains 18 hyperbolic orbits (2.2\%), the lowest fraction 
of all datasets quoted by \citet{Hajdukova}, and their eccentricities do not exceed 1.05. It is a consequence of the
better precision of our data. Only two orbits remain hyperbolic within the three sigma error. These, and perhaps some others,
may be truly hyperbolic but their slight hyperbolicity was very probably acquired by processes within the
Solar System. We therefore confirm the non-detection of interstellar meteoroids in the centimeter to decimeter size range.
A larger sample of precise fireball orbits must be collected to set limits on their frequency.

Radiants and orbits of meteor showers have been directly compared to other sources in Sect.~\ref{showersection},
and are not discussed here. Similarly,
the frequency of meteorite falls was compared with the literature in Sect.~\ref{meteoritesection}.

\section{Conclusions}

The purpose of this extended work was the presentation and analysis of data on 824 fireballs observed
by the digital cameras of the European Fireball Network in 2017-2018. Arguably, this is the most precise set
of fireball orbits, atmospheric trajectories, and photometric data published so far. Despite covering only two years of data,
a number of new findings were obtained. The main emphasis was on the combination of orbital and physical properties,
partly also supplemented with spectral data, with the aim of revealing source regions of various types of meteoroids.
The most important findings are:
\begin{enumerate}

\item The aphelion distance, $Q$, is a better indicator of asteroidal origin of Earth-crossing meteoroids
than the Tisserand parameter. 
The boundary is not strict, but sporadic meteoroids with $Q<4.9$ AU are more likely to be asteroidal
than cometary. We call such orbits classical asteroidal.

\item There are orbits containing dominantly asteroidal (i.e.,\ physically strong) material even with $Q>4.9$ AU.
These are orbits with either high inclination ($i \ga 40\degr$) or high eccentricity ($e \ga 0.9$), and thus low perihelion distance. In both cases, the semimajor axis is $a<5$ AU. We call these orbits excited 
and speculate that the material originated
in the outer asteroidal belt. Part of this population is the Machholz complex containing the meteoroid streams
$\delta$ Aquariids and Quadrantids.

\item All other orbits contain mostly cometary meteoroids. They can be divided into short-period or Jupiter-family
orbits ($a<5$ AU, $Q\ge4.9$ AU, $i\le40\degr$, and $e\le0.9$), and long-period or Halley-type orbits ($a\ge5$ AU,
any inclination and eccentricity).

\item There are important intrusions of cometary material into classical asteroidal orbits connected with comet 2P/Encke,
forming the Taurid complex, and comet 169P/NEAT, forming the Capricornid complex. Other cometary meteoroids
can be encountered on asteroidal orbits as well.

\item Similarly, some meteoroids of asteroidal origin are encountered on Jupiter-family orbits.

\item On the other hand, material of high strength is rarely encountered on Halley-type orbits. 
If so, it probably represents strong inclusions contained in comets.

\item Parent bodies of meteoroid streams are not homogeneous and contain material with a relatively wide
range of strengths. There is a trend, which is pronounced especially in Taurids but it is probably also present in other cometary
streams, that large meteoroids are weaker than small ones \citep[see also][]{Taurid_phys}.

\item There are, nevertheless, differences among average physical properties of meteoroids from different
streams. The differences are more evident when large meteoroids are compared (with masses of at least tens of grams).
Capricornids and Leonids are among the weakest, and Geminids and $\eta$ Virginids are among the strongest
meteoroids.

\item Centimeter-sized iron meteoroids in our sample were encountered only on classical asteroidal orbits, 
some of them with high inclinations or low perihelia. We note that
\citet{Vojacek} detected two small iron meteoroids on excited orbits ($a\sim3.5$ AU, $i\sim 63\degr$)
and \citet{Abe_Gem} observed three iron Geminids.

\item A couple of sodium-poor meteoroids were detected on excited orbits (with high inclination),
and one sodium-poor but iron-rich meteoroid was detected on a Jupiter-family orbit. 
All three were physically strong.

\item No meteoroid with perihelion distance, $q$, lower than 0.07 AU was detected. Interestingly, all seven
sporadic meteoroids with $q<0.11$ had longitudes of perihelia, $\tilde{\omega}$, between 190\degr\ and 290\degr.
There were also six $\delta$ Aquariids with $95\degr<\tilde{\omega}<115\degr$. All meteoroids with
low perihelia were relatively strong. 

\item No clearly interstellar meteoroid was detected. A few of the observed meteoroids 
may have been accelerated to hyperbolic orbits within the Solar System.

\item There seems to be an excess of meteoroids in 1:1 and 3:2 mean motion resonances with Jupiter.
All these resonant meteoroids have inclinations higher than 10\degr. 

\item The orbits of Geminid meteoroids can be divided into two groups, forming a core and a wing of the stream.
The core has a semimajor axis similar to Phaethon but a somewhat larger perihelion distance. The wing
meteoroids have semimajor axes  larger than Phaethon and perihelia similar to Phaethon. The wing meteoroids have slightly
larger inclinations, on average, than those in the core.

\item The central part of the Perseid meteoroid stream is concentrated in inclination around the
inclination of the parent comet 109P/Swift-Tuttle but contains some meteoroids with markedly lower
perihelion distances. On the other hand, the outer part (encountered well before the shower maximum)
has larger spread in inclination but a lower range of perihelion distances.

\item As already pointed out by \citet{Spurny_Taurids} and \citet{Egal}, the Taurid meteoroid stream
contains several cataloged asteroids. We suggest that there may be a  similar situation with the $\alpha$
Capricornid stream.

\item Meteor showers $\nu$ Draconids (220 NDR) and $\xi^2$ Capricornids (623 XCS), 
which are currently on the working list, were confirmed. 
Another detected shower best corresponds to the Southern October $\delta$ Arietids (28 SOA).
August $\mu$ Draconids (470 AMD) were probably detected as well.

\item Fireballs with end heights below 32 km and terminal velocities below 7.5 km s$^{-1}$ were found to
be candidates for meteorite falls. Nevertheless, an analysis of the light curve and deceleration 
toward the end is needed in each individual case to confirm that significant mass has fallen.

\end{enumerate}
Some of these findings will need confirmation or deeper elaboration using more data and more focused studies.
In any case, it is encouraging that the digital cameras of the European Fireball Network are providing
good data, enabling us to better understand the population of centimeter- to decimeter-sized meteoroids in the 
vicinity of the Earth. As noted in Paper I, observations are continuing and more data will be published in the future.

\begin{acknowledgements}
This work was supported by grant no.\ 19-26232X from Czech Science Foundation and
Praemium Academiae of the Czech Academy of Sciences, which provided funds for digitization of the part
of the European Fireball Network. 
The institutional research plan is RVO:67985815.
\end{acknowledgements}

\begin{appendix}

\section{Additional tables}
\label{atables}

\begin{table}[h]
\caption{Fireballs possibly associated with the Northern or Southern $\delta$ Cancrids (96 NCC and 97 SCC).
Selected orbital elements and the pressure factor are given. Orbital elements from the literature are given at the bottom.}
\label{NCC}
\begin{small}
\begin{tabular}{llllll}
\hline\hline  \noalign{\smallskip}
Code & $q$ & $e$ & $i$ & $\Omega$ & $P\!f$   \\
\hline \hline  \noalign{\smallskip}
  EN070117\_235400 & 0.365& 0.774&  0.57& 287.45&    1.0     \\
  EN200117\_181106 & 0.546& 0.753&  2.77& 120.69&    1.3     \\
  EN200117\_192130 & 0.520& 0.768&  2.03& 120.75&    1.6     \\
  EN200117\_234318 & 0.528& 0.775&  0.09& 299.71&    2.0     \\
  EN290117\_033628 & 0.521& 0.806&  1.44& 129.25&   0.22     \\
  \hline \noalign{\smallskip}
  $\delta$ Cancrids [1] & 0.45 & 0.80 & 0 & 296 \\
  NCC [2] & 0.410 &     0.814 & 2.7 &   290.0   \\
  SCC [2] & 0.430 &     0.811 & 4.7 &   109.3 \\ 
\hline \hline
\end{tabular}
\end{small}
\tablefoot{[1] \citet{Cook}, [2] \citet{CAMS}}
\end{table}

\begin{table}[h]
\caption{Fireballs associated with the $\sigma$ Hydrids (16 HYD), July $\gamma$ Draconids  (184 GDR),
 and $\kappa$ Ursae Majorids (445 KUM).}
\label{HYD}
\begin{small}
\begin{tabular}{llllll}
\hline \hline  \noalign{\smallskip}
Code & $q$ & $e$ & $i$ & $\Omega$ & $P\!f$   \\
\hline \hline  \noalign{\smallskip}
EN071217\_025105 & 0.234& 0.973&126.92&  74.94&   0.40     \\
EN071217\_230112 & 0.236& 0.999&127.89&  75.79&   0.40     \\
  \hline \noalign{\smallskip}
HYD [1] & 0.237 & 0.978 &       126.5 & 79.1 \\
HYD [3] & 0.257 & 0.985 &        128.7 &        76.5      \\
HYD [4] & 0.249 & 0.970 & 128.3 & 78.47 \\
\hline \hline \noalign{\smallskip}
EN280717\_235801 & 0.980& 0.999& 41.72& 125.85&   0.42     \\
EN290717\_212641 & 0.980& 0.993& 40.19& 126.70&   0.39     \\
\hline \noalign{\smallskip}
GDR [2] & 0.978  & 0.972 & 40.24 &      124.66 \\
GDR [3] & 0.977  & 0.967 &40.3  & 124.7 \\
GDR [4] & 0.978 & 0.966 &40.28 &  125.62 \\
\hline \hline \noalign{\smallskip}
EN051118\_215131 & 0.987& 0.981&129.98& 223.17&   0.032 \\
EN061118\_190859 & 0.988& 0.986&129.38& 224.06&   0.038 \\
\hline \noalign{\smallskip}
KUM [3] & 0.988 & 1.000 &        129.6 & 224.0 \\
KUM [4] & 0.983 & 0.890 & 129.0 & 224.49 \\
\hline \hline \noalign{\smallskip}
\end{tabular}
\end{small}
\tablefoot{[1] \citet{Jopek2003}, [2] \citet{GDR}, [3] \citet{CAMS}, [4] \citet{Vida_GMN}.}
\end{table}

\begin{table}[h]
\caption{Fireballs possibly associated with the $\nu$ Draconids (220 NDR).}
\label{NDR}
\begin{small}
\begin{tabular}{llllll}
\hline \hline  \noalign{\smallskip}
Code & $q$ & $e$ & $i$ & $\Omega$ & $P\!f$   \\
\hline \hline  \noalign{\smallskip}
EN100917\_225415 & 1.006& 0.665& 33.35& 168.17&   0.30     \\
EN090918\_000545 & 1.007& 0.675& 35.00& 166.03&   0.41     \\
EN110918\_231100 & 1.005& 0.700& 27.53& 168.91&   0.099 \\
EN170918\_010734 & 0.998& 0.665& 24.92& 173.86&   0.30     \\
EN180918\_030212 & 1.002& 0.606& 30.17& 174.91&    1.4     \\
  \hline \noalign{\smallskip}
NDR [1] & 1.004 & 0.654 &  28.8 & 171.7 \\
\hline \hline \noalign{\smallskip}
\end{tabular}
\end{small}
\tablefoot{[1] \citet{CAMS2}.}
\end{table}

\begin{table}
\caption{Fireballs possibly associated with the Southern $\delta$ Piscids (216 SPI),
Southern October $\delta$ Arietids (28 SOA), or $\xi$ Arietids (624 XAR).}
\label{SPI}
\begin{small}
\begin{tabular}{llllll}
\hline \hline  \noalign{\smallskip}
Code & $q$ & $a$ & $i$ & $\Omega$ & $P\!f$   \\
\hline \hline  \noalign{\smallskip}
EN150918\_231506 & 0.234& 2.31&  8.36& 352.80&    1.8     \\
EN290917\_030848 & 0.243& 1.85 &  7.61&   5.94&   0.83     \\
EN081018\_224959 & 0.234& 1.75&  7.17&  15.35&   0.51     \\
EN101018\_023945 & 0.256& 1.93&  6.34&  16.50&   0.79     \\
EN131018\_020534 & 0.282& 1.44&  4.97&  19.45&   0.67     \\
EN151017\_012737 & 0.262& 1.71&  6.14&  21.66&   0.57     \\
  \hline \noalign{\smallskip}
SPI [1] & 0.251 & 1.45 &        5.6     &       353.9   \\
SOA [1] &  0.286 & 1.75 &       5.7 &   15.4 \\
XAR [1] &       0.312 & 1.86 &  5.7  &  24.4     \\
\hline \hline \noalign{\smallskip}
\end{tabular}
\end{small}
\tablefoot{[1] \citet{CAMS}.}
\end{table}

\begin{table}
\caption{Fireballs possibly associated with the August $\mu$ Draconids (470 AMD) and
$\xi^2$ Capricornids (623 XCS).}
\label{AMD}
\begin{small}
\begin{tabular}{llllll}
\hline \hline  \noalign{\smallskip}
Code & $q$ & $e$ & $i$ & $\Omega$ & $P\!f$   \\
\hline \hline  \noalign{\smallskip}
EN140817\_193344 & 1.013& 0.658& 35.79& 141.95&   0.29     \\
EN240817\_235428 & 1.010& 0.657& 34.78& 151.75&   0.19     \\
EN260817\_185317 & 1.006& 0.677& 31.97& 153.48&    1.1     \\
  \hline \noalign{\smallskip}
AMD [1] & 1.011 & 0.654 &        30.3 & 145.4   \\
AMD [2] & 1.012 & 0.631 &   29.5 &      144.4  \\
AMD [3] & 1.009 & 0.648 & 33.8 &        149.5  \\
\hline \hline \noalign{\smallskip}
EN190717\_002210 & 0.463& 0.796&  7.47& 116.32&   0.069 \\
EN190717\_011014 & 0.471& 0.805&  8.02& 116.35&   0.18    \\
EN190718\_223059 & 0.476& 0.805&  7.94& 116.96&   0.058 \\
  \hline \noalign{\smallskip}
XCS [3] & 0.509 & 0.786 &7.6 &  119.7       \\
\hline \hline \noalign{\smallskip}
\end{tabular}
\end{small}
\tablefoot{[1] \citet{Rudawska}, [2] \citet{Kornos}, [3] \citet{CAMS}.}
\end{table}

\begin{table}
\caption{Orbit of asteroid 2019 DN and three fireballs with similar orbits.}
\label{2019DN}
\begin{small}
\begin{tabular}{lllllll}
\hline \hline  \noalign{\smallskip}
Code & $a$ & $q$ & $i$ & $\omega$ & $\Omega$ & $D_{\rm SH}$   \\
\hline \hline  \noalign{\smallskip}
2019 DN                 &2.43  & 1.01   &  2.9   &162.2  &357.0 \\
EN240317\_212628 &2.41   &0.99    &1.0    &167.6   & 4.2  &0.045 \\
EN310317\_015155  &2.20  &1.00    &1.3    &175.9   &10.3  &0.050\\
EN040318\_224701  &2.34  &0.99    &1.9    &177.0   &344.0 & 0.035\\
\hline \hline \noalign{\smallskip}
\end{tabular}
\end{small}
\end{table}

\end{appendix}

\end{document}